\documentclass[
    aps,
    prl,
    reprint,
    superscriptaddress,
    amsmath,
    amssymb,
    longbibliography,
    nobibnotes
]{revtex4-2}
\usepackage{bbm}
\usepackage{multirow}
\usepackage{siunitx}

\usepackage{minted}
\usepackage{bm}
\usepackage{amsmath}
\usepackage{amssymb}
\usepackage{bm}
\usepackage{graphicx}
\usepackage[dvipsnames]{xcolor}
\usepackage{hyperref}
\usepackage{bibunits}
\defaultbibliographystyle{apsrev4-2}
\defaultbibliography{collection}

\hypersetup{
    colorlinks=true,
    linkcolor=BrickRed,
    citecolor=Violet,
    urlcolor=RoyalBlue
}

\definecolor{darkgreen}{rgb}{0,0.4,0}
\definecolor{darkred}{rgb}{0.4,0,0}
\definecolor{darkblue}{rgb}{0,0,0.4}
\definecolor{lightblue}{rgb}{0.6,0.6,0.9}

\renewcommand{\theequation}{\arabic{equation}}

\begin{document}
\begin{bibunit}

\title{Odd Viscosity-Driven Phase Separation in Chiral Active Mixtures}
\author{Xiaoyi Wu} 
\affiliation{Department of Physics, 
University of California San Diego,
La Jolla, CA 92093} 
\author{Tzer Han Tan} 
\affiliation{Department of Physics, 
University of California San Diego,
La Jolla, CA 92093} 
\author{Hongbo Zhao}
\email{hongbo@ucsd.edu}
\affiliation{Department of Physics, 
University of California San Diego,
La Jolla, CA 92093} 
\affiliation{Department of Chemistry, University of California San Diego, La Jolla, CA 92093, USA}

\begin{abstract}

Binary mixtures of active particles spinning in opposite directions can spontaneously separate into homochiral domains, yet the mechanism behind this robust demixing remains poorly understood. Here, we establish a two-fluid hydrodynamic theory that generalizes odd viscosity to multicomponent systems. We show that cross parity-violating viscosities, which couple the stress of one chiral species to the strain rate of another, can destabilize the homogeneous mixture, while odd viscosity within each homochiral domain governs the interfacial stability and dynamics.
Microscopic measurements of stress and velocity correlations yield viscosity estimates that predict spontaneous demixing followed by domain coarsening.
These findings identify odd viscosity as a route to programming phase behavior and domain morphology in chiral active mixtures.

\end{abstract}

\maketitle

Active matter consumes energy to generate motion and forces, giving rise to emergent phenomena absent in their passive counterparts. 
For example, self-propelled active Brownian particles can undergo motility-induced phase separation (MIPS) and form dense clusters~\cite{Cates2015}. Mixtures of active and passive Brownian particles~\cite{Wysocki2016,Stenhammar2015} and active nematics exerting differential stresses can demix~\cite{Bhattacharyya2023,Bhattacharyya2024}. In these systems, phase separation disappears when nonequilibrium driving is switched off, as their equilibrium thermodynamics favors a homogeneous or mixed state.

Remarkably, chiral active matter, where the microscopic constituents generate handed motion such as rotation~\cite{Liebchen2022}, also provides a distinct route to phase separation. In both circling swimmers that self-propel and rotate and spinners that only rotate, mixtures of oppositely rotating particles can demix into homochiral domains where the particles share the same handedness. In mixtures of circling swimmers, phase separation occurs through alignment, flocking, and MIPS~\cite{Levis2019,Ventejou2021,Ai2023}. In mixtures of spinners, phase separation emerges robustly across diverse experimental and computational systems, in the presence or absence of hydrodynamic interactions (wet or dry systems), inertial versus overdamped dynamics, and externally torque-driven versus torque-free spinners~\cite{Nguyen2014,Guo2025,Yeo2015,Scholz2018, Deb2026}.
Yet, despite this remarkable generality, the mechanism underlying phase separation in spinner mixtures remains poorly understood, especially compared to other examples of active phase separation above~\cite{Cates2025}, and a theory is still lacking.
 
In systems of interacting spinners, extensive work in the past decade has now established that the broken parity and time-reversal symmetry allows non-dissipative and parity-violating components in the viscosity tensor $\eta_{ijkl}$, known as odd viscosity, to emerge. In two-dimensional isotropic systems, the odd viscosity tensor takes the form $\eta_\mathrm{o} (\epsilon_{ik}\delta_{jl}+\epsilon_{jl}\delta_{ik})$, where $\epsilon_{ij}$ is the two-dimensional Levi-Civita tensor~\cite{Khain2022,Fruchart2023, Banerjee2017, Markovich2024,Markovich2021}. 
Subsequent studies have revealed that odd viscosity has a profound impact on interfacial phenomena and phase separation, giving rise to the formation of bubble phases devoid of spinners~\cite{Caprini2025,Marconi2026}, phase separation in externally sheared spinner fluids~\cite{Yu2024}, and odd surface waves~\cite{Soni2019,Caporusso2024,Neville2026,Abanov2018,Souslov2019}. What these cases have in common is that odd viscosity couples shear flow to normal stresses, which can redistribute density and alter interfacial dynamics. 
Motivated by these findings, we generalize the odd viscosity to multicomponent fluids. In this work, we first show the emergence of phase separation in a mixture from a microscopic model, develop a continuum two-fluid model, then investigate how odd viscosity $\eta_\mathrm{o}$ controls the late-stage interfacial dynamics, and lastly show through microscopic fluctuation correlations that a parity-violating viscosity $\eta_\mathrm{o}'$ that couples the stress and strain rate of different components underlies the onset of demixing.

\emph{Microscopic model.---}
We consider a two-dimensional binary mixture of particles spinning at equal and opposite angular velocities $\pm \Omega$. They are governed by overdamped dynamics
\begin{equation} \label{eq:microscopic}
\sum_{j}\big(\bm{F}_{ij}+\bm{F}_{ij}^{\perp}\big)- \Gamma \bm{v}_i= 0,
\end{equation} where $\Gamma$ is the damping coefficient and $\mathbf{v}_i$ is the velocity of particle $i$. This model can be extended to include inertia by introducing $m\dot{\bm{v}}_i$ on the right-hand side, where $m$ is the particle mass~\cite{Han2021,Caprini2025,Nguyen2014}. The particles interact through a pairwise steric repulsion $\mathbf{F}_{ij}$ from the Weeks-Chandler-Andersen (WCA) potential with energy scale $\epsilon$, and a drag force $\bm{F}_{ij}^{\perp}$ proportional to their relative surface velocity transverse to the line of centers. When rotational motion dominates translation, this velocity is proportional to $\Omega_i+\Omega_j$, where $\Omega_i$ is the spin rate of particle $i$. Here, we use hydrodynamic drag with a distance dependence predicted by lubrication theory~\cite{Tan2022} (Supplemental Material Sec.~\ref{sec:SI_microscopic_model}), although the collective behavior is insensitive to the specific mechanism and distance dependence of the drag~\cite{Nguyen2014,Guo2025,Yeo2015,Scholz2018, Deb2026}.

\begin{figure}
    \centering    \includegraphics[width=1.0\linewidth]{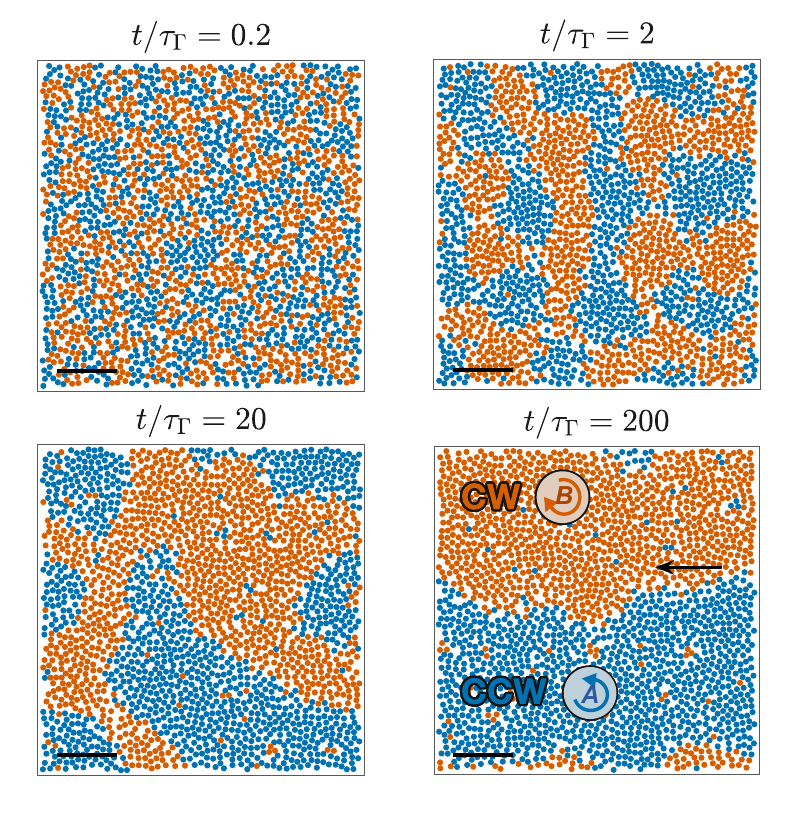}
    \caption{Phase separation in a binary chiral mixture. Time evolution of an overdamped mixture of $N_A=N_B=1075$ disks of diameter $\sigma$ in agent-based simulations with periodic boundary conditions at a packing fraction of $0.6$. At this density, the characteristic interparticle spacing is $1.14\sigma$. Species $A$ spins counterclockwise ($\Omega_A=\Omega >0$, blue) and species $B$ clockwise ($\Omega_B =-\Omega<0$, red), with a transverse to steric force ratio $f_0\Omega\sigma/\epsilon = 10^2$, where $f_0$ is the rotational drag coefficient. Time is measured in units of the overdamped relaxation time $\tau_{\Gamma} =\Gamma\sigma^2/\epsilon$. An initially homogeneous mixture spontaneously separates into homochiral domains. As the domains coarsen, their interfaces become nearly planar and support unidirectional traveling waves. (Supplemental Video S1). The arrow in the final snapshot indicates the direction of wave propagation. Scale bar: $10\sigma$.}
    \label{fig:1}
\end{figure}

In an overdamped and homochiral system where all particles spin at the same angular velocity, the system remains homogeneous (Fig.~\ref{fig:S1}). In a heterochiral mixture, however, the two species spontaneously segregate into homochiral domains (Fig.~\ref{fig:1}). The domains subsequently coarsen, forming nearly planar interfaces along which waves propagate unidirectionally. This remarkable phase separation in the absence of attractive pair interactions motivates a continuum theory of the active mixture.

\emph{Continuum model.---}
For a binary mixture, the density of each component $n_\alpha$, where the component is labeled by $\alpha = A,B$, follows the continuity equation
\begin{equation} \label{eq:continuity}
    \partial_t n_\alpha + \nabla\cdot(\mathbf{u}_\alpha n_\alpha) = 0,
\end{equation}
where $\mathbf{u}_\alpha$ is the continuum velocity of component $\alpha$.
The momentum balance of each component satisfies~\cite{Bhattacharyya2023,Bhattacharyya2024,Bhattacharyya2025,Bruinsma2014,Rautu2026,Marconi2026}
\begin{equation}
   \nabla \cdot \bm\sigma_\alpha - n_\alpha \nabla \mu_\alpha = \Gamma \mathbf{u}_\alpha+\Gamma_f\mathbf{u}_{\alpha\beta}.~\label{eq:momentum_balance}
\end{equation}
The total stress $\bm\sigma_\alpha=\bm \sigma_{\nu,\alpha} + \bm\sigma_{\mathrm o,\alpha} + \bm\sigma_{\tau,\alpha}$ consists of the conventional and odd viscous stresses $\bm\sigma_{\nu,\alpha}$ and $\bm\sigma_{\mathrm o,\alpha}$ and a torque-induced stress $\bm\sigma_{\tau,\alpha}=\eta_R \Omega_{\alpha}\bm \epsilon n_\alpha$ that captures the rotational drag between particles, with $\eta_R$ the rotational viscosity.
$\mu_\alpha$ is the chemical potential of $\alpha$, 
and $\Gamma_f$ is the interspecies drag coefficient acting on the relative velocity $\mathbf{u}_{\alpha\beta}\equiv\mathbf{u}_\alpha-\mathbf{u}_\beta$ that exchanges momentum between A and B.

\emph{Late-stage interfacial dynamics.---}
We first use the theory to examine interfacial stability in the late stages of phase separation, when each domain consists almost entirely of a single component and obeys the corresponding one-component hydrodynamic equations. 
As observed in Fig.~\ref{fig:1}, the total density is nearly uniform throughout the system due to steric repulsion. We therefore approximate each component as incompressible with a constant density $n_0$, with $\nabla\cdot \mathbf{u}_\alpha=0$, and $n_0\mu_\alpha$ acting as the Lagrange multiplier that enforces this constraint.  
At the interface, we impose velocity continuity $\mathbf{u}_A=\mathbf{u}_B$, which governs the interfacial dynamics.
Within each domain, the conventional and odd viscous stresses of each component reduce to their one-component forms, with $\nabla\cdot \bm\sigma_{\nu,\alpha}=\eta_\alpha \nabla^2\textbf{u}_\alpha$ and $\nabla \cdot \bm\sigma_{\mathrm o,\alpha}={\eta}_\mathrm{o,\alpha}\bm{\epsilon}\nabla^2\mathbf{u}_\alpha$ \cite{Banerjee2017,Markovich2021,Markovich2024,Fruchart2023,Soni2019}, where $\eta_\alpha$ and $\eta_{o,\alpha}$ are the shear and odd viscosities of component $\alpha$, respectively.
Because A and B spin at opposite angular velocities $\Omega_{A} = -\Omega_{B} =\Omega$ and the shear and odd viscosities are parity even and odd, we have $\eta_A=\eta_B=\eta$ and $\eta_{\mathrm{o},A}=-\eta_{\mathrm{o},B}=\eta_\mathrm{o}$.
Lastly, the momentum balance of the two components combined (Eq.~\ref{eq:momentum_balance}) requires continuity of traction across the interface.

\begin{figure}
    \centering   \includegraphics[width=1.0\linewidth]{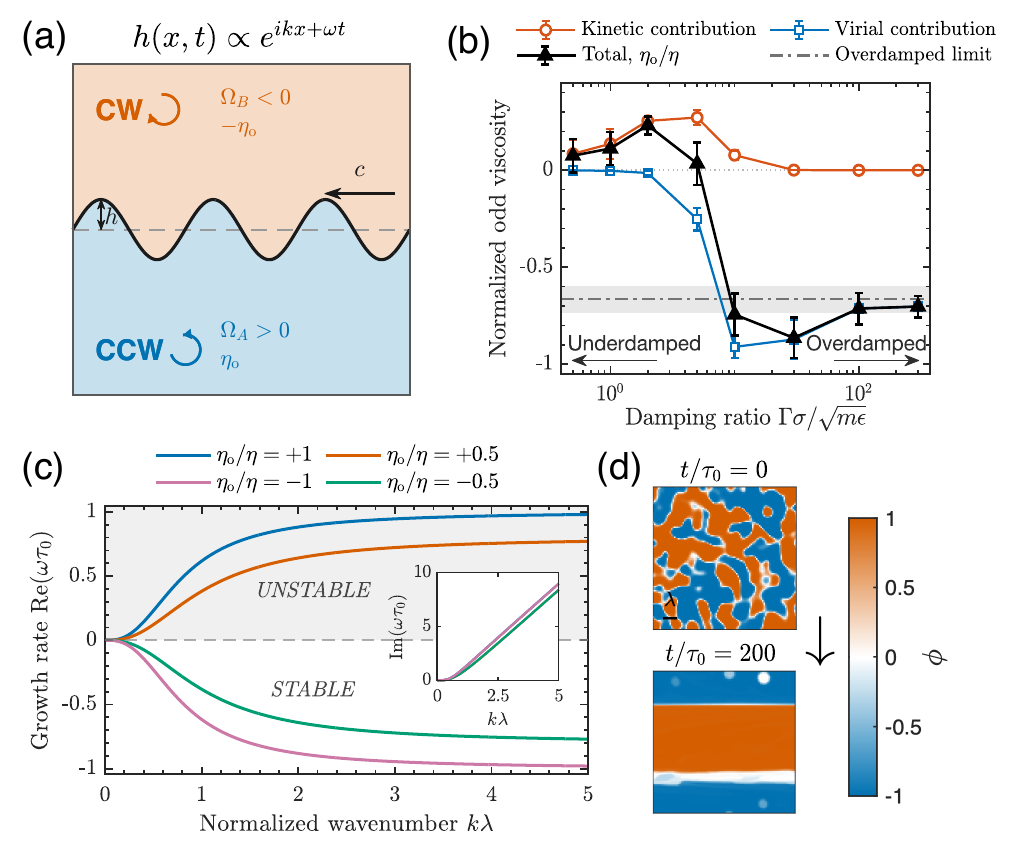}
    \caption{Odd viscosity stabilizes interfaces between homochiral domains at late stages of phase separation. (a) Schematic of the interface between species $A$ and $B$, where the phase velocity $c=-\mathrm{Im}{\omega}/k$. (b) Odd viscosity $\eta_{\mathrm{o}}/\eta$ at varying damping ratio (dimensionless friction-to-inertia ratio), calculated using Green--Kubo relation and averaged over five independent runs. $\Omega>0$. Increasing damping suppresses the kinetic contribution $\eta_\mathrm{o}^{\text{kin}}/\eta$ relative to the virial $\eta_\mathrm{o}^{\text{vir}}/\eta$ (Supplemental Material, Sec.~\ref{sec:one_component_Gree-Kubo}), reversing the sign of the odd viscosity. The gray dashed line and shaded band denote the overdamped Green--Kubo result $\eta_{\mathrm{o}}/\eta = -0.66 \pm 0.07$ ($95\%$ confidence interval). (c) Growth rate of interfacial instability $\mathrm{Re}{\omega}$ for different values of $\eta_{\mathrm{o}}/\eta$. 
    Inset: $\mathrm{Im}{\omega}$ determines the direction of wave propagation; curves with equal values of  $|\eta_\mathrm{o}/\eta|$ overlap. 
    (d) Snapshots of the incompressible two-fluid model initialized with random, fully segregated A and B domains when $\eta_\mathrm{o}\Omega<0$. 
    The characteristic length and time scales are $\lambda=\sqrt{\eta/\Gamma}$, $\tau_0=\eta/(\eta_R\Omega n_0)$. 
    The color indicates the composition field $\phi=(n_A-n_B)/(n_A+n_B)$.}
    \label{fig:2}
\end{figure}

In this sharp interface limit, we analyze the linear stability of a planar interface separating two homochiral domains at steady state by introducing a small interfacial perturbation $h\sim e^{ikx+\omega t}$ (Supplemental Material, Sec.~\ref{sec:s4d}), where $x$ is the coordinate along the interface, and $k$ is the wavenumber (Fig.~\ref{fig:2}(a)).
In the long wavelength limit $|k\lambda|\ll1$, the dispersion relation scales as
\begin{equation}
\mathrm{Im}{\omega} \sim -\frac{\eta_R \Delta \Omega}{4\eta}(k\lambda)^3,\quad
\mathrm{Re}{\omega} \sim \frac{\eta_R \Delta \eta_\mathrm{o}\Delta \Omega}{2\eta^{2}}|k\lambda|^3,
\end{equation}
where $\Delta \Omega = \Omega_B -\Omega_A$, $\Delta \eta_\mathrm{o}= \eta_{\mathrm{o},B}-\eta_{\mathrm{o},A}$, and  $\lambda =\sqrt{\eta/\Gamma}$ is the hydrodynamic screening length. The imaginary part of the dispersion relation $\mathrm{Im} \omega(k)$ reveals a unidirectional traveling wave. In the configuration illustrated in Fig.~\ref{fig:2}(a), the wave travels leftward, consistent with the interfacial waves observed in our microscopic simulations (Supplementary Video S1). 
The real part of the dispersion relation $\mathrm{Re} \omega(k)$, which is proportional to $\eta_{\mathrm{o}}\Omega$, controls the stability of the interface. 
This is because odd viscosity couples shear flow in the two domains to normal stresses at the interface, acting as either a positive or a negative interfacial tension between the two domains~\cite{Soni2019}, depending critically on the sign of $\eta_\mathrm{o} \Omega$. If $\eta_\mathrm{o}\Omega<0$, the interface is stable and the homochiral domains stay separated. If $\eta_\mathrm{o}\Omega>0$, the two domains interpenetrate, effectively mixing A and B (Fig.~\ref{fig:2}(c)).

Previous microscopic simulations have reported both positive and negative values of $\eta_\mathrm{o} \Omega$ \cite{Han2021,Hargus2020,Liao2019,Mecke2023,Yu2024, Eren2026}. An established approach for determining viscosity is the Green–Kubo relation, which connects the viscosity tensor to the time integral of the stress correlation function \cite{Han2021,Hargus2020,Hargus2025}: $\eta_{ijkl}\propto \int_0^\infty \langle \sigma_{ij}(t)\sigma_{kl}(0)\rangle dt$, where $\sigma_{ij}(t)$ is the fluctuating stress measured from the microscopic simulations based on the Irving-Kirkwood formula\cite{Han2021}. Here, we apply this approach to our microscopic model in a homochiral system with added inertia and systematically vary the damping-to-inertia ratio (Supplemental Material, Sec.~\ref{sec:one_component_Gree-Kubo}).
We find that increasing damping reverses the sign of $\eta_{\mathrm{o}}\Omega$, as the virial contribution, which originates from the interparticle forces, overtakes the kinetic contribution from the particle momentum (Fig.~\ref{fig:2}(b)). 
In the underdamped regime, $\eta_\mathrm{o}\Omega>0$, consistent with previous work~\cite{Han2021,Hargus2020,Mecke2023}. 
In the strongly overdamped and inertialess regime that we focus on in this work, $\eta_{\mathrm{o}}\Omega<0$~\cite{Yu2024}.

This has important implications. First, $\eta_\mathrm{o}\Omega<0$ explains the stability of the planar interface observed at late times in the microscopic simulation.
Moreover, the stability of the interface goes beyond the linear regime. A simulation of the two-component model (Eqs.~\ref{eq:continuity} and~\ref{eq:momentum_balance}) under the incompressibility assumption shows that when the system starts from a demixed state where the two components are segregated into separate domains of random geometry, the domains coarsen over time until the interface eventually becomes flat (Fig.~\ref{fig:2}(d)), in good agreement with the microscopic simulation.
In contrast, when there is no odd viscosity, a regime studied by previous works of passively phase-separating systems driven by external torque~\cite{Spellings2015,Sabrina2015,Maji2025}, the interface continuously deforms and does not flatten. When $\eta_\mathrm{o}\Omega>0$, the interface quickly expands and enhances mixing (Fig.~\ref{fig:S100}).
Second, if we remove the incompressibility constraint and allow density to vary, $\eta_\mathrm{o}\Omega<0$ suppresses density fluctuations in each phase, as confirmed by linear stability analysis of a homogeneous density field and simulations (Supplemental Material Sec.~\ref{sec:2}, Fig.~\ref{fig:S10})~\cite{Marconi2026}. This is consistent with the homogeneity of homochiral domains in our overdamped simulations (Fig.~\ref{fig:1} and Fig.~\ref{fig:S2}). In underdamped systems, $\eta_\mathrm{o}\Omega>0$ can trigger density instability and the formation of particle depleted bubbles~\cite{Caprini2025,Marconi2026}.

While odd viscosity explains the coarsening of homochiral domains, an unresolved question is what initially triggers the demixing into homochiral domains. 
Even in a macroscopically homogeneous mixture, local composition fluctuations generate regions enriched in opposite chiralities. We propose that the coupled odd viscous response at the boundaries of these regions gives rise, upon coarse-graining, to an additional cross parity-violating viscosity between the two components (Supplemental Material Sec.~\ref{sec:coarse_graining}). We therefore extend the continuum theory to include this response, examine its effect on the stability of the homogeneous mixture, and measure the corresponding transport coefficient from microscopic fluctuation correlations.

\begin{figure*}[t]
    \centering    \includegraphics[width=0.95\textwidth]{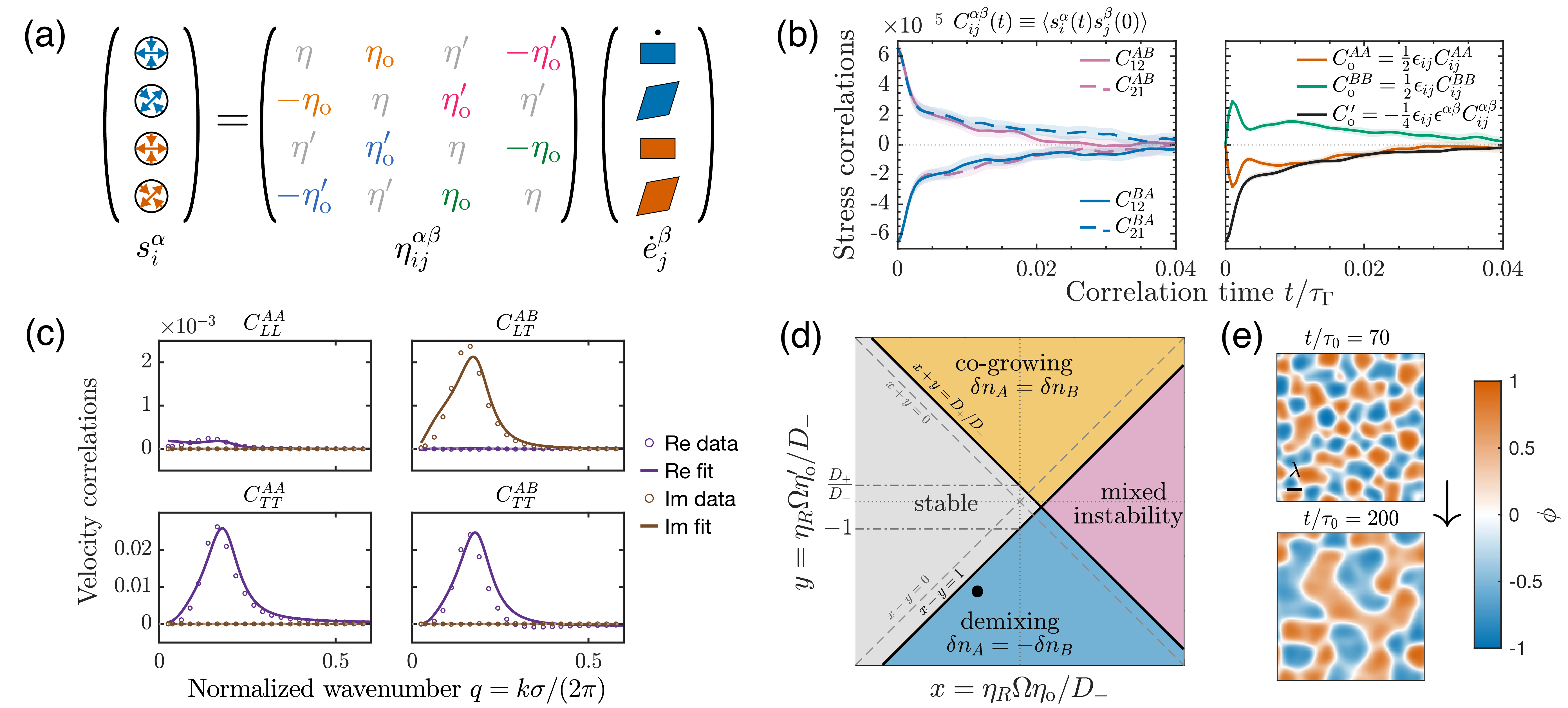}
    \caption{Cross parity-violating viscosity initiates bulk demixing. (a)~The relationship between shear stress and shear strain rate in a counterrotating mixture, illustrating the symmetry-constrained viscosity tensor. (b)~Stress correlations in a homogeneous mixture. Left: odd stress correlations between A and B satisfy parity-exchange symmetry. Right: combined intraspecies and interspecies odd stress correlations. Their temporal integrals are proportional to $\eta_\mathrm{o}$ and $\eta_\mathrm{o}'$, respectively. The correlations are averaged over 80 trajectories and measured at a packing fraction of 1.1. (c)~Real and imaginary parts of the velocity correlations $C_{LL}^{AA}$, $C_{LT}^{AB}$, $C_{TT}^{AA}$, and $C_{TT}^{AB}$ measured from microscopic simulations at the frequency $\omega\tau_\Gamma=15.7$, compared with the best fits from the fluctuating hydrodynamic theory. (d)~Linear stability diagram in the $\eta_{\mathrm{o}},\eta_{\mathrm{o}}'$ plane, showing the stable, demixing ($\delta n_A=-\delta n_B$ mode unstable), co-growing ($\delta n_A=\delta n_B$ mode unstable), and mixed instability (both modes unstable) regimes. 
    (e)~Snapshots of the two-fluid model simulation with parity-violating viscosities indicated by the black dot in (d), starting from homogeneous densities with small perturbations.} 
    \label{fig:3}
\end{figure*}

\emph{Onset of demixing.---}
Generalizing the viscosity tensor to include cross responses and omitting bulk viscosity, the divergence of viscous stresses becomes
\begin{equation}
    \nabla\cdot \bm{\sigma}_{\nu,\alpha}= {\eta}_{\alpha\beta}\nabla^2\mathbf{u}_\beta,\quad \nabla\cdot \bm\sigma_{\mathrm{o},\alpha}=\eta_{\mathrm{o},\alpha\beta}\bm{\epsilon}\nabla^2\mathbf{u}_\beta.
    \label{eq:generalize}
\end{equation}
A mixture with spinners of opposite chirality possesses a combined parity-exchange symmetry (mirror reflection accompanied by $A \leftrightarrow B$ exchange). This parity-exchange symmetry in an equal composition mixture constrains the odd viscosity tensor in species space to
\begin{equation}
    \bm{\eta}_{\mathrm{o}} =
    \begin{pmatrix}
        \eta_{\mathrm{o}} & -\eta_{\mathrm{o}}' \\
        \eta_{\mathrm{o}}' & -\eta_{\mathrm{o}}
    \end{pmatrix},
    \label{eq:oddmatrix}
\end{equation}
and shear viscosity tensor to ${\eta}_{\alpha\beta}
=\eta\,\delta_{\alpha\beta}
+\eta'(1-\delta_{\alpha\beta})$, where $\eta^\prime$ is the cross shear viscosity, and $\eta_{\mathrm{o}}^\prime$ is the cross parity-violating viscosity. 
Figure~\ref{fig:3}(a) shows the generalized viscosity tensor coupling the two independent shear components of stress and strain rate of both species, $s_i^\alpha = \eta_{ij}^{\alpha\beta} \dot{e}_j^\beta$, where $(s_1^\alpha,s_2^\alpha)=(\sigma_{\alpha,xx}-\sigma_{\alpha,yy},\,\sigma_{\alpha,xy}+\sigma_{\alpha,yx})/2$ and $(\dot e_1^\alpha,\dot e_2^\alpha)=(\partial_x u_{\alpha,x}-\partial_y u_{\alpha,y},\,\partial_x u_{\alpha,y}+\partial_y u_{\alpha,x})$. 
We note that while the odd viscosity of the same species $\eta_{\mathrm{o}}$ is nondissipative, the cross parity-violating viscosity $\eta_\mathrm{o}'$ can contribute to viscous dissipation.

Based on the generalized viscosity tensor and Eqs.~\ref{eq:continuity}-\ref{eq:momentum_balance}, we can analyze the linear stability of the homogeneous mixture $n_A=n_B=n_0$ against small density perturbations (Supplemental Material Sec.~\ref{sec:SI_bulk_instability_binary}). 
This requires expanding the chemical potential with respect to density perturbations, $\delta \mu_\alpha = n_\alpha^{-1} K_{\alpha\beta}\delta n_\beta - \kappa \nabla^2 \delta n_\alpha$, where $K_{\alpha\beta}=n_\alpha \partial \mu_\alpha /\partial n_\beta$, and $\kappa$ is the coefficient of the lowest-order energy penalty for density gradients. Because of species exchange symmetry, we have $K_{AA} =K_{BB}=K$ and $K_{AB}=K_{BA}=K_c$, and the two density eigenmodes in the linear stability analysis are the in-phase (total density) mode $\delta n_A = \delta n_B$ and out-of-phase (composition) mode $\delta n_A = - \delta n_B$. We define $K_\pm = K \pm K_c$.
The total density mode becomes unstable when
\begin{equation} \label{eq:co-growing}
    \eta_R(\eta_{\mathrm{o}}+\eta_{\mathrm{o}}^\prime)\Omega
> D_+  \equiv \Big(\sqrt{(\eta-\eta^\prime)K_+}+\sqrt{\kappa n_0(\Gamma+2\Gamma_f)}\Big)^2,
\end{equation}
causing the densities of A and B to co-grow into dense and dilute regions.
The composition mode becomes unstable when
\begin{align}
    \eta_R(\eta_{\mathrm{o}}-\eta_{\mathrm{o}}^{\prime})\Omega> D_-\equiv(\sqrt{(\eta+\eta^{\prime}) K_-}+\sqrt{\kappa n_0 \Gamma})^2,
    \label{eq:demixing}
\end{align}
leading to demixing. These stability criteria are illustrated in the mixture phase diagram (Fig.~\ref{fig:3}(d)). Without cross parity-violating viscosity, odd viscosity of the same species $\eta_\mathrm{o}$ alone cannot explain the demixing.
The blue region in Fig.~\ref{fig:3}(d) delineates the conditions for demixing at constant total density.

Next, we generalize the stress correlation from a single component to a mixture, $C_{ij}^{\alpha\beta}(t)=\langle s_i^{\alpha}(t)s_j^{\beta}(0)\rangle$, to look for signatures of $\eta_\mathrm{o}$ and $\eta_\mathrm{o}'$.
Rotational and parity-exchange symmetries imply $C_{12}^{AA}(t)=-C_{21}^{AA}(t)=-C_{12}^{BB}(t)=C_{21}^{BB}(t)$, and that these correlations are odd functions of the time lag $t$, while $C_{12}^{AB}(t) = C_{21}^{BA}(t) = - C_{21}^{AB}(t) = -C_{12}^{BA}(t)$ are even functions of $t$. 
We measure these stress correlations in a homogeneous heterochiral mixture by performing simulations at a high packing density (Fig.~\ref{fig:S5}). The ensemble-averaged results in Fig.~\ref{fig:3}(b) satisfy the symmetry relations within statistical uncertainty.

Generalizing the Green--Kubo relation to a mixture, $\smash[t]{\eta_{ij}^{\alpha\beta}} \propto \int_0^\infty \smash[t]{C_{ij}^{\alpha\beta}(t)}dt$ (Supplemental Material Sec.~\ref{sec:green-kubo-mixture}), we calculate $\eta_\mathrm{o}$ and $\eta_\mathrm{o}'$ based on the symmetry conditions: 
$\eta_\mathrm{o}=\frac{1}{4}\epsilon_{ij} (\eta_{ij}^{AA}-\eta_{ij}^{BB})$ and $\eta_{\mathrm{o}}^{\prime}=-\frac{1}{4}\epsilon_{ij}\smash[t]{\epsilon^{\alpha\beta}}\smash[t]{\eta_{ij}^{\alpha\beta}}$, where $\epsilon^{AB}=-\epsilon^{BA} = 1$. With $\Omega>0$, we find $\eta_\mathrm{o}/\eta= -0.064 \pm0.007$, $\eta_\mathrm{o}^{\prime}/\eta = -0.086 \pm0.008$ within 95\% confidence intervals.
The result $\eta_\mathrm{o}'<\eta_\mathrm{o}<0$ supports the necessary sign condition for demixing with spatially uniform total density based on Eqs.~\ref{eq:co-growing}-\ref{eq:demixing}.

To test the proposed Green-Kubo relation for the active mixture, we extend the hydrodynamic model to include fluctuating stress $\bm\sigma^R_\alpha$ and force, and compare its predictions for velocity fluctuations with those measured in simulations (Supplemental Material Sec.~\ref{sec:velocity_correlation}). We use a minimal model for the fluctuating stress based on the required symmetry and Green--Kubo relation, $\langle \sigma_{ij}^{R,\alpha}(\mathbf{r},t) \sigma_{kl}^{R,\beta}(\mathbf{r}',t') \rangle \propto \delta(\mathbf{r}-\mathbf{r}') ( \eta_{ijkl,\text{sym}}^{\alpha\beta} \delta(t-t') + \eta_{ijkl,\text{anti}}^{\alpha\beta} \psi(t-t') )$, where $\psi(t)$ is an odd function of time. As mentioned above, the symmetric part of the stress correlation $\eta_{ij,\text{sym}}^{\alpha\beta} = \eta_{ji,\text{sym}}^{\beta\alpha}$ is even under the reversal of time lag, and the antisymmetric part $\eta_{ij,\text{anti}}^{\alpha\beta} = - \eta_{ji,\text{anti}}^{\beta\alpha}$ is odd. We can then calculate the correlations of velocity fluctuations in the homogeneous mixture using $\langle \hat{u}_\gamma^\alpha(\mathbf{k},\omega) \hat{u}_\delta^{*\beta}(\mathbf{k}',\omega')\rangle=C_{\gamma\delta}^{\alpha\beta}(k,\omega)\delta(\mathbf{k}-\mathbf{k}')\delta(\omega-\omega')$, where $\gamma,\delta\in \{L,T\}$, $\hat{u}_L^\alpha=\mathbf{\hat{k}}\cdot\mathbf{\hat{u}}_\alpha$ and $\hat{u}_T^\alpha=(\bm\epsilon\mathbf{\hat k})\cdot\mathbf{\hat{u}}_\alpha$ are the longitudinal and transverse Fourier amplitudes of the velocity fields, respectively, and $\mathbf{\hat{k}}$ is the unit wavevector. The parity-exchange symmetry requires that $C_{\gamma\gamma}^{AA}=C_{\gamma\gamma}^{BB}$ and $C_{\gamma\gamma}^{AB}=C_{\gamma\gamma}^{BA}$ are real, while $C_{LT}^{AA}=-C_{LT}^{BB}$ and $C_{LT}^{AB}=-C_{LT}^{BA}$ are complex. A subset of the correlations are shown in Fig.~\ref{fig:3}(c). The longitudinal velocity fluctuations, which contribute to density changes, are much weaker than the transverse velocity fluctuations, which contribute to shear flow. We then fit the theory to the velocity correlations from microscopic simulations and find good agreement between the two (Fig.~\ref{fig:3}(c)). We use bootstrapping to obtain uncertainty estimates for the fitted parameters (Supplemental Material Sec.~\ref{sec:velocity_correlation}), obtaining $\eta_\mathrm{o}/\eta=-0.02\pm0.02$ and $\eta_\mathrm{o}'/\eta=-0.1\pm0.04$ within 95\% confidence interval. This is consistent with the results from stress correlations that $\eta_\mathrm{o}'<\eta_\mathrm{o}<0$.
A continuum simulation of the generalized two-fluid model with parity-violating viscosities in the blue region in Fig.~\ref{fig:3}(d) ($\eta_\mathrm{o}'<\eta_\mathrm{o}<0$) shows that, starting from uniform densities, the two components demix and the domain coarsen over time (Fig.~\ref{fig:3}(e)), while if $\eta_\mathrm{o}'=0$, they stay mixed (Supplemental Material, Sec.~\ref{sec:binary_compressible_simulation}). Together, the microscopic measurements and continuum simulations support a mechanism of bulk demixing mediated by the cross parity-violating viscous transport.

\emph{Discussion.---}
Generalizing odd viscous hydrodynamics to a two-component system, we developed a continuum theory for the spontaneous demixing of heterochiral mixtures and the subsequent coarsening of the homochiral domains. 
It reveals how phase separation can emerge directly from nonequilibrium transport coefficients, without an imposed double-well free energy that drives demixing~\cite{Pisegna2025, Spellings2015, Nguyen2014, Maji2025, Sabrina2015,Joanny2007,Callan-Jones2011}.
Our analysis of stress and velocity correlations further provides a microscopic foundation for the generalized viscosity tensor and an understanding of fluctuating hydrodynamics of nonequilibrium mixtures.
We find that the sign and relative magnitudes of the parity-violating viscosities play a critical role in the stability of the interface and density fluctuations. This implies that microscopic parameters like inertia and spinning speeds, which can change the viscosity ratio~\cite{Eren2026}, can in turn alter the phase behavior. Indeed, different microstructures have been recently reported in mixtures with varying compositions and angular velocity ratios~\cite{Guo2025}.
Our extension of the classical two-fluid model to chiral active mixtures~\cite{Bhattacharyya2023,Bhattacharyya2024,Bhattacharyya2025,Bruinsma2014,Rautu2026,Callan-Jones2011,Joanny2007} hence opens avenues for exploring new instabilities and patterns in different regions of the phase diagram.

An intriguing direction is to explore whether odd viscosity gives rise to effective interfacial tensions that govern the contact angles at triple junctions between phases (Supplemental Material, Fig.~\ref{fig:S7}). More broadly, this may establish a generalized framework for capillary phenomena in chiral active mixtures, where interfacial dynamics are governed by nonequilibrium stresses~\cite{Zhao2026}.
Experimental realizations using geometrically chiral particles driven by oscillating electric or acoustic fields, or chiral robots driven by internal motors \cite{Sabrina2018,Ma2015,Scholz2021,Bo2026} could provide a platform to explore the theoretical phase diagram of chiral active mixtures and provide a new route to separating racemic mixtures into homochiral components.

\begingroup
\renewcommand{\addcontentsline}[3]{}
\begin{acknowledgments}
\emph{Acknowledgments.---} H.Z. acknowledges support from Society of Hellman fellowship. T.T. acknowledges support from UC MRPI Active Matter Hub. T.T. acknowledges support from DARPA DSO Pitch Day Award HR0011-26-3-E052.
\end{acknowledgments}

\putbib
\end{bibunit}

\endgroup

\clearpage
\onecolumngrid 

\begin{center}
  {\large\bfseries Supplemental Material: Odd Viscosity-Driven Phase Separation in Chiral Active Mixtures}\\[1em]
  {\normalsize Xiaoyi Wu,$^{1}$ Tzer Han Tan,$^{1}$ and Hongbo Zhao$^{1,2}$}\\[0.5em]
  {\small\itshape
  $^{1}$Department of Physics, University of California San Diego, La Jolla, CA 92093, USA\\
  $^{2}$Department of Chemistry, University of California San Diego, La Jolla, CA 92093, USA\par}
\end{center}
\setcounter{equation}{0}
\setcounter{figure}{0}
\setcounter{table}{0}
\setcounter{page}{1}
\setcounter{section}{0}
\renewcommand{\theequation}{S\arabic{equation}}
\renewcommand{\thefigure}{S\arabic{figure}}
\renewcommand{\thetable}{S\arabic{table}}
\renewcommand{\thesection}{S\arabic{section}}
\renewcommand{\bibnumfmt}[1]{[S#1]}
\renewcommand{\citenumfont}[1]{S#1}

\vspace{1em}

\hypersetup{
    colorlinks=true,
    linkcolor=RoyalBlue,
    citecolor=RoyalBlue,
    urlcolor=RoyalBlue
}

\renewcommand{\thesection}{S\arabic{section}}
\numberwithin{equation}{section}
\renewcommand{\thefigure}{S\arabic{figure}}

\setcounter{secnumdepth}{3}

\tableofcontents

\begin{bibunit}
\section{Microscopic Model}\label{sec:1}
In this section, we introduce our microscopic model, and then use the Green--Kubo relations to extract the transport coefficients from microscopic stress correlations.

\subsection{Particle model of chiral spinners}
\label{sec:SI_microscopic_model}
We consider a microscopic two-dimensional chiral spinners subjected to a purely repulsive Weeks-Chandler-Andersen (WCA) potential and a transverse pairwise force due to friction in the transverse direction.  The WCA potential $U^\textrm{WCA}$ and the pairwise force $F_{ij}^\textrm{WCA}$ between particle $i$ and $j$ follow
\begin{align}
    U^{\mathrm{WCA}}(r)&=
    \begin{cases}
    \displaystyle
    4\epsilon\bigg[\bigg(\frac{\sigma}{r}\bigg)^{12}
    -\bigg(\frac{\sigma}{r}\bigg)^6+\frac14\bigg],
    & r\leq 2^{1/6}\sigma,\\[6pt]
    0, & r>2^{1/6}\sigma.
    \end{cases}
    ,\\ 
    \bm{F}_{ij}^{\mathrm{WCA}} &=
    \begin{cases}
    \displaystyle
    \frac{24\epsilon}{r_{ij}}
    \bigg[2\bigg(\frac{\sigma}{r_{ij}}\bigg)^{12}
    -\bigg(\frac{\sigma}{r_{ij}}\bigg)^6\bigg]\hat{\bm r}_{ij},
    & r_{ij}\leq 2^{1/6}\sigma,\\[6pt]
    \bm 0, & r_{ij}>2^{1/6}\sigma.
    \end{cases}
\end{align}
where particle diameter $\sigma = 1$ sets the WCA cutoff at $2^{1/6}\sigma$, $r$ is the distance between particles, $\epsilon$ is the characteristic energy scale, $\mathbf{r}_{ij}=\mathbf{r}_i-\mathbf{r}_j$, $r_{ij}=|\mathbf{r}_{ij}|$, $\hat{\mathbf{r}}_{ij}=\mathbf{r}_{ij}/r_{ij}$. $\mathbf{r}_i$ is the coordinates of particles $i$.
Particle $i$ spins at an angular velocity of $\Omega_i$, and interacts with one another hydrodynamically. When the particles are close, the drag between them is approximated by the lubrication theory, which is proportional to the relative velocity between the adjacent surfaces in the direction transverse to the line connecting the centers of the spinners. When the spinning speed is much larger than the translational motion of the spinners $\Omega_i \sigma \gg |\mathbf{v}_i|$,
the hydrodynamic drag between a pair of spinners follows,
\begin{equation}
    \bm{F}_{ij}^{\perp} =\frac{1}{2}f_0 (\Omega_i+\Omega_j)\ln{\frac{d_c}{r_{ij}}}(\hat{\bm{z}}\times\hat{\bm{r}}_{ij}),\qquad r_{ij}<d_c=3\sigma,
\end{equation} 
where $f_0$ is the drag coefficient due to rotational motion. Beyond the distance cutoff $d_c$, the drag vanishes.

The equations of motion of the spinners are described by the overdamped dynamics,
\begin{gather}
    \dot{\bm{x}}_i = \bm{v}_i,\\
    0 = \sum_{j}\big(\bm{F}_{ij}^{\text{WCA}}+\bm{F}_{ij}^{\perp}\big)-\Gamma \bm{v}_i .\label{eq:1.4}
\end{gather}
For inertial spinners, the left-hand side of Eq.~\ref{eq:1.4} becomes $m\dot{\bm{v}}_i$, where $m$ denotes the particle mass.

\begin{figure}
    \centering
    \includegraphics[width=1.0\linewidth]{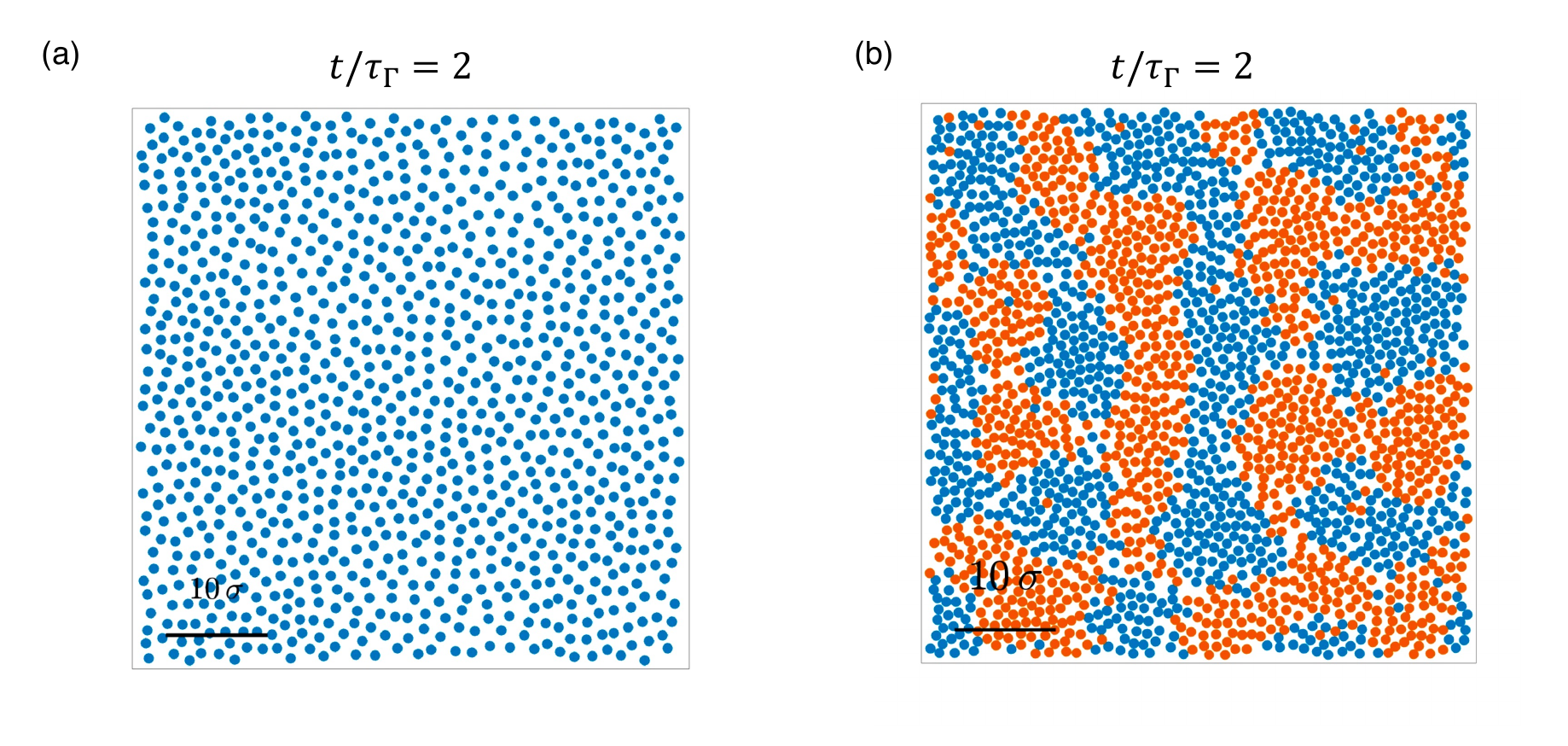}
    \caption{\textbf{Homogeneity of one-component spinner and demixing of a binary mixture.} \textbf{(a)}~A one-component system of $N_{A}=1075$ counterclockwise spinners (blue) remains homogeneous. \textbf{(b)}~A binary mixture with $N_{A}=N_B=1075$ at $t/\tau_\Gamma=2$ undergoes demixing.}
    \label{fig:S1}
\end{figure}

For agent-based simulations, we choose $f_0 =1$, $\epsilon=0.01$, and $\Omega=1$ in simulation units, unless otherwise specified. In our convention, $\Omega_A=+\Omega>0$ denotes counterclockwise spinners, shown in blue, while $\Omega_B=-\Omega<0$ denotes clockwise spinners, shown in red (Fig.~\ref{fig:S1}). For single-species simulations, we use $\Omega_i=+\Omega$. 
Particles are initially randomly placed on a hexagonal lattice with $N_x=43$ rows and $N_y=50$ columns in a rectangular domain with periodic boundary conditions applied. The lattice spacing is $a=2^{1/6}a_0$, with
$a_0=1.1$ unless otherwise specified, corresponding to a packing density $\rho\approx 0.6$ on a fully occupied lattice.
Our microscopic simulations show contrasting behavior in a purely overdamped chiral active system: a one-component chiral fluid remains stable, whereas a mixture of counterrotating spinners undergoes robust phase separation, as shown in Fig.~\ref{fig:S1}.

\begin{figure}
    \centering  \includegraphics[width=1.0\linewidth]{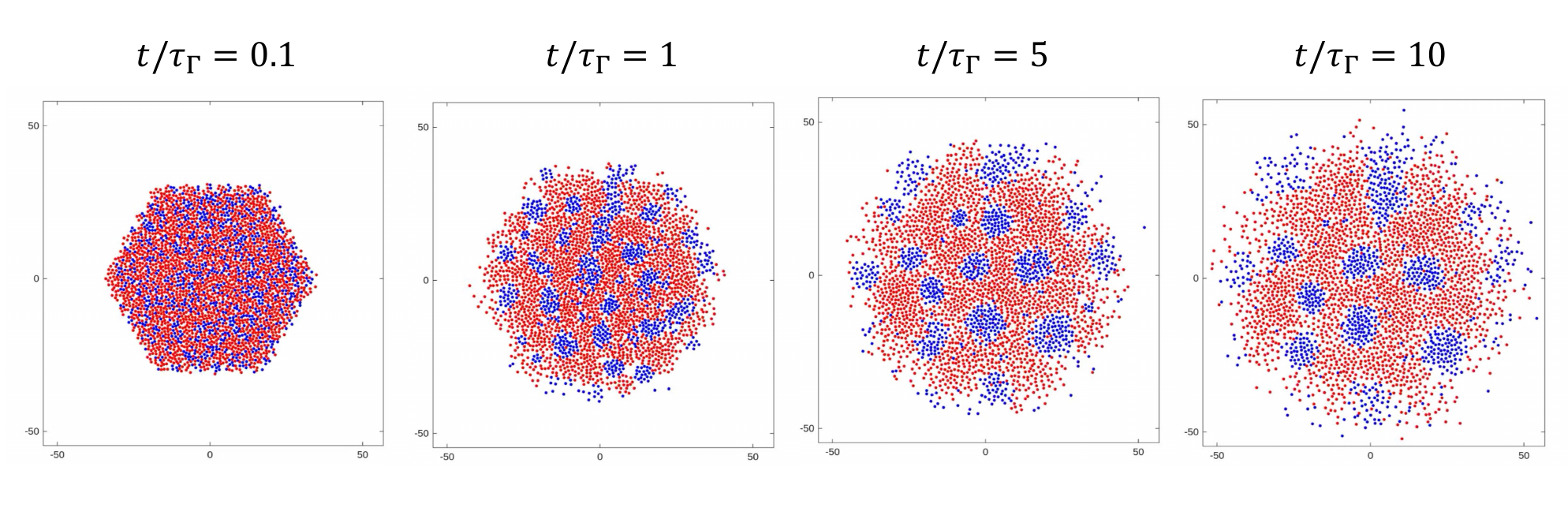}
    \caption{\textbf{Robust phase separation at an unequal composition in the overdamped binary mixture.} Time evolution snapshots for a mixture with 30\% CCW and 70\% CW spinners in an open-space simulation.}
    \label{fig:S2}
\end{figure}

We present additional results for an overdamped chiral active mixture with 30\% of CCW spinners and 70\% of CW spinners in an open space in Fig.~\ref{fig:S2}, showing robustness of phase separation at compositions other than the equal composition mixture that we focus on in this work. 

\subsection{Green--Kubo relations} \label{sec:one_component_Gree-Kubo}
The fluctuating stress in an active chiral fluid can be related to odd viscosities via Green-Kubo relations~\cite{Han2021}.
In agent-based simulations, the Irving--Kirkwood stress tensor $\bm{\sigma}$ is measured through
\begin{align}
    \bm{\sigma} &= -\frac{1}{V}\biggl[\sum_{i=1}^N m\bm{v}_i\bm{v}_i +\sum_{i<j}(\bm{F}_{ij}^{\text{WCA}}+\bm{F}_{ij}^{\perp})\bm{r}_{ij}\biggr] \label{eq:IK}\\
    &= \bm{\sigma}^{\text{kin}} +\bm{\sigma}^{\text{vir}},
\end{align}
where $\bm{v}_i\bm{v}_i$ and $\bm{F}_{ij}\bm{r}_{ij}$ denote dyadic products, and $V=L_xL_y =L^2$ is the size of the two dimensional simulation box. Note that $L_x \approx N_x a$ and $L_y \approx (\sqrt{3}/2)N_y a$ in a hexagonal lattice and thus $L_x\approx L_y\approx 53$ in our default setup.
We split the stress tensor into the kinetic contribution $\bm{\sigma}^{\text{kin}}$, the first term in Eq.~\ref{eq:IK} arising from particle momentum, and the configurational or virial contribution $\bm{\sigma}^{\text{vir}}$ from interparticle forces, to investigate the physical origins of measured viscosities. 
The two independent components of the symmetric traceless stress, transforming in the ($\ell=2$) irreducible representation, are defined as
\begin{align}
    s_1 = \frac{1}{2}(\mathbf{\sigma}_{xx} - \mathbf{\sigma}_{yy}),\qquad s_2 = \frac{1}{2}(\mathbf{\sigma}_{xy} + \mathbf{\sigma}_{yx}),\label{eq:gk}
\end{align}
and are referred to as shear stress 1 and shear stress 2 in Ref.~\cite{Fruchart2023}. 
Due to rotational symmetry, the correlations of the stress components satisfy $\langle s_1(t)s_1(0)\rangle = \langle s_2(t)s_2(0)\rangle$, and $\langle s_1(t)s_2(0)\rangle=-\langle s_2(t)s_1(0)\rangle$.
Hence we define the auto correlation function and the cross correlation function of these stress components, 
\begin{align}
    C_{\text{auto}} &= \frac{1}{2}\big(\langle s_1(t)s_1(0)\rangle + \langle s_2(t)s_2(0)\rangle\big),\\
    C_{\text{cross}} &= \frac{1}{2}\big(\langle s_1(t)s_2(0)\rangle - \langle s_2(t)s_1(0)\rangle\big).
\end{align}
We also define the kinetic and virial components of the shear stresses, for example $s_1^\textrm{kin} = \frac{1}{2}(\mathbf{\sigma}^\textrm{kin}_{xx} - \mathbf{\sigma}^\textrm{kin}_{yy})$, and the corresponding correlations functions $C_{\text{auto}}^\textrm{kin} = \frac{1}{2}\big(\langle s_1^\textrm{kin}(t)s_1^\textrm{kin}(0)\rangle + \langle s_2^\textrm{kin}(t)s_2^\textrm{kin}(0)\rangle\big)$, etc. Note that $C_\text{auto} \neq C_\text{auto}^\text{kin}+C_\text{auto}^\text{vir}$, because $C_\text{auto}$ also involves correlations between the kinetic and virial stresses.

\begin{figure}
    \centering
    \includegraphics[width=1.0\linewidth]{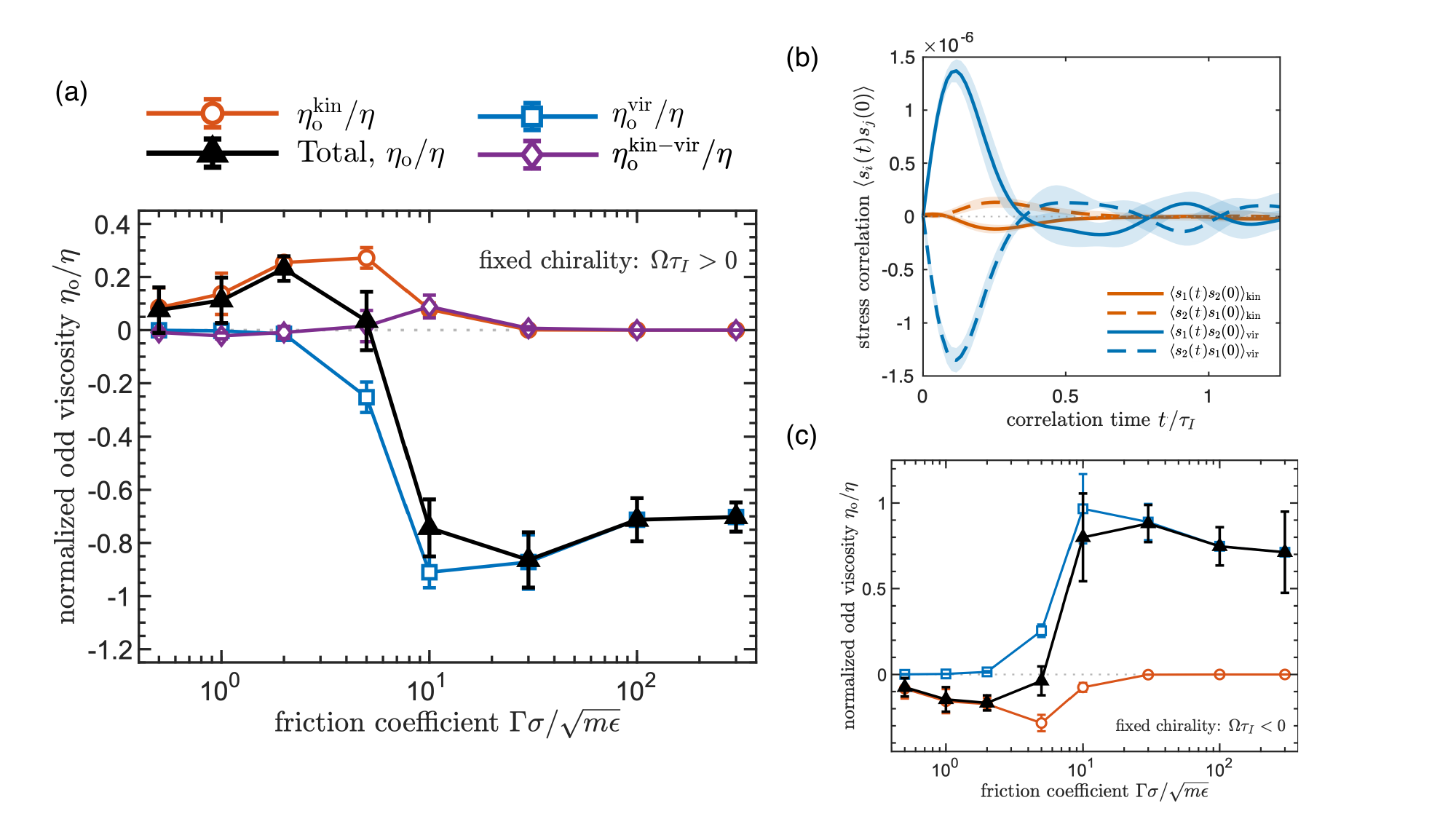}
    \caption{\textbf{Damping-induced sign reversal of the odd viscosity.} \textbf{(a)}~Dependence of $\eta_{\mathrm{o}}/\eta$ on the damping ratio $\tilde\Gamma$ for CW spinners with $\Omega\tau_I>0$, including the kinetic--virial cross contribution (purple), which remains small compared to the kinetic and virial contributions $\eta_\mathrm{o}^\text{kin}$ and $\eta_\mathrm{o}^\text{vir}$ in determining the sign of the odd viscosity. Damping coefficient is nondimensionalized by the inertial timescale $\tau_I = \sigma \sqrt{m/\epsilon}$ through $\tilde{\Gamma} = \Gamma\tau_I/m = \Gamma \sigma/\sqrt{m\epsilon}$, the ratio of $\tau_I$ to the momentum relaxation time $m/\Gamma$. \textbf{(b)}~Stress cross correlation functions $\langle s_i(t)s_j(0)\rangle$ for $\Omega\tau_I<0$ at $\tilde{\Gamma} = 10$, resolved into kinetic and virial contributions; the virial contribution dominates in the strongly damped regime. \textbf{(c)}~$\eta_{\mathrm{o}}/\eta$ versus $\tilde\Gamma$ for $\Omega\tau_I<0$. The kinetic contribution (red) is progressively suppressed with increasing damping, yielding $\eta_{\mathrm{o}}/\eta>0$ in the overdamped limit. Reversing the spinner chirality reverses the sign of $\eta_{\mathrm{o}}$, consistent with $\eta_{\mathrm{o}}\Omega<0$ in the overdamped regime for both chiralities. Values are reported as mean $\pm$ 95\% confidence intervals over five independent trajectories.}
    \label{fig:S3}
\end{figure}

The shear viscosity and odd viscosity are related to the correlation functions via 
\begin{align}
    \eta &= \frac{V}{k_BT_{\mathrm{eff}}}\int C_{\text{auto}}(t) dt\\
    \eta_{\mathrm{o}} &= \frac{V}{k_BT_{\text{eff}}}\int C_{\text{cross}}(t)dt,
\end{align}
where $k_B$ is the Boltzmann constant. All velocity fluctuations are generated by the deterministic potential and transverse forces and there is no thermal noise. At finite inertia, the kinetic energy of the velocity fluctuations defines an effective temperature $T_{\mathrm{eff}}=m\langle |\mathbf{v}|^2\rangle/2$~\cite{Loi2008}. 
We report the dimensionless ratio $\eta_{\mathrm{o}}/\eta$ where $T_\text{eff}$ cancels. As shown in Fig.~3(b), the odd viscous stress correlation signals are zero-valued at $t=0$, and is a result of the antisymmetry upon time reversal: $\langle s_1(t)s_2(0)\rangle=-\langle s_1(-t) s_2(0)\rangle$~\cite{Han2021}.
We also report the kinetic and virial contribution to the odd viscosity defined by
\begin{align}
        \eta_{\mathrm{o}}^{\text{kin}} &= \frac{V}{k_BT_{\text{eff}}}\int C_{\text{cross}}^{\text{kin}}(t)dt \\
    \eta_{\mathrm{o}}^{\text{vir}} &= \frac{V}{k_BT_{\text{eff}}}\int C_{\text{cross}}^{\text{vir}}(t)dt.
\end{align}
as well as the contribution to the odd viscosity by the cross correlation between the kinetic and virial stresses,
\begin{equation}
    \eta_{\mathrm{o}}^{\text{kin-vir}} = \frac{V}{k_BT_{\text{eff}}}\int C_{\text{cross}}^{\text{kin-vir}}(t)dt,
\end{equation}
where 
\begin{equation}
    C_{\text{cross}}^{\text{kin-vir}} = \frac{1}{2}\big(\langle s_1^\text{kin}(t)s_2^\text{vir}(0)\rangle + \langle s_1^\text{vir}(t)s_2^\text{kin}(0)\rangle - \langle s_2^\text{kin}(t)s_1^\text{vir}(0)\rangle - \langle s_2^\text{vir}(t)s_1^\text{kin}(0)\rangle \big).
\end{equation}
Overall, $\eta_\mathrm{o} = \eta_\mathrm{o}^\text{kin} + \eta_\mathrm{o}^\text{vir} + \eta_\mathrm{o}^{\text{kin-vir}}$. 
We show in Fig.~\ref{fig:S3}(a) that the cross kinetic-virial contribution to the odd viscosity $\eta_\mathrm{o}^{\text{kin-vir}}$ is not the dominating contribution to the sign of $\eta_{\mathrm{o}}/\eta$ across all damping ratios that we calculate.

Fig.~\ref{fig:S3}(a) shows $\eta_{\mathrm{o}}/\eta$ as a function of the damping ratio $\tilde\Gamma$. When the damping is weak relative to inertia, the kinetic contribution dominates and $\eta_{\mathrm{o}}>0$, consistent with the sign predicted by kinetic theory for chiral gases~\cite{Fruchart2022}; during collision between particles, the transverse forces due to the rotational motion produce deflections that generate a transverse momentum flux that contributes to a positive odd viscosity. Increasing $\tilde\Gamma$ suppresses the kinetic part, leaving stress to be transmitted through instantaneous interparticle forces, for which the transverse force biases the contact directions and produces a transverse response of the opposite sign. The two contributions cancel near $\tilde\Gamma\approx 5$, beyond which the virial contribution gradually plateaus: purely overdamped simulations give $\eta_{\mathrm{o}}/\eta = -0.66 \pm 0.07$, 
consistent with the negative sign measured under imposed Lees-Edwards shear in the overdamped regime~\cite{Yang2021}. 
The positive values for $\eta_\mathrm{o}\Omega$ predicted and measured for underdamped chiral gases~\cite{Fruchart2022, Han2021} and the negative values reported for overdamped spinners~\cite{Yang2021,Yu2024} represent two limiting regimes of a broader odd-viscous response governed by the balance between inertia and dissipation.

The reported single-species odd viscosities in Fig.~\ref{fig:S3} and main text Fig.~2(c) are measured at a volume fraction of 10\%, where the fluid remains homogeneous and is therefore suitable for Green--Kubo analysis. For each damping coefficient, the measurements are averaged over five independent trajectories. The first half of each trajectory is discarded to allow the system to reach steady state, and correlation functions are computed from the second half. The Green--Kubo integration window is chosen based on the decay of the total shear stress autocorrelation, and the integrals are evaluated using the trapezoidal rule. Statistical uncertainties and shaded regions reported in this work are estimated by taking two-sided $95\%$ confidence interval following student-t distribution.

\section{One-Component Bulk Model and Stability Analysis}\label{sec:2}
We introduce a minimal continuum model for a one-component overdamped chiral fluid. Motivated by hydrodynamic descriptions of chiral active fluids~\cite{Soni2019,Marconi2026,Neville2026}, the model combines mass conservation, momentum balance, viscous stress, odd viscous stress, and torque that arise from microscopic spinning dynamics. We first examine the bulk dynamics and stability of the one-component fluid, before extending the framework to binary mixtures of opposite chirality in Sec.~{\ref{sec:3}}.

For a two-dimensional single-component chiral fluid in the limit of negligible inertia, the density $n(\mathbf{r},t)$ and velocity field $\mathbf{u}(\mathbf r, t)$ are described by equations of the general form
\begin{gather}
  \partial_t n + \nabla \cdot (n \mathbf{u}) = 0, \label{eq:one_sum1}\\
  \nabla \cdot (\bm\sigma_v + \bm\sigma_o + \bm\sigma_\tau) - n\nabla\mu - \Gamma \mathbf{u} = 0, \label{eq:one_sum2}
\end{gather}
where $\Gamma>0$ is the substrate friction coefficient, 
$\mu = \delta F/\delta n$ is the chemical potential, where $F$ is the free energy. To describe spatially varying density fields, we allow the constitutive coefficients to depend on $n$. The viscous, odd viscous, and active torque stresses are
\begin{gather}
    \sigma_{ij,v} = \eta(n)\left(\partial_i u_j+\partial_j u_i
       -\frac{2}{d}\delta_{ij}\partial_k u_k\right)
       +\zeta(n)\delta_{ij}\partial_k u_k,\label{eq:01}\\
    \sigma_{ij,o} = \eta_\mathrm{o}(n) \left( \epsilon_{ik}\partial_k u_j + \epsilon_{jk} \partial_i u_k \right)
  = \eta_\mathrm{o}(n) \left( \epsilon_{ik} \partial_j u_k + \epsilon_{jk} \partial_k u_i \right),\label{eq:02}\\
  \sigma_{ij,\tau} = \eta_R \Omega n \epsilon_{ij},\label{eq:03}
\end{gather}
where $\zeta$ is the bulk viscosity and $\epsilon_{ij}$ is the two-dimensional Levi-Civita tensor, the dimension of space is $d=2$.
We assume that the shear, bulk, and odd viscosities $\eta(n)$, $\zeta(n)$, and $\eta_\mathrm{o}(n)$ depend linearly on density $n$. 
 The divergence of the odd viscous stress becomes 
\begin{equation}
f_{i,o} = \partial_j \sigma_{ij,\mathrm{o}}
=\eta_\mathrm{o}(n)\epsilon_{ik}\partial_{jj}u_k
+\eta_\mathrm{o}'(n)(\partial_j n)
\left(\epsilon_{ik}\partial_j u_k
+\epsilon_{jk}\partial_k u_i\right). \label{eq:odd_deriv}
\end{equation}
Therefore, for perturbations about a homogeneous steady state with $n=n_0$ and $\bm u=0$, the only linear-order contribution in Eq.~\ref{eq:odd_deriv} is $\eta_{\mathrm{o}} (n_0) \bm\epsilon \nabla^2\mathbf{u}$. All remaining contributions are nonlinear in the perturbations. The density-dependent terms are used in the full continuum simulations but do not modify the bulk linear stability analysis below.

\subsection{Continuum model and bulk stability}\label{sec:s2.1}
Omitting the nonlinear terms for the linear stability analysis, Eq.~\ref{eq:one_sum2} becomes
\begin{equation}
    \eta\nabla^2\mathbf{u}+\eta_R\Omega\bm{\epsilon}\nabla n-\Gamma\mathbf{u}-n\nabla\mu +\eta_{\mathrm{o}}\bm{\epsilon}\nabla^2\mathbf{u}+\zeta\nabla(\nabla\cdot\mathbf{u})=0. \label{eq:2}
\end{equation}
We write the chemical potential corresponding to a square-gradient free energy as $\mu =\mu_h(n)-\kappa\nabla^2 n$, where $\mu_h$ is the homogeneous contribution and $\kappa>0$ penalizes density gradients. The passive force density is therefore $ -n \nabla \mu =-n\nabla \mu_{h}(n) + \kappa n\nabla(\nabla^2n)$. Linearizing this term about $n=n_0$ gives $-K\nabla\delta n+\kappa n_0\nabla(\nabla^2\delta n)$, where $K\equiv(n\partial{\mu_h}/\partial{n})|_{n_{0}}>0$. The condition $K>0$ ensures that the passive contribution alone does not destabilize the homogeneous state.

We linearize about the homogeneous steady state $n=n_0$, $\mathbf{u}=0$ by introducing weak perturbations of the form $n = n_0 +\delta \hat{n} e^{i\bm{k}\cdot\bm{r}+\omega t}$, $\mathbf{u} = \delta\hat{\mathbf{u}} e^{i\bm{k}\cdot\bm{r}+\omega t}$. To linear order, from Eq.~\ref{eq:one_sum1} and \ref{eq:2},
\begin{gather}
\omega\delta\hat{n}+in_0\bm{k}\cdot\delta\hat{\mathbf{u}}=0,\label{eq:2.3}\\
-\eta k^2\delta\hat{\mathbf{u}}+i\eta_R\Omega\epsilon\bm{k}\delta\hat{n}-\Gamma\delta\hat{\mathbf{u}}-iK\bm{k}\delta\hat{n}-i\kappa n_0  k^2 \bm{k} \delta\hat{n}-\eta_{\mathrm{o}} k^2\epsilon\delta\hat{\mathbf{u}}-\zeta \bm{k}(\bm{k}\cdot\delta\hat{\mathbf{u}})=0.
\end{gather}
For simplicity we set $\zeta = 0$. The momentum equation can then be written in matrix form as
\begin{align}
\begin{pmatrix}
    -\eta k^2-\Gamma & -\eta_{\mathrm{o}} k^2\\
    \eta_{\mathrm{o}} k^2 & -\eta k^2-\Gamma
\end{pmatrix}\delta\hat{\mathbf{u}}
+
\begin{pmatrix}
    -iK-i\kappa n_0 k^2 & i\eta_R\Omega\\
    -i\eta_R\Omega & -iK - i\kappa n_0 k^2
\end{pmatrix} \begin{pmatrix}
    k_x\\
    k_y
\end{pmatrix} \delta\hat{n} = 0.
\end{align}
Solving for the velocity perturbation gives
\begin{align}
    \delta\hat{\mathbf{u}} = \frac{1}{(\eta k^2+\Gamma)^2+\eta_{\mathrm{o}}^2 k^4}
    \begin{pmatrix}
        -\eta k^2-\Gamma & \eta_{\mathrm{o}} k^2\\
        -\eta_{\mathrm{o}} k^2 & -\eta k^2-\Gamma
    \end{pmatrix}
    \begin{pmatrix}
        (iK +i\kappa n_0 k^2)k_x - i\eta_R\Omega k_y\\
        i\eta_R\Omega k_x +(iK+i\kappa n_0 k^2)k_y
    \end{pmatrix}\delta\hat{n}.\label{eq:2.6}
\end{align}
Substituting Eq.~\ref{eq:2.6} into Eq.~\ref{eq:2.3} gives the dispersion relation
\begin{align}
    \omega = n_0\frac{\eta_{\mathrm{o}} \Omega\eta_R k^4-(\eta k^2+\Gamma)(K k^2+\kappa n_0 k^4)}{(\eta k^2+\Gamma)^2+\eta_{\mathrm{o}}^2 k^4}.\label{eq:5}
\end{align}

The homogeneous state is unstable if $\omega(k)>0$ for any physical wavenumber. Since the denominator of Eq.~\ref{eq:5} is strictly positive, the instability condition is determined by its numerator, which requires 
\begin{equation}
    \eta_R\eta_{\mathrm{o}}\Omega>(\sqrt{\eta K}+\sqrt{\kappa n_0\Gamma})^2. \label{eq:one_instability}
\end{equation}
In the limit of $\kappa = 0$,  Eq.~\ref{eq:one_instability} reduces to $\eta_R\eta_{\mathrm{o}}\Omega>\eta K$, consistent with the instability condition in Ref.~\cite{Marconi2026}.

\subsection{Compressible continuum simulations}~\label{sec:one_cont}
Based on the overdamped bulk model introduced in Eqs.~\ref{eq:one_sum1} and~\ref{eq:one_sum2}, we implement continuum simulations of a compressible one-component chiral active fluid in a periodic box domain. The simulations are initialized from a nearly homogeneous state with mean number density $n_0=0.5$ and spatial random noise of amplitude $0.01$. For the nonlinear dynamics, we retain the linear density dependence of the odd, shear, and bulk viscosities introduced above.

\begin{figure}
    \centering
    \includegraphics[width=1.0\linewidth]{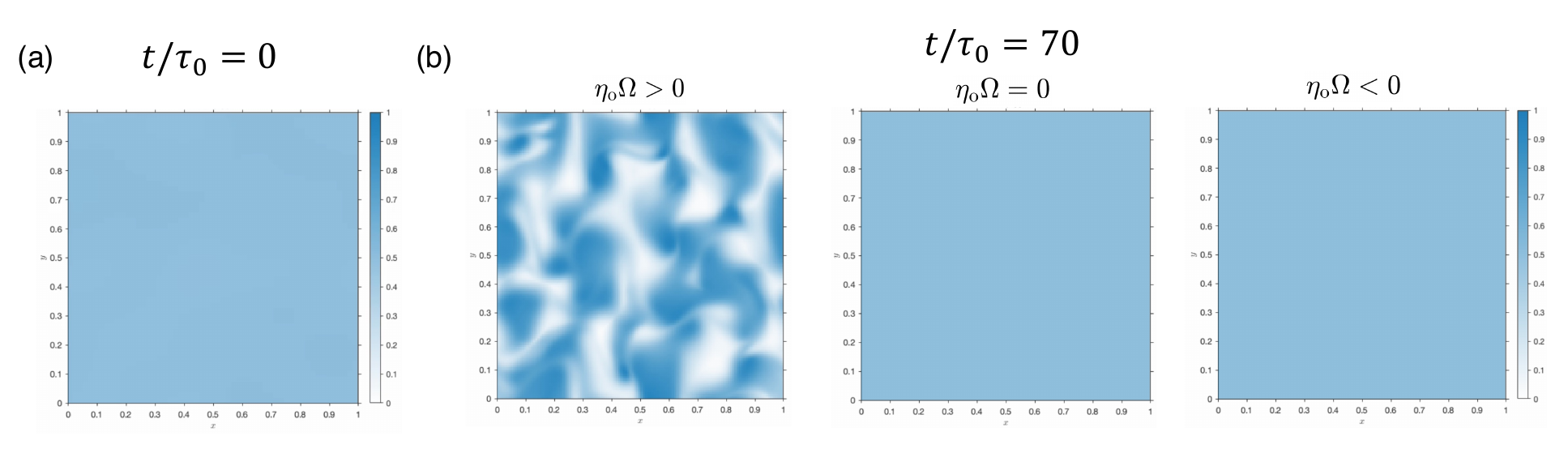}    \caption{\textbf{Compressible continuum simulations of one-component bulk stability.} \textbf{(a)}~ Initial condition at $t/\tau_0= 0$, consisting of a homogeneous state with mean density $n_0=0.5$ and spatial white noise of amplitude $0.01$. \textbf{(b)}~Density fields at $t/\tau_0=70$ for $\eta_{\mathrm{o}}/\eta=2$, $0$, and $-2$, corresponding to $\eta_\mathrm{o}\Omega>0$, $\eta_\mathrm{o}\Omega=0$, and $\eta_\mathrm{o}\Omega<0$ respectively under the counterclockwise $\Omega>0$ convention. The $\eta_{\mathrm{o}}/\eta=2$ case develops a bulk instability, whereas the $\eta_{\mathrm{o}}/\eta=0$ and $-2$ cases remain homogeneous. The color bar denotes the number density $n$, ranging from $0$ to $1$.}
    \label{fig:S10}
\end{figure}

We nondimensionalize the force balance equations using a characteristic density $n_0$, length $\lambda$, and velocity $u_0$. Balancing the viscous force with characteristic scale $\eta u_0/\lambda^2$, where $\eta$ denotes the viscosity at a reference density of 1, against substrate friction of scale $\Gamma u_0$, gives $\lambda=\sqrt{\eta/\Gamma}$; while balancing these terms with the active torque force of scale $\eta_R\Omega n_0/\lambda$, gives $u_0=\eta_R\Omega n_0/\sqrt{\eta\Gamma}$. The corresponding characteristic time scale is therefore
$\tau_0=\lambda/u_0
=\eta/(\eta_R\Omega n_0)$.
In the simulations, we set $\lambda=0.1$ and $u_0=1$, such that $\tau_0=0.1$, and discretize the domain on a $128$ by $128$ grid. We use a simple free energy model based on the ideal entropy of mixing, $f(n)=A[\frac{n}{\bar{n}}\ln \frac{n}{\bar{n}}+(1-\frac{n}{\bar{n}})\ln(1-\frac{n}{\bar{n}})]$, where we set the dimensionless ratio $A/\eta_R\Omega n_0=1$, and $\bar{n}=1$ is the highest density allowed. With these scales and all other dimensionless parameters fixed, the parameter varied in Fig.~\ref{fig:S10} is the odd-to-shear viscosity ratio $\eta_\mathrm{o}/\eta$, whose value is indicated in each panel.

The agent-based simulations in Fig.~\ref{fig:S1} show no bulk density instability for a one-component collection of identical overdamped chiral spinners. This behavior is consistent with the instability condition in Eq.~{\ref{eq:one_instability}}, together with the sign of the odd viscosity obtained from the Green--Kubo measurements in Fig.~{\ref{fig:S3}} and Fig.~1 of the main text. In our convention, counterclockwise spinning corresponds to $\Omega>0$. In the overdamped regime, the measured odd viscosity has the opposite sign to $\Omega$, such that $\eta_{\mathrm{o}}\Omega<0$ and the instability criterion therefore cannot be satisfied. To illustrate the role of this sign, Fig.~\ref{fig:S10} compares the nonlinear evolution for $\eta_{\mathrm{o}}/\eta=2$, $0$, and $-2$, with all other dimensionless parameters held fixed. For $\Omega>0$, these correspond to positive, zero, and negative $\eta_{\mathrm{o}}\Omega$, respectively. Bulk instability is observed for $\eta_{\mathrm{o}}\Omega=2$, whereas the homogeneous state remains stable for $\eta_{\mathrm{o}}\Omega= 0$ and $-2$.

The one-component bulk model, together with the sign reversal of the odd viscosity reported in the main text (that $\eta_\mathrm{o}\Omega<0$ in the overdamped limit), therefore explains the absence of a bulk instability in the overdamped one-component agent-based simulations shown in Fig.~\ref{fig:S1}.

\section{Free Surface Interfacial Model}\label{sec:6}
To set the stage for the discussion of the stability of the interface between two homochiral domains that rotate in opposite directions in Sec.~\ref{sec:4}, we consider a simpler scenario: the interface between a homochiral domain with constant and uniform density, which is hence incompressible, and vacuum\cite{Soni2019}. The analysis is based on the one-component continuum model introduced in Eqs.~\ref{eq:one_sum1} and~\ref{eq:one_sum2} in the incompressible limit. 
We take the density and spinning frequency to be spatially uniform within the fluid and zero outside. The density-dependent transport coefficients therefore take constant bulk values $\eta$ and $\eta_\mathrm{o}$ inside the fluid and vanish elsewhere. Incompressibility eliminates the contribution from bulk viscosity. To isolate the effect of the odd viscosity, the interface has no passive surface tension.

\subsection{Linear stability analysis}
Applying the same constitutive relations as in Sec.~\ref{sec:2}, the equations governing incompressible chiral fluids in the domain $D$ are
\begin{gather}
    \nabla\cdot\mathbf u=0,\label{eq:incompressible1}\\
    \eta\nabla^2 \mathbf u +\eta_{\mathrm{o}}\bm \epsilon\nabla^2\mathbf u +\eta_R \bm\epsilon \nabla \Omega -\nabla p -\Gamma \mathbf u=0.\label{eq:incompressible2}
\end{gather}
The density in the domain remains constant.
The passive force that arises from the free energy $-n\nabla \mu$ is replaced by $-\nabla p$, where $p$ is the pressure, the Lagrange multiplier that enforces the incompressibility condition $\nabla\cdot \mathbf{u}=0$.
The constant bulk density has been absorbed into the definition of $\eta_R$. For spatially uniform $\Omega$, the torque term vanishes in the bulk but remains in the interfacial force balance. 

Using incompressibility, we obtain $\bm \epsilon \nabla^2\mathbf{u}=\nabla \omega$, where $\omega=\epsilon_{ij}\partial_i u_j$ is the vorticity. The pressure and odd viscous terms can therefore be rewritten as $-\nabla p+\eta_{\mathrm{o}}\nabla \omega= -\nabla(p-\eta_{\mathrm{o}}\omega)$. Defining $q=p-\eta_{\mathrm{o}}\omega$, the bulk equation reduces to $\eta\nabla^2 \mathbf u  -\nabla q -\Gamma \mathbf u=0$. Taking the divergence and curl of the force balance gives
\begin{equation}
    \nabla^2 q=0,\qquad
    \lambda^2\nabla^2\omega -\omega=0,
\end{equation}
where $\lambda=\sqrt{\eta/\Gamma}$ is the hydrodynamic screening length. Thus for spatially uniform $\eta_\mathrm{o}$, odd viscosity drops out of the bulk flow equation and enters the dynamics only through the interfacial force balance as in Ref.~\cite{Soni2019}.

The stress tensor follows Eqs.~\ref{eq:01}-\ref{eq:03} and can then be written as
\begin{equation}
    \sigma_{ij} = -q\delta_{ij}+\eta_R\epsilon_{ij}\Omega+\eta(\partial_i u_j +\partial_j u_i)+2\eta_{\mathrm{o}}\epsilon_{jk}\partial_{k}u_i.
\end{equation}

At the surface of the domain $\partial D$, we neglect surface tension and impose the stress balance 
\begin{equation}
    \bm{\sigma}\cdot \hat{\bm{n}}|_{\partial D} = 0,
\end{equation}
together with the kinematic boundary condition for the domain shape, described by $f(x,y,t)=0$:
\begin{equation} \label{eq:kinematics}
    \frac{Df}{Dt} = \partial_t f+\mathbf{u}\cdot\nabla f=0.
\end{equation}

We describe the domain occupied by the chiral fluid by $D = \{(x,y) |y<g(x,t)\}$, with vacuum above the interface $y=g(x,t)$. 
Hence Eq.~\ref{eq:kinematics} becomes $\partial_t g=u_y - u_x \partial_x g$.
For the unperturbed flat interface $y=g(x,t)=0$, the steady flow is translationally invariant along $x$ and takes the form 
\begin{equation}
    \bar{\bm u}(y)=(\bar{u}(y),0), \qquad \bar{u}(y)=-\frac{\eta_R}{\eta}\Omega\lambda e^{y/\lambda},
\end{equation}
where we use bar to denote the base state.
We then introduce a small interfacial perturbation, $\tilde{g}(x,t)=\hat{g} e^{ikx+s t}$, and linearize the boundary force balance conditions about the flat interface, which gives $\tilde{\bm{\sigma}}(x,0)\hat{\bm{y}}
= \tilde{g}_x(x)\bar{\bm{\sigma}}(x,0)\hat{\bm{x}}-\tilde{g}(x)\bar{\bm{\sigma}}_y(x,0)\hat{\bm{y}}$, where we use tilde to denote first-order perturbations and use subscripts to denote partial derivatives.
In the following derivation, we use $\tilde{u}$ and $\tilde{v}$ to denote the $x$ and $y$ components of the velocity perturbation $\delta \mathbf{u}$, respectively.
Using the base state solution, the $x$ and $y$ components of the boundary force boundary conditions for the first-order perturbations reduce to
\begin{align*}
    \eta\tilde{u}_y+\eta\tilde{v}_x-2\eta_\mathrm{o}\tilde{u}_x
    &=
    2\tilde{g}_x\eta_\mathrm{o}\bar{u}_y-\tilde{g}\eta\bar{u}_{yy},\\
    -\tilde q-2\eta_\mathrm{o}\tilde{v}_x+2\eta\tilde{v}_y
    &= \tilde{g}_x(-\eta_R\Omega+\eta\bar{u}_y),
\end{align*}
where all quantities are evaluated at $y=0$.
Note that subscripts denote partial derivatives.
The general solution to Eqs.~\ref{eq:incompressible1}-\ref{eq:incompressible2} for the first-order perturbations is
\begin{align*}
    \tilde q &= Ae^{|k|y}e^{ikx+s t},\\
    [\tilde{u},\tilde{v}]
    &=
    \left[
    -\frac{ik}{\Gamma}Ae^{|k|y}+Be^{\chi y},
    -\frac{|k|}{\Gamma}Ae^{|k|y}
    -\frac{ik}{\chi}Be^{\chi y}
    \right]e^{ikx+s t}.
\end{align*}
where $\chi = \sqrt{k^2+\lambda^{-2}}$ and A and B are unknown coefficients to be determined from boundary conditions. Substitution into the linearized boundary force balance conditions gives
\begin{align}
    &\frac{-2ik|k|\eta-2k^2\eta_\mathrm{o}}{\Gamma}A
    +\left(
    \eta\chi+\eta\frac{k^2}{\chi}
    -2ik\eta_\mathrm{o}
    \right)B=
    \frac{\Omega\eta_R}{\eta}
    \left(
    -2ik\eta_\mathrm{o}+\frac{\eta}{\lambda}
    \right)\hat{g}, \label{eqn:bc_1}
    \\
    &\left(
    -1+\frac{2ik|k|\eta_\mathrm{o}-2\eta k^2}{\Gamma}
    \right)A
    +
    \left(
    -\frac{2\eta_\mathrm{o}k^2}{\chi}
    -2ik\eta
    \right)B
    \nonumber =
    \left(
    -2ik\eta_R\Omega
    \right)\hat{g}. \label{eqn:bc_2}
\end{align}
The linearized kinematic boundary condition is $\tilde{g}_t=\tilde{v}_y - \bar{u}\tilde{g}_x$ gives
\begin{equation}
    \left(
    s
    -ik\frac{\eta_R\Omega}{\eta}\lambda
    \right)\hat{g}
    =
    -\frac{|k|}{\Gamma}A
    -\frac{ik}{\chi}B. \label{eqn:kinematics}
\end{equation}
Combining Eqs.~\ref{eqn:bc_1}-\ref{eqn:kinematics} yields the dispersion relation $s(k)$. In the long and short wavelength limits of the dispersion relation,
\begin{gather*}
     \mathrm{Im} [s(k)] \sim \begin{cases}
  \frac{2\eta_R\Omega}
    {\eta}(k\lambda)^3 & \text{if $|k\lambda|\ll1$} \\
  \frac{ \eta_R\Omega}{\eta} k\lambda& \text{if $|k\lambda|\gg1$}
\end{cases}
\end{gather*}
\begin{gather*}
    \mathrm{Re} [s(k)] \sim \begin{cases}
  \frac{2\eta_\mathrm{o}\eta_R\Omega}{\eta^2}(|k|\lambda)^3 & \text{if $|k\lambda|\ll1$} \\
    \frac{\eta_R\eta_\mathrm{o}\Omega}{\eta^2+\eta_\mathrm{o}^2} & \text{if $|k\lambda|\gg1$}
\end{cases}
\end{gather*}
The sign of the growth rate is therefore controlled by $\eta_\mathrm{o}\Omega$: $\eta_\mathrm{o}\Omega>0$ destabilizes the free surface, whereas $\eta_\mathrm{o}\Omega<0$ suppresses interfacial perturbations.

\subsection{Incompressible continuum simulations}
\label{sec:incompressible_single_component}
We now examine the nonlinear interfacial dynamics of the overdamped one-component chiral fluid in the incompressible limit by solving the continuum equations Eq.~\ref{eq:incompressible1}-\ref{eq:incompressible2} in the continuum simulations. We use the same characteristic length and time scales, $\lambda$ and $u_0$, and the same $128$ by $128$ grid on a periodic domain as in the compressible simulations above. The initial density field is generated from a spatially correlated random field with correlation length $1.5\lambda$ and mapped onto the bounded interval $0\leq n\leq 1$. This produces an initially segregated configuration between the odd viscous fluid where $n$ is 1 and the vaccum where $n$ is 0 with irregular interfaces between them. This allows us to study the evolution of the interface of a viscous fluid. When solving the advection equation for the density field, to preserve the sharpness of the interface at which the density gradient is large, we use THINC (Tangent of Hyperbola for Interface Capturing) scheme to capture the interface motion and suppress numerical smearing at the interfaces. 

\begin{figure}
    \centering
    \includegraphics[width=1.0\linewidth]{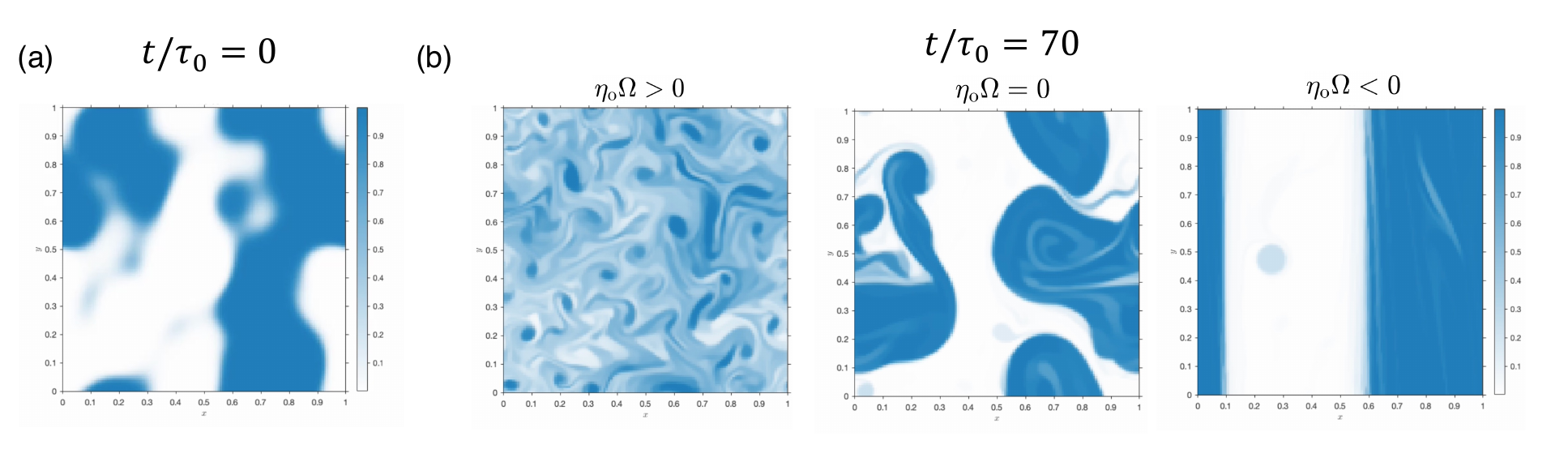}
    \caption{\textbf{Incompressible continuum simulations of one-component interfacial stability.} \textbf{(a)}~ Initial configuration at $t/\tau_0=0$, generated from a Gaussian random field with a Gaussian covariance kernel. \textbf{(b)}~Density fields at $t/\tau_0=70$ for $\eta_{\mathrm{o}}/\eta=2$, $0$, and $-2$, corresponding to $\eta_\mathrm{o}\Omega>0$, $\eta_\mathrm{o}\Omega=0$, and $\eta_\mathrm{o}\Omega<0$ respectively for $\Omega>0$. The color bar denotes the density $n$, ranging from $0$ to $1$.}
    \label{fig:S100}
\end{figure}

With all other dimensionless parameters fixed, Fig.~\ref{fig:S100} compares the nonlinear evolution for $\eta_{\mathrm{o}}/\eta=2$, $0$, and $-2$ at $t/\tau_0=70$. In our simulations where $\Omega>0$, these cases correspond to $\eta_{\mathrm{o}}\Omega>0$, $\eta_{\mathrm{o}}\Omega=0$, and $\eta_{\mathrm{o}}\Omega<0$ respectively. The three cases exhibit qualitatively distinct interfacial dynamics. For $\eta_{\mathrm{o}}/\eta=2$, the interfaces become unstable and develop finer interconnected structures. For $\eta_{\mathrm{o}}/\eta=0$, the interfaces remain curved and continue to evolve while the domain continues to deform and rotate. In contrast, for $\eta_{\mathrm{o}}/\eta=-2$, interfacial deformations are strongly suppressed, and the system relaxes towards a slab-like configuration with nearly planar interfaces. 
The latter case of $\eta_\mathrm{o}\Omega<0$ corresponds to the sign of the odd viscosity obtained from the Green--Kubo measurements in the overdamped regime. 

\section{Interfacial Stability Analysis for the Binary Mixture}\label{sec:4}
We next generalize the one-component continuum model to a binary chiral active mixture with densities $n_{\alpha}$ and velocity fields $\mathbf u_\alpha$, where $\alpha\in \{A,B\}$ labels the two species. Extending Eqs.~\ref{eq:one_sum1} and \ref{eq:one_sum2} to a two-component system,
\begin{gather}
  \partial_t n_\alpha + \nabla \cdot (n_\alpha \mathbf{u}_\alpha) = 0,\label{eq:binary1} \\
  \nabla \cdot (\sigma_{v,\alpha} + \sigma_{o,\alpha} + \sigma_{\tau,\alpha}) - n_\alpha \nabla\mu_\alpha - \Gamma_\alpha \mathbf{u}_\alpha - \Gamma_f (\mathbf{u}_\alpha - \mathbf{u}_\beta) = 0,\label{eq:binary2}
\end{gather}
where $\beta\neq \alpha$. The viscous, odd viscous, and torque stresses have the same constitutive form as in Sec.~\ref{sec:2}, which is applied separately to each component. The chemical potentials $\mu_\alpha =\delta F/\delta n_{\alpha}$ are obtained from the binary free energy $F[n_A,n_B]$ and determine the passive thermodynamic force densities. The term proportional to $\Gamma_f$ represents interspecies drag and penalizes relative motion between the two components. 
In the main text, we noted at the late stages of phase separation, the two components are nearly completely segregated into complementary domains and that the density remains nearly uniform. In this section, we focus on the limit where the components are incompressible and there exists a sharp interface between domains, and perform full simulations and analysis of the stability of the interface.

\subsection{Bulk model and boundary conditions}
To analyze the interfacial stability, we consider the late-stage configuration in which the two species are well segregated by a sharp interface. Away from the interface, each phase is dominated by a single component with uniform density. We treat each bulk phase as an incompressible one-component chiral fluid, as in Eqs.~\ref{eq:incompressible1} and \ref{eq:incompressible2}.

We consider a horizontal interface that separates the upper and lower phases, denoted by $+$ and $-$, respectively, and write
\begin{gather*}
    \mathbf{u}(x,y,t)=\begin{cases}
  \mathbf{u}^+(x,y,t) & \text{if $y>g(x,t)$} \\
  \mathbf{u}^-(x,y,t) & \text{if $y<g(x,t)$}
\end{cases},\qquad
\Omega=\begin{cases}
  \Omega^+ & \text{if $y>g(x,t)$} \\
  \Omega^- & \text{if $y<g(x,t)$}
\end{cases}.
\end{gather*}
Here $g(x,t)$ denotes the interfacial height. The interface is $\partial D=\{(x,y)|f(x,y,t)=0\}$, where $f(x,y,t)=y-g(x,t)$. We allow $|\Omega^+|\neq |\Omega^-|$, so that the analysis is not restricted to the exactly counter-rotating phases, though we assume that the shear viscosity $\eta$ is the same everywhere. Within each phase, $\Omega^\pm$ and $\eta_{\mathrm{o}}^{\pm}$ are spatially uniform. Defining $q^\pm =p^\pm-\eta_\mathrm{o}^\pm \omega^\pm$, the bulk equations reduce to 
\begin{gather}
    \nabla\cdot\mathbf{u}^\pm = 0,\\
    \eta\nabla^2 \mathbf u^\pm  -\nabla q^\pm -\Gamma \mathbf u^\pm=0.
\end{gather}
Consequently, 
\begin{equation}
    \nabla^2 q^\pm=0,\qquad\lambda^2\nabla^2\omega^\pm -\omega^\pm=0,
\end{equation}
with $\lambda=\sqrt{\eta/\Gamma}$ the hydrodynamic screening length defined above. As in the one-component case, odd viscosity is absorbed into the effective pressure in the incompressible bulk and enters only through the interfacial stress. The stress in each phase is 
\begin{equation}
    \sigma_{ij}^{\pm} = -q^{\pm}\delta_{ij}+\eta_R\epsilon_{ij}\Omega^{\pm}+\eta(\partial_i u_j^{\pm} +\partial_j u_i^{\pm})+2\eta_{\mathrm{o}}^{\pm}\epsilon_{jk}\partial_k u_i^\pm.
\end{equation}
To isolate the effect of chiral stresses on interfacial stability, we neglect capillary stress so the interfacial stress condition reduces to continuity of traction
\begin{equation}
    (\bm{\sigma}^+ -\bm{\sigma}^-)\cdot \hat{\bm{n}}|_{\partial D} = 0,
\end{equation} 
together with continuity of velocity
\begin{equation}
    \tilde{\mathbf{u}}^+|_{\partial D}=\tilde{\mathbf{u}}^-|_{\partial D} = \mathbf{u}|_{\partial D},
\end{equation} and the kinematic condition $\partial_t f+\mathbf{u}\cdot\nabla f=0$ for the evolution of the boundary.

\subsection{Steady state solutions}
We first determine the steady, unperturbed two-phase base state with a planar interface about which the interfacial stability analysis is performed. We consider two semi-infinite half spaces separated by a flat interface at $y=0$. The base flow is translationally invariant along the interface and time independent, with the velocity fields vanishing as $y \to \pm \infty$. Together with the bulk equations, these conditions give exponentially decaying solutions away from the interface. 
Similar to Sec.~\ref{sec:3}, we use a hat to denote the base state flow $\mathbf{u}^\pm(x,y)=(\bar{u}^\pm(x),0)$.

For the flat steady interface, the kinematic boundary condition is automatically satisfied, whereas the stress boundary conditions reduce to
\begin{equation}
    \eta_R(\Omega^+ - \Omega^-) - \eta(\omega^+ - \omega^-) = 0,
    \qquad
    -(p^+ - p^-) +(\eta_{\mathrm{o}}^+\omega^+-\eta_{\mathrm{o}}^-\omega^-) = 0.
\end{equation}
These conditions are satisfied by the base state vorticity solutions
\begin{equation}
    \bar{\omega}^+(y) = \frac{\eta_R}{2\eta}(\Omega^+ - \Omega^-)e^{-y/\lambda},\qquad
    \bar{\omega}^-(y) = \frac{\eta_R}{2\eta}(\Omega^- - \Omega^+)e^{y/\lambda},
\end{equation}
and the corresponding tangential velocity solutions
\begin{equation}
    \bar{u}^+(y) = \frac{\eta_R}{2\eta}(\Omega^+ - \Omega^-)\lambda e^{-y/\lambda},\qquad
    \bar{u}^-(y) = \frac{\eta_R}{2\eta}(\Omega^+ - \Omega^-)\lambda e^{y/\lambda}.
\end{equation}

As a consistency check, when $\Omega^- = -\Omega^+$, corresponding to a configuration with mirror symmetry, the steady-state velocity field $\bar{u}$ reduces to the known solution for a chiral fluid next to a vacuum \cite{Soni2019}. This correspondence can be understood through the method of mirror images, where the opposite side of the interface is replaced by a mirror image fluid to enforce the free-slip boundary condition in the free surface problems~\cite{Wang2013}.

\subsection{Linearized bulk equations and boundary conditions}

We allow the two chiral bulk phases to have different odd viscosities $\eta_{\mathrm{o}}^\pm$. An infinitesimal, zero-mean perturbation $y=\tilde{g}(x,t)=\hat{g} e^{ikx+st}$ is introduced to the flat interface described above.
We use the same symbols as in Sec.~\ref{sec:3}. We use $\tilde{u}$ and $\tilde{v}$ to denote the $x$ and $y$ components of the perturbed velocity, use bar to denote the base state quantities, and tilde to denote first-order perturbations. The perturbed velocity, vorticity, stress, and pressure are:
\begin{align*}
[u^{\pm}(x,y), v^{\pm}(x,y)]&=[\bar{u}^{\pm}+\tilde{u}^{\pm}(x,y), \tilde{v}^{\pm}(x,y)],\\
\omega^{\pm} &= \bar{\omega}^{\pm}(y)+\tilde{\omega}^{\pm}(x,y),\\
\bm{\sigma}^{\pm} &= \bar{\bm{\sigma}}^{\pm}(y)+\tilde{\bm{\sigma}}^{\pm}(x,y),\\
q^{\pm}&=\tilde q^{\pm}(x,y).
\end{align*}
We next linearize the interfacial boundary conditions.
Linearizing the stress balance equation to first order in $\epsilon$ gives:
\begin{align*}
    [\tilde{\bm{\sigma}}^+(x,0)-\tilde{\bm{\sigma}}^-(x,0)]\hat{\bm{y}} = \tilde{g}_x(x)[\bar{\bm{\sigma}}^+(0)-\bar{\bm{\sigma}}^-(0)]\hat{\bm{x}}-\tilde{g}(x)[\bar{\bm{\sigma}}^+_y(0)-\bar{\bm{\sigma}}^-_y(0)]\hat{\bm{y}},
\end{align*}
where the subscripts denote partial derivatives. Equivalently,
\begin{gather*}
 \begin{bmatrix} \tilde{\sigma}_{xy}^+(x,0)- \tilde{\sigma}_{xy}^-(x,0) \\ \tilde{\sigma}_{yy}^+(x,0)- \tilde{\sigma}_{yy}^-(x,0)  \end{bmatrix}
 =
   \tilde{g}_x(x)
  \begin{bmatrix} \bar{\sigma}_{xx}^+(0) - \bar{\sigma}_{xx}^-(0) \\ \bar{\sigma}_{yx}^+(0)- \bar{\sigma}_{yx}^-(0) \end{bmatrix} \\ -\tilde{g}(x)\partial_y \begin{bmatrix} \bar{\sigma}_{xy}^+(0) - \bar{\sigma}_{xy}^-(0) \\ \bar{\sigma}_{yy}^+(0)- \bar{\sigma}_{yy}^-(0) \end{bmatrix}.
\end{gather*}

Given that $\tilde{\omega}^{\alpha}=\partial_x \tilde{v}^{\alpha}-\partial_y \tilde{u}^{\alpha}$ and $\bar{\omega}^{\alpha}=-\partial_y \bar{u}^{\alpha}$, the simplified normal and tangential stress boundary equations are
\begin{gather} \label{eqn:interface_LSA_force_BC}
    \eta(\tilde{u}_y^+ - \tilde{u}_y^-)+\eta(\tilde{v}_x^+ - \tilde{v}_x^-)-2(\eta_{\mathrm{o}}^+\tilde{u}_x^+-\eta_{\mathrm{o}}^-\tilde{u}_x^-) = 2\tilde{g}_x(\eta_{\mathrm{o}}^+\bar{u}_y^+-\eta_{\mathrm{o}}^-\bar{u}_y^-)-\tilde{g}\eta(\bar{u}_{yy}^+ - \bar{u}_{yy}^-),\\
    -(\tilde q^+-\tilde q^-)-2(\eta_{\mathrm{o}}^+\tilde{v}_x^+-\eta_{\mathrm{o}}^-\tilde{v}_x^-)+2\eta(\tilde{v}_y^+ - \tilde{v}_y^-) = \tilde{g}_x[-\eta_R (\Omega^+-\Omega^-)+\eta(\bar{u}_y^+-\bar{u}_y^-)].
\end{gather}
Across the perturbed interface, the linearized kinematic boundary condition then becomes $\tilde{v}^-(x,0)=\tilde{g}_t+\tilde{g}_x\bar{u}^-(0)$.
The normal and tangential velocity continuity equations $\mathbf{u}^+|_{\partial D}=\mathbf{u}^-|_{\partial D}$, after linearization, give:
\begin{align} \label{eqn:interface_LSA_velocity}
    \tilde{u}^+(x,0) + \tilde{g}\partial_y \bar{u}^+(0) &= \tilde{u}^-(x,0) + \tilde{g}\partial_y \bar{u}^-(0),\\
    \tilde{v}^+(x,0) &= \tilde{v}^-(x,0).
\end{align}

\subsection{Interfacial dispersion relation}\label{sec:s4d}
Since the system is translationally invariant along $x$, the first-order perturbation of the velocity and $q$ have the form of $\tilde{u}(x,y,t)=\hat{u}(y)e^{ikx+st}$, $\tilde{v}(x,y,t)=\hat{v}(y)e^{ikx+st}$, $\tilde{q}(x,y,t)=\hat{q}(y)e^{ikx+st}$. Projecting the linearized bulk equation to a Fourier mode with a wavevector $k$, we have 
\begin{gather*}
    \eta(\partial_{yy}-k^2)\begin{bmatrix}
        \hat{u},\hat{v}
    \end{bmatrix}^{\pm} - \Gamma \begin{bmatrix}
        \hat{u},\hat{v}
    \end{bmatrix}^{\pm} -\begin{bmatrix}
        ik, \partial_y
    \end{bmatrix} \hat q^{\pm} = \bm{0},\\
    ik \hat{u}^\pm + \partial_y\hat{v}^\pm = 0,\\
    (\partial_{yy}-k^2)\hat q^{\pm}=0.
\end{gather*}

Solving the bulk equations above for $\hat{u}(y)$, $\hat{v}(y)$, and $\hat{q}(y)$ so that the velocity field vanishes at $\pm \infty$ gives
\begin{align} \label{eqn:interface_LSA_bulk}
    \tilde q^+ &= A^+ e^{-|k|y}e^{ikx+s t},\\
    \tilde q^- &= A^- e^{|k|y}e^{ikx+s t},\\
    [\tilde{u},\tilde{v}]^+ &= \left[-\frac{ik}{\Gamma}A^+e^{-|k|y}+B^+e^{-\chi y}, \frac{|k|}{\Gamma}A^+e^{-|k|y}+\frac{ik}{\chi}B^+e^{-\chi y}\right]e^{ikx+s t}, \\
    [\tilde{u},\tilde{v}]^- &= \left[-\frac{ik}{\Gamma}A^-e^{|k|y}+B^-e^{\chi y}, -\frac{|k|}{\Gamma}A^-e^{|k|y}-\frac{ik}{\chi}B^-e^{\chi y}\right] e^{ikx+s t},
\end{align}
where $\chi = \sqrt{k^2+\lambda^{-2}}$. 
Because the base state interfacial velocity is $\bar u_0=\bar{u}^{\pm}(0)=\frac{\eta_R}{2\eta}(\Omega^+ - \Omega^-)\lambda$. The kinematic boundary condition becomes
\begin{equation*} \label{eqn:interface_LSA_kinematics}
    (s +ik\bar u_0)\hat{g} = -\frac{|k|}{\Gamma}A^- - \frac{ik}{\chi}B^- = \frac{|k|}{\Gamma}A^+ + \frac{ik}{\chi}B^+.
\end{equation*}
Plugging the bulk solution Eq.~\ref{eqn:interface_LSA_bulk} into the interfacial force balance Eq.~\ref{eqn:interface_LSA_force_BC} and velocity continuity Eq.~\ref{eqn:interface_LSA_velocity} and combine that with the kinematic boundary condition Eq.~\ref{eqn:interface_LSA_kinematics}, we obtain a system of linear equations
\begin{gather*}
    \begin{bmatrix}
        \frac{2(i\eta k|k|-\eta_{\mathrm{o}}^+ k^2)}{\eta}\lambda^2 & \frac{2(i\eta k|k|+\eta_{\mathrm{o}}^-k^2)}{\eta}\lambda^2 & -\frac{\eta k^2}{\chi}-2ik\eta_{\mathrm{o}}^+ - \eta\chi & -\frac{\eta k^2}{\chi}+2ik\eta_{\mathrm{o}}^- - \eta\chi & 2ik\frac{\bar u_0}{\lambda}(\eta_{\mathrm{o}}^+ +\eta_{\mathrm{o}}^-)\\
        -\frac{2(i\eta_{\mathrm{o}}^+ k|k| +\eta k^2)}{\eta}\lambda^2 - 1 & \frac{2(\eta k^2-i\eta_{\mathrm{o}}^- k|k|)}{\eta}\lambda^2 +1 & -2\eta ik + \frac{2k^2}{\chi}\eta_{\mathrm{o}}^+ & 2\eta ik+\frac{2k^2}{\chi}\eta_{\mathrm{o}}^- & \frac{4ik\eta\bar u_0}{\lambda}\\
        \frac{|k|}{\Gamma} & \frac{|k|}{\Gamma} & \frac{ik}{\chi} & \frac{ik}{\chi} & 0 \\
        -\frac{ik}{\Gamma} & \frac{ik}{\Gamma} & 1 & -1 & -\frac{2 \bar u_0}{\lambda} \\
        0 & \frac{|k|}{\Gamma} & 0 & \frac{ik}{\chi} & s+ik\bar u_0
    \end{bmatrix}
    \begin{bmatrix}
        A^+ \\ A^- \\ B^+ \\ B^- \\ \hat{g}
    \end{bmatrix} =0 .
\end{gather*}
Setting the determinant of the matrix to zero gives the interfacial dispersion relation. Its asymptotic analytical form in the long and short wavelength limits are
\begin{gather*}
     \mathrm{Im} [s(k)] \sim \begin{cases}
  -\frac{\eta_R(\Omega^+-\Omega^-)}{4\eta}(\lambda k)^3 & \text{if $|k\lambda|\ll1$} \\
  -\bar u_0 k +\frac{4\eta^2\bar u_0 k}
{\left[4\eta^2+(\eta_{\rm o}^{+}-\eta_{\rm o}^{-})^2\right]\lambda|k|}& \text{if $|k\lambda|\gg1$}
\end{cases}
\end{gather*}
\begin{gather*}
    \mathrm{Re} [s(k)] \sim \begin{cases}
  \frac{\eta_R(\Omega^+-\Omega^-) (\eta_\mathrm{o}^+-\eta_\mathrm{o}^-)}{2\eta^2}(\lambda|k|)^3 & \text{if $|k\lambda|\ll1$} \\
    \frac{2\eta(\eta_{\rm o}^{+}-\eta_{\rm o}^{-})\bar u_0}
{\lambda\left[4\eta^2+(\eta_{\rm o}^{+}-\eta_{\rm o}^{-})^2\right]} & \text{if $|k\lambda|\gg1$}
\end{cases}
\end{gather*}
The imaginary part of the full dispersion relation is odd in $k$. Since the phase velocity $c\equiv v_{\text{ph}}=-\mathrm{Im}{[s(k)]}/k$, modes with $\pm k$ propagate in the same direction, corresponding to unidirectional interfacial waves.

We now specialize to two exactly counter-rotating phases $\Omega^+=-\Omega^-$, with equal and opposite odd viscosities $\eta_\mathrm{o}^+=-\eta_\mathrm{o}^-$. In this case, the long wavelength growth rate is proportional to the common product $\eta_\mathrm{o}^\alpha \Omega^\alpha$ in either phase. Thus, $\eta_\mathrm{o}^\alpha \Omega^\alpha<0$ gives a negative growth rate and stabilizes the interface, whereas $\eta_\mathrm{o}^\alpha \Omega^\alpha>0$ destabilizes it. The sign measured with the Green--Kubo method in the overdamped regime therefore corresponds to the stable branch. When $\Omega^+=-\Omega^-$, the system has mirror symmetry, and the instability growth rate is the same as the one-component free surface result in both the long and short wavelength limits. The propagation rate at the binary interface is reduced by the presence of the second chiral fluid at long wavelengths and acquires a frequency offset at short wavelengths, while retaining the same odd parity in $k$.

For comparison with the agent-based simulations, we assign the upper half plane to species $B$, which rotates clockwise, and the lower half plane to species A, which rotates counterclockwise. With counterclockwise rotation defined as positive, this gives $\Omega^+ = \Omega_B<0$ and $\Omega^- = \Omega_A>0$. The resulting dispersion relation therefore predicts a stable interface with unidirectional propagation in the $-x$ direction, consistent with the late stage dynamics observed in the agent-based simulations.

For the dispersion curves in main text Fig.~2, we set $\eta=\Gamma=1$, $\eta_R=2$, $\Omega^+ = -1$, and $\Omega^- = 1$, so that $\lambda =\sqrt{\eta/\Gamma}=1$. We assign odd viscosities of equal magnitude and opposite sign to the two counter-rotating phases, $\eta_\mathrm{o}\equiv\eta_\mathrm{o}^-=-\eta_\mathrm{o}^+$, and vary the normalized odd viscosity across $\eta_\mathrm{o}/\eta = 1, 0.5,-0.5,-1$ to study the effect of odd viscosity sign reversal on interfacial stability. The dispersion relation is evaluated over $0<k\lambda\leq 5$. 

\begin{figure}
    \centering
    \includegraphics[width=1.0\linewidth]{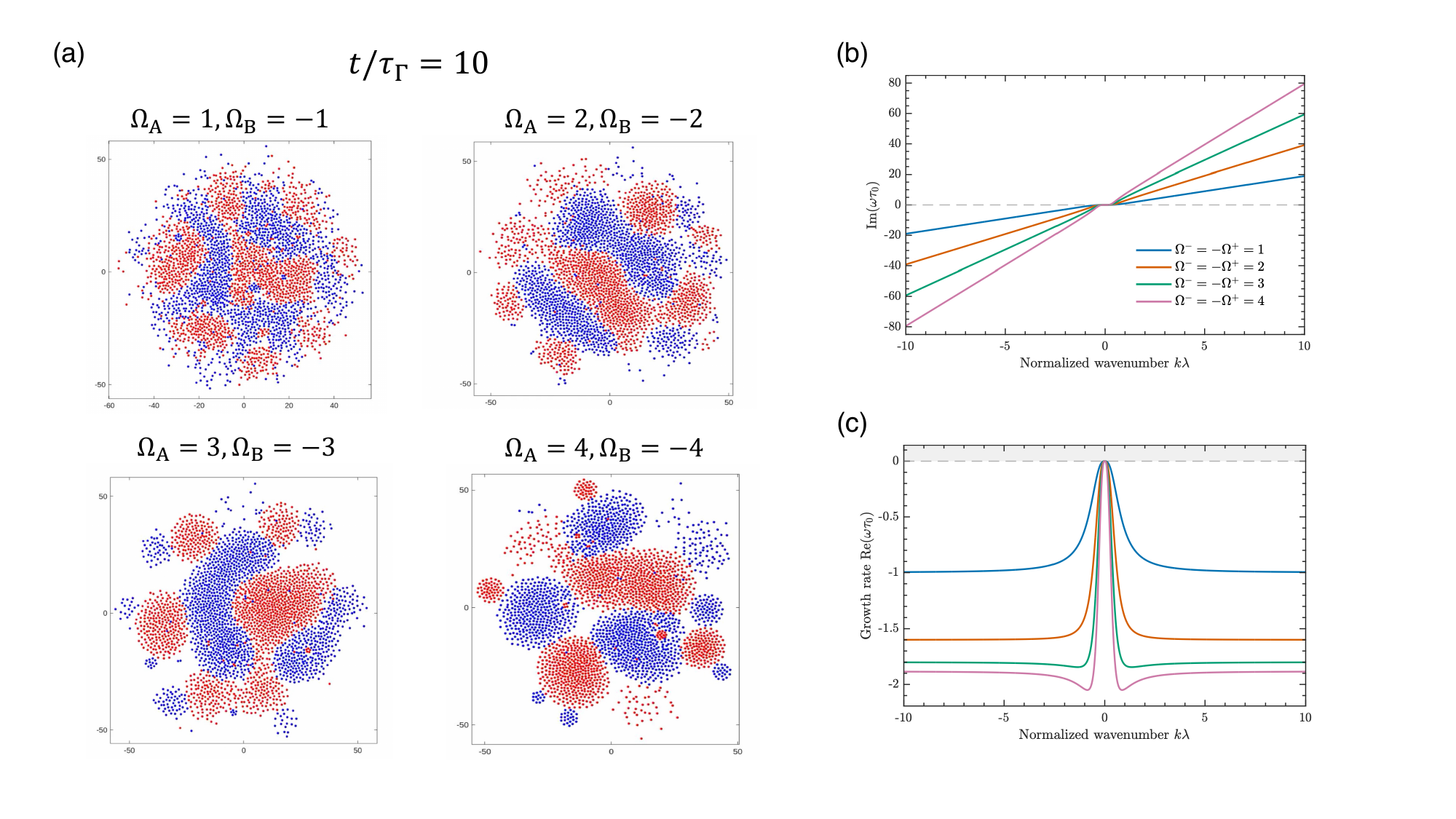}
    \caption{\textbf{Effect of spinning frequency on interfacial dynamics of a binary chiral mixture.} \textbf{(a)}~ Agent-based simulations at $t/\tau_\Gamma=10$ for $|\Omega|=1,2,3,4$ showing progressively clearer interfaces and stronger segregation between spinners that rotate in opposite directions with increasing $|\Omega|$. \textbf{(b)}~Imaginary part of the analytical interfacial dispersion $\mathrm{Im}{\omega}$, showing unidirectional propagation for all $|\Omega|$. \textbf{(c)}~Growth rate $\mathrm{Re}{\omega}$, which becomes more negative as $|\Omega|$ increases, indicating enhanced interfacial stability. The trend is consistent with the agent-based simulation results in panel (a).}
    \label{fig:S7}
\end{figure}

In Fig.~\ref{fig:S7}(b-c), we fix $\eta_\mathrm{o}^+ =-1$ and $\eta_\mathrm{o}^- =1$, consistent with the signs of the odd viscosities measured in the overdamped regime. Then we investigate the effect of increasing $|\Omega|$ on the interfacial waves between counter-rotating chiral fluids. The analytical dispersion predicts unidirectional propagation for all $|\Omega|$, while the growth rate becomes increasingly negative as $|\Omega|$ increases. This trend is consistent with the stronger suppression of interfacial fluctuations, faster coarsening, and decreasing contact between the two phase-separated domains with increasing $|\Omega|$ observed in the late stages of particle simulations (Fig.~\ref{fig:S7}(a)).

\subsection{Incompressible continuum simulations}
\label{sec:incompressible_two_fluid}
We simulate the incompressible two-fluid model using the same nondimensionalization and THINC scheme for capturing moving interfaces as in the free surface interfacial simulations in Sec.~\ref{sec:incompressible_single_component}. The two components have equal shear viscosities, $\eta_A=\eta_B\equiv\eta$, and equal substrate friction coefficients, $\Gamma_A=\Gamma_B\equiv \Gamma$. Since the interspecies friction occurs when both species are present, we assume the following density dependence for the interspecies friction coefficient
\begin{equation}
\Gamma_f(n_A,n_B)
=\Gamma_{f0}\frac{n_An_B}{n_A+n_B},
\end{equation}
with $\Gamma_{f0} n_0/\Gamma=50$, so that relative motion is strongly suppressed within the mixed interfacial region, providing an approximation to the velocity continuity condition imposed in the sharp interface model. In the sharp interface limit, the interspecies drag terms cancel upon summing the two force balance equations, and integration of the overall force balance across the interface yields continuity of the traction. Consistent with the one-component continuum simulations, the viscosities depend linearly on the particle densities. The two species are assigned equal magnitude and opposite sign for spinning frequencies and odd viscosity coefficients, with $\eta_\mathrm{o} \Omega <0$ corresponding to the sign measured in the overdamped particle simulations.

\begin{figure}
    \centering
    \includegraphics[width=1.0\linewidth]{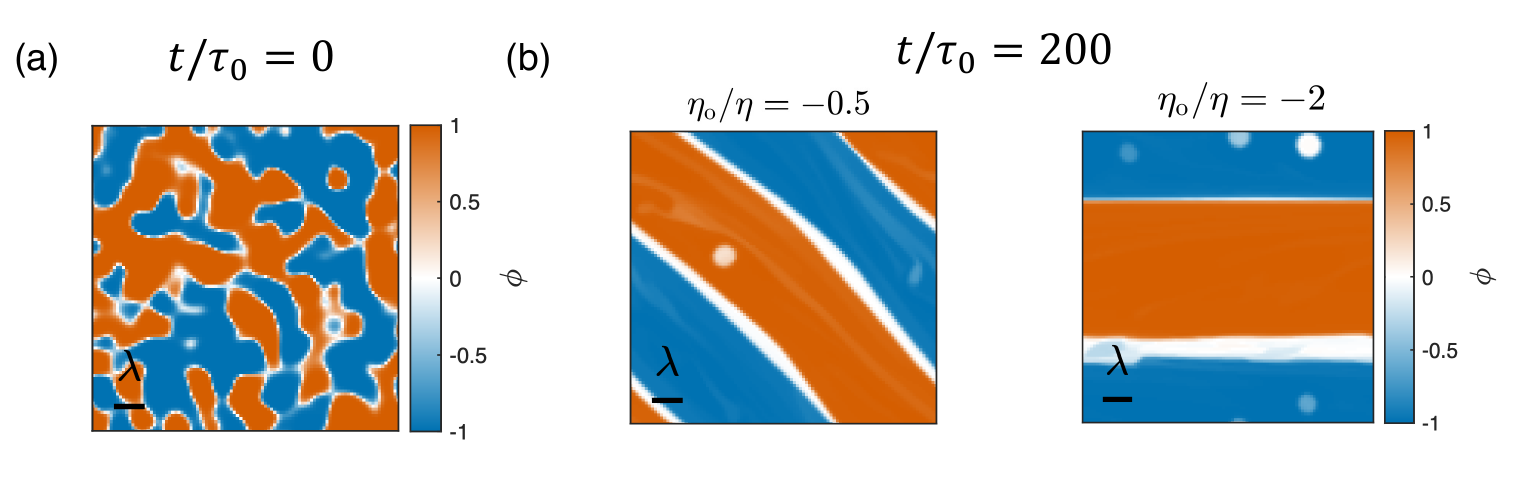}
    \caption{\textbf{Incompressible continuum simulations of binary interfacial dynamics.} \textbf{(a)}~ Initial configuration at $t/\tau_0=0$, generated from a Gaussian random field with a Gaussian covariance kernel. \textbf{(b)}~Composition fields at $t/\tau_0=200$ for $\eta_\mathrm{o}/\eta = -0.5$ and $\eta_\mathrm{o}/\eta = -2$ respectively for $\Omega>0$. The color bar denotes the composition $\phi=(n_A-n_B)/(n_A+n_B)$, ranging from $-1$ to $1$, where $n_A$ is the density of counterclockwise spinning chiral fluid, and $n_B$ is denotes the clockwise one.}
    \label{fig:S101}
\end{figure}

The initial condition is generated from a spatially correlated random composition field with $n_B =1-n_A$, producing a segregated configuration with irregular interfaces and large density gradients near the interface. As shown in Fig.~\ref{fig:S101}, increasing $|\eta_\mathrm{o}|/\eta$ when $\eta_\mathrm{o} \Omega <0$ stabilizes the interfaces and promotes coarsening of the phase-separated domains, consistent with the late-stage dynamics from the agent-based simulations and the interfacial stability analysis above.

\subsection{Coarse-grained origin of the cross parity-violating response}
\label{sec:coarse_graining}
In the previous sections, the interfacial linear stability analysis and continuum simulations show that, for $\eta_\mathrm{o} \Omega<0$, interfaces between oppositely rotating components are stabilized when relative slip is suppressed, and the macroscopically defined domains formed by the two species coarsen over time. We note that, even in a macroscopically homogeneous mixture, fluctuations of the local composition still generate microdomains enriched in counterclockwise ($A$) or clockwise ($B$) spinners at smaller length scales. A natural question is then whether the same physical mechanism of odd viscosity-driven interfacial stability can remain relevant for local structure approaching the microscopic particle length scale, and how its averaged effect can enter the continuum description of a macroscopically mixed state.

Two observations motivate such a connection. First, suppression of relative motion between spinners of opposite chirality is observed in mixed regions, providing a local analogue of the velocity continuity condition used in the sharp interface theory. Cooperative particle motion persists even for microdomains approaching the particle length scale and is localized near interfaces, as reported in Ref.~\cite{Nguyen2014}.
Second, the response to an interfacial deformation in our continuum theory is itself spatially localized. The velocity field perturbations decay away from the interface over the length scales $k^{-1}$ and $\chi^{-1}=(k^2+\lambda^{-2})^{-1}<k^{-1}$, with both decay lengths of order $|k|^{-1}$ for $|k\lambda|\gg1$. Thus, when we zoom in on the microdomains, where the boundary between A and B domains fluctuates at a length scale on the order of particle size and not exceeding the size of the microdomain, the interfacial fluctuations then produce a corresponding localized response close to the interface and within the microdomain. While this does not imply that the sharp interface continuum theory remains quantitative at the scale of microdomains whose sizes are on the order of a few particles, the parity-violating forces, which are the microscopic mechanism that give rise to odd viscosity, are still at play at this length scale and couple shear flow between particles to a normal force. Near the interface between microdomains, the response to the interface fluctuation is local, and the averaged effect of the nonlinear response is a force that stabilizes the undulations, whereas the linear response to infinitesimal interfacial perturbations averages to zero.

Indeed, if we consider the coarse-grained continuum model where there is no A-B crosstalk, based on Sec.~\ref{sec:s2.1}, $\eta_\mathrm{o}\Omega<0$ suppresses the density variation. Near the interface between the microdomains, particle A will penetrate into B-rich microdomains, and vice versa. Therefore, the cross-species coupling from the interaction at the interface between microdomains is crucial for the phase separation. A possible microscopic origin of the cross-species coupling is as follows. As a B-rich microdomain shears past neighboring A particles, steric contacts can bias the orientations of nearby A-A pairs. Because the spinners in each A-A pair exert a spin-dependent transverse force on one another, this orientational bias can contribute to a normal-stress difference in A proportional to the shear rate in B. Upon coarse-graining, such a response force, arising from the interaction between A and B near the interface, couples the shear flow of one component to the normal stress of another component.

In summary, we propose the following physical picture. Composition fluctuations generate local $A$-$B$ interfaces, whose stability is controlled by the odd viscous response of the constituent chiral fluids. We hypothesize that the nonlinear cross-species response to the locally generated flows at these interfaces modifies the constitutive relation of the mixture upon coarse-graining. At the continuum level, this additional response is represented by the parity violating coefficient $\eta_\mathrm{o}^{\prime}$, which characterizes the transverse response of one component to the flow of the other. In the following section, we independently measure this transport coefficient from the microscopic stress, examine its consequences in the bulk stability analysis, and compare the resulting predictions with numerical simulations. 

\section{Binary Mixture Bulk Model and Stability Analysis}\label{sec:3}
\subsection{Continuum model for the binary mixture}
We now extend the two-fluid model in Equations~\ref{eq:binary1} and~\ref{eq:binary2} to include cross species viscous couplings between the two velocity fields. As discussed above, these effective couplings can arise from the response to fluctuations of the microdomain interfaces in the chiral mixture. The Eqs.~\ref{eq:01}-\ref{eq:03} are therefore generalized to
\begin{align}
    \sigma_{ij,v,\alpha}&=\eta_{\alpha\beta} \left(\partial_i u_{j,\beta} + \partial_j u_{i,\beta} - \frac{2}{d} \delta_{ij} \partial_k u_{k,\beta} \right) + \zeta_{\alpha\beta} \delta_{ij} \partial_k u_{k,\beta}, \label{eq:multi_viscous_stress} \\
    \sigma_{ij,o,\alpha} &= \eta_{\mathrm{o},\alpha\beta} \left( \epsilon_{ik} \partial_j u_{k,\beta} + \epsilon_{jk} \partial_k u_{i,\beta} \right), \label{eq:multi_odd_stress} \\
    \sigma_{ij,\tau,\alpha} &= \eta_{R,\alpha} \Omega_\alpha  n_\alpha \epsilon_{ij}, \label{eq:multi_torque}
\end{align}
to include cross species couplings.

For a compressible equal composition binary mixture of counter-rotating chiral active fluids, we introduce the cross species shear viscosity $\eta^{\prime}$ and transverse viscosity $\eta_\mathrm{o}^{\prime}$.
The following quantities satisfy the combined parity and species exchange symmetry. 
For species $A$ representing the counterclockwise spinners and species $B$ the clockwise spinners, we write $\Omega_A =-\Omega_B\equiv \Omega$ and $\eta_{\mathrm{o},AA}=-\eta_{\mathrm{o},BB} \equiv \eta_{\mathrm{o}}$, $\eta_{\mathrm{o},AB}=-\eta_{\mathrm{o},BA}=-\eta_{\mathrm{o}}'$. Omitting the bulk viscosity $\eta_{\alpha\beta}$, the bulk model then becomes
\begin{gather}
    \partial_t n_A + \nabla\cdot(n_A\mathbf{u}_A) = 0, \\
    \partial_t n_B + \nabla\cdot(n_B\mathbf{u}_B) = 0, \\
    \eta\nabla^2\mathbf{u}_A+\eta^{\prime}\nabla^2\mathbf{u}_B+\eta_R\nabla\cdot(\mathbf{\Omega}n_A)-\Gamma\mathbf{u}_A-n_A \nabla \mu_A  +\eta_{\mathrm{o}}\mathbf{\epsilon}\nabla^2\mathbf{u}_A-\eta_{\mathrm{o}}^{\prime}\mathbf{\epsilon}\nabla^2\mathbf{u}_B - \Gamma_f (\mathbf{u}_A - \mathbf{u}_B)=0,~\label{eq:mixture1}\\
    \eta\nabla^2\mathbf{u}_B+\eta^{\prime}\nabla^2\mathbf{u}_A-\eta_R\nabla\cdot(\mathbf{\Omega}n_B)-\Gamma\mathbf{u}_B -n_B \nabla \mu_B  -\eta_{\mathrm{o}}\mathbf{\epsilon}\nabla^2\mathbf{u}_B+\eta_{\mathrm{o}}^{\prime}\mathbf{\epsilon}\nabla^2\mathbf{u}_A - \Gamma_f (\mathbf{u}_B - \mathbf{u}_A)=0.~\label{eq:mixture2}
\end{gather}
The passive force density is written in terms of the chemical potential as $ -n_{\alpha} \nabla \mu_{\alpha}$, where $\mu_\alpha=\mu_{h,\alpha}(n_A, n_B)-\kappa \nabla^2 n_\alpha$ and $\mu_{h,\alpha}$ is the homogeneous contribution. 

\subsection{Bulk instability of the binary mixture}
\label{sec:SI_bulk_instability_binary}
We consider perturbations about the homogeneous steady state of the form $n_\alpha =n_{0,\alpha}+\delta \hat{n}_\alpha e^{i\bm{k}\cdot\bm{r}+s t}$ and $\textbf{u}_\alpha =\delta \hat{\textbf{u}}_\alpha e^{i\bm{k}\cdot\bm{r}+s t}$, where $s$ is the complex growth rate. 
Linearizing about the homogeneous state $(n_{0,A},n_{0,B})$ gives passive force density $-n_{\alpha} \nabla \mu_{\alpha} = - \left( K_{\alpha\beta} \nabla \delta n_\beta \right) +\kappa n_{0,\alpha} \nabla(\nabla^2\delta n_\alpha)$ to linear order, where $K_{\alpha\beta}=(n_\alpha \partial{\mu_{h,\alpha}}/\partial{n_\beta})|_{\left(n_{0,A},n_{0,B}\right)}$. We assume $K$ to be positive definite and $\kappa>0$, so that the passive term introduces no thermodynamic instability to the homogeneous state. 

To linear order, the continuity and force balance equations become
\begin{gather}
    s \delta\hat n_A
    + i n_{0,A}\bm{k}\cdot\delta\hat{\mathbf{u}}_A = 0,\\
    s \delta\hat n_B
    + i n_{0,B}\bm{k}\cdot\delta\hat{\mathbf{u}}_B = 0,\\
    0=-\big[(\eta k^2+\Gamma+\Gamma_f)\mathbb{I}+\eta_{\mathrm{o}}k^2\epsilon\big]\delta\hat{\mathbf{u}}_A
    -\big[(\eta'k^2-\Gamma_f)\mathbb{I}
    -\eta_{\mathrm{o}}'k^2\epsilon\big]\delta\hat{\mathbf{u}}_B+i\big[\eta_R\Omega\epsilon
    -(K_{11}+\kappa n_{0,A}k^2)\mathbb{I}\big]
    \bm{k}\delta\hat n_A
    -iK_{12}\bm{k}\delta\hat n_B,\\
    0=
    -\big[(\eta k^2+\Gamma+\Gamma_f)\mathbb{I}
    -\eta_{\mathrm{o}}k^2\epsilon\big]\delta\hat{\mathbf{u}}_B
    -\big[(\eta'k^2-\Gamma_f)\mathbb{I}
    +\eta_{\mathrm{o}}'k^2\epsilon\big]\delta\hat{\mathbf{u}}_A -i\big[\eta_R\Omega\epsilon
    +(K_{22}+\kappa n_{0,B}k^2)\mathbb{I}\big]
    \bm{k}\delta\hat n_B
    -iK_{21}\bm{k}\delta\hat n_A.
\end{gather}

To simplify the notation, we define $\alpha = \eta k^2 +\Gamma+\Gamma_f$, $\beta = \eta_{\mathrm{o}} k^2$, $\alpha^{\prime} = \eta^{\prime}k^2-\Gamma_f$, $\beta^{\prime} = \eta_{\mathrm{o}}^{\prime}k^2$, $\gamma_A = -K_{11} -\kappa n_{0,A}k^2$, $\gamma_B =  -K_{22} -\kappa n_{0,B}k^2$, $\nu = \eta_R \Omega$.
The linearized force balance equations can then be written as
\begin{align}
    \begin{pmatrix}
        -\alpha \mathbb{I} -\beta \epsilon & -\alpha^{\prime} \mathbb{I}+\beta^{\prime} \epsilon\\
        -\alpha^{\prime} \mathbb{I}-\beta^{\prime}\epsilon & -\alpha \mathbb{I} +\beta \epsilon 
    \end{pmatrix}
    \begin{pmatrix}
        \delta \hat{\mathbf{u}}_A\\
        \delta \hat{\mathbf{u}}_B
    \end{pmatrix} = -
    \begin{pmatrix}
        (\gamma_A \mathbb{I}+\nu\epsilon) i\bm{k} & -K_{12} i\bm{k}\\
        -K_{21}i\bm{k} & (\gamma_B\mathbb{I}-\nu\epsilon)i\bm{k}
    \end{pmatrix}
    \begin{pmatrix}
        \delta \hat{n}_A \\
        \delta\hat{n}_B
    \end{pmatrix}.
\end{align}
Schur complement gives the inverse of a matrix $M = \begin{pmatrix}
    A & B\\
    C & D
    \end{pmatrix}$:
\begin{align}
    M^{-1} = \begin{pmatrix}
    S^{-1} & -S^{-1}BD^{-1}\\
    -D^{-1}CS^{-1} & D^{-1}+D^{-1}CS^{-1}BD^{-1}
    \end{pmatrix},
\end{align} where $S=A-BD^{-1}C$. Also note that for $A = a\mathbb{I}+b\epsilon$, $A^{-1}=\frac{a\mathbb{I}-b\epsilon}{a^2+b^2}$.
Then we have:
\begin{align}
    \begin{pmatrix}
        \delta \hat{\mathbf{u}}_A\\
        \delta \hat{\mathbf{u}}_B
    \end{pmatrix} = -\frac{1}{(\alpha^{\prime 2}-\alpha^2)+(\beta^{\prime 2}-\beta^2)}
    \begin{pmatrix}
        \alpha \mathbb{I} -\beta \epsilon & -\alpha^{\prime}\mathbb{I}  +\beta^{\prime} \epsilon\\
        -\alpha^{\prime}\mathbb{I} -\beta^{\prime}\epsilon & \alpha \mathbb{I} +\beta \epsilon 
    \end{pmatrix}
    \begin{pmatrix}
        (\gamma_A \mathbb{I}+\nu\epsilon) i\bm{k} & -K_{12} i\bm{k}\\
        -K_{21}i\bm{k} &(\gamma_B\mathbb{I}-\nu\epsilon)i\bm{k}
    \end{pmatrix}
    \begin{pmatrix}
        \delta \hat{n}_A \\
        \delta\hat{n}_B
    \end{pmatrix}.
\end{align}
Plugging this equation into the linearized continuity equations gives:
\begin{align}
    s 
    \begin{pmatrix}
        \delta \hat{n}_A\\
        \delta \hat{n}_B
    \end{pmatrix}=\frac{1}{(\alpha^2-\alpha^{\prime 2})+(\beta^2-\beta^{\prime 2})}
    \begin{pmatrix}
        n_{0,A}(\alpha \gamma_A +\beta \nu+\alpha^{\prime}K_{21})k^2 & n_{0,A}(-\alpha^{\prime}\gamma_B+\beta^{\prime}\nu-\alpha K_{12}) k^2\\
        n_{0,B}(-\alpha^{\prime}\gamma_A+\beta^{\prime}\nu -\alpha K_{21}) k^2 & n_{0,B}(\alpha \gamma_B +\beta \nu+\alpha^{\prime} K_{12})k^2
    \end{pmatrix}
    \begin{pmatrix}
        \delta \hat{n}_A \\
        \delta \hat{n}_B
    \end{pmatrix},
\end{align} since $\bm{k}\cdot \mathbb{I}\bm{k}=k^2$, $\bm{k}\cdot \mathbb{\epsilon}\bm{k}=0$.
Solving this eigenproblem gives the dispersion relation and instability condition.

For the equal composition mixture considered here, we have $n_{0,A}=n_{0,B}=n_0$, and in addition
$K_{AA}=K_{BB}\equiv K$ and
$K_{AB}=K_{BA}\equiv K_c$ because of species exchange symmetry.
Linearizing the equations about the homogeneous mixture
decouples the density fluctuations into symmetric and antisymmetric
eigenmodes. For the composition mode $\delta n_A = -\delta n_B$ and total-density mode $\delta n_A = \delta n_B$, the growth rates are:

\begin{subequations}\label{eq:disp}
\begin{align}
s_+ &= \frac{n_0k^2\big[-\big((\eta-\eta^\prime)k^2+\Gamma+2\Gamma_f\big)\big(K+K_c+\kappa n_0k^2\big)+(\eta_{\mathrm{o}}+\eta_{\mathrm{o}}^\prime)\,\eta_R\Omega\,k^2\big]}{\big((\eta-\eta^\prime)k^2+\Gamma+2\Gamma_f\big)\big((\eta+\eta^\prime)k^2+\Gamma\big)+\big(\eta_{\mathrm{o}}^2-\eta_{\mathrm{o}}^{\prime2}\big)k^4},
\label{eq:disp_plus}\\[4pt]
s_- &= \frac{n_0k^2\big[-\big((\eta+\eta^\prime)k^2+\Gamma\big)\big(K-K_c+\kappa n_0k^2\big)+(\eta_{\mathrm{o}}-\eta_{\mathrm{o}}^\prime)\,\eta_R\Omega\,k^2\big]}{\big((\eta-\eta^\prime)k^2+\Gamma+2\Gamma_f\big)\big((\eta+\eta^\prime)k^2+\Gamma\big)+\big(\eta_{\mathrm{o}}^2-\eta_{\mathrm{o}}^{\prime2}\big)k^4}.
\label{eq:disp_minus}
\end{align}
\end{subequations}
Maximizing over all wavenumbers $k$ gives the instability condition for each mode:
\begin{subequations}\label{eq:thresholds}
\begin{align}
\text{co-growing }(\delta n_A=\delta n_B):&\quad
(\eta_{\mathrm{o}}+\eta_{\mathrm{o}}^\prime)\eta_R\Omega
> D_+  \equiv \Big(\sqrt{(\eta-\eta^\prime)K_+}+\sqrt{\kappa n_0(\Gamma+2\Gamma_f)}\Big)^2
\label{eq:thr_plus},\\
\text{demixing }(\delta n_A=-\delta n_B):&\quad
(\eta_{\mathrm{o}}-\eta_{\mathrm{o}}^\prime)\eta_R\Omega
> D_-\equiv\Big(\sqrt{(\eta+\eta^\prime)K_-}+\sqrt{\kappa n_0\Gamma}\Big)^2.
\label{eq:thr_minus}
\end{align}
\end{subequations}
We note that positive definiteness of the $K_{\alpha\beta}$ matrix requires $K_+\equiv K+K_c>0$ and $K_-\equiv K-K_c>0$. 
The dissipative shear response is also described by a symmetric Onsager matrix $\bm{\eta} =
    \begin{pmatrix}
        \eta & \eta' \\
        \eta'& \eta
    \end{pmatrix}$, whose positive definiteness requires $\eta>|\eta^\prime|$. The conditions above define the linear stability diagram in main Fig.~3(c) in terms of the dimensionless coordinates
$x=\eta_R\Omega \eta_{\mathrm{o}}/D_-$ and $y=\eta_R\Omega \eta_{\mathrm{o}}^{\prime}/D_-$. The composition mode $\delta n_A=-\delta n_B$ is unstable for $x-y>1$, whereas the total-density mode $\delta n_A=\delta n_B$ is unstable for $x+y>D_+/D_-$. Both modes are unstable where these two conditions overlap; the homogeneous state is linearly stable when neither condition is satisfied. For the phase diagram, we set $\eta=K=\kappa=n_0=\Gamma =1$, $K_c =\Gamma_f=0$, and $\eta^{\prime}=0.5$, such that $K_+ = K_- =1$ and $\eta > \eta^{\prime}>0$ satisfies the positive definiteness condition. These parameters give $D_+/D_- = 0.59$ which sets the instability boundary $x+y=D_+/D_-$.

\subsection{Green--Kubo relations for the multicomponent viscosity matrix}
\label{sec:green-kubo-mixture}
In this section, we obtain the Green-Kubo relation for a multicomponent viscosity tensor. We first start from the derivation of the Irving-Kirkwood total virial stress involving all particles, and then obtain the expression for the virial stresses of each component. Next, we derive the Green-Kubo relation for a mixture at thermodynamic equilibrium, and then discuss the assumptions made to obtain the closure for the active chiral systems that we study.

We first derive the stresses of each component based on the microscopic interaction forces between particles. 
Let $\mathbf r_{ij}=\mathbf r_i-\mathbf r_j$, let
$\mathbf F_{ij}=\mathbf F_{ij}^{\mathrm{WCA}}+\mathbf F_{ij}^{\perp}$
be the force on $i$ due to $j$, $\mathbf F_{ij}=-\mathbf F_{ji}$. 
We first consider the total stress resulting from the interactions between all particles. 
The interaction force density $f_\mu$, where $\mu$ is the spatial index, is
\begin{equation}
 f_\mu(\mathbf x)=\sum_i\sum_{j\ne i}F_{ij,\mu}\delta(\mathbf{x}-\mathbf{r}_i)
 =\sum_{i<j} F_{ij,\mu}(\delta(\mathbf{x}-\mathbf{r}_i)-\delta(\mathbf{x}-\mathbf{r}_j)).
 \label{eq:mixgk-force}
\end{equation}
Define the following normalized distribution localized along the straight line from
$\mathbf b$ to $\mathbf a$,
\begin{equation}
 b_{\mathbf a\mathbf b}(\mathbf x)=
 \int_0^1 d\lambda\,\delta(\mathbf x-\mathbf b-\lambda(\mathbf a-\mathbf b)).
\end{equation}
Its spatial integral is $\int{b_{\mathbf a\mathbf b}(\mathbf x)d\mathbf x}=1$. 
We have the following identity
\begin{equation} \label{eq:mixgk-bond}
  \delta(\mathbf{x}-\mathbf a)-\delta(\mathbf{x}-\mathbf b) = \int_0^1 d\lambda\,\frac{d}{d\lambda}\delta(\mathbf x-\mathbf b-\lambda(\mathbf a-\mathbf b))
 =-(a_\nu-b_\nu)\partial_\nu b_{\mathbf a\mathbf b}(\mathbf x).
\end{equation}
Therefore, the force density can also be written as
\begin{align*}
  f_\mu(\mathbf x)&=\sum_{i<j}F_{ij,\mu}(\delta(\mathbf{x}-\mathbf{r}_i)-\delta(\mathbf{x}-\mathbf{r}_j))
  =-\sum_{i<j}F_{ij,\mu}r_{ij,\nu}\partial_\nu b_{\mathbf r_i\mathbf r_j}(\mathbf x) \\
  = &- \partial_\nu\sum_{i<j}F_{ij,\mu}r_{ij,\nu}b_{\mathbf r_i\mathbf r_j}(\mathbf x) = \sum_{i<j}\partial_\nu\sigma_{\mu\nu}^{ij}.
  \label{eq:mixgk-force2}
\end{align*}
where
\begin{equation}
 \sigma_{\mu\nu}^{ij}(\mathbf x)
 =-F_{ij,\mu}r_{ij,\nu}b_{\mathbf r_i\mathbf r_j}(\mathbf x).
 \label{eq:mixgk-ik}
\end{equation}
The integral of the stress over the simulation box gives the virial stress
\begin{equation}
 \overline{\sigma}_{\mu\nu}=-\frac{1}{V}\sum_{i<j}F_{ij,\mu}r_{ij,\nu}.
 \label{eq:mixgk-virial-total}
\end{equation}
where $V$ is the volume of the system.

For simplicity of notation, we define
$\delta_{\mathbf a}(\mathbf x)=\delta(\mathbf x-\mathbf a)$ and
$\delta_i=\delta_{\mathbf r_i}$.
For an unlike pair $i\in A$, $j\in B$, we define the midpoint between them 
$\mathbf m_{ij}=(\mathbf r_i+\mathbf r_j)/2$ and write the contribution of particle $i$ and $j$ to the force density as
\begin{align*}
 \mathbf F_{ij}\delta_i&=\mathbf F_{ij}(\delta_i-\delta_{\mathbf{m}_{ij}})
 +\mathbf F_{ij}\delta_{\mathbf{m}_{ij}},\\
 -\mathbf F_{ij}\delta_j&=\mathbf F_{ij}(\delta_{\mathbf{m}_{ij}}-\delta_j)
 -\mathbf F_{ij}\delta_{\mathbf{m}_{ij}} .
\end{align*}
Applying Eq.~\eqref{eq:mixgk-bond} we can rewrite the above as 
\begin{align*}
 \mathbf F_{ij}\delta_i&=-\nabla\cdot \boldsymbol\sigma_A^{ij} + \mathbf q_A^{ij},\\
 -\mathbf F_{ij}\delta_j&=-\nabla\cdot \boldsymbol\sigma_B^{ij} + \mathbf q_B^{ij} .
\end{align*}
where
\begin{align}
 \boldsymbol\sigma_A^{ij}(\mathbf x)
 &=-\mathbf F_{ij}\otimes(\mathbf r_i-\mathbf m_{ij})
 b_{\mathbf r_i\mathbf m_{ij}}(\mathbf x),\\
 \boldsymbol\sigma_B^{ij}(\mathbf x)
 &=-\mathbf F_{ij}\otimes(\mathbf m_{ij}-\mathbf r_j)
 b_{\mathbf m_{ij}\mathbf r_j}(\mathbf x),\\
 \mathbf q_A^{ij}&=-\mathbf q_B^{ij}=\mathbf F_{ij}\delta_{\mathbf{m}_{ij}},
\end{align}
where $\otimes$ denote the outer product of two vectors.
The sum is defined to be
\begin{align}
  \boldsymbol{\sigma}_A &= \sum_{i\in A,j\neq i}\boldsymbol\sigma_A^{ij} \qquad
  \boldsymbol{\sigma}_B = \sum_{i\in B,j\neq i}\boldsymbol\sigma_B^{ij} \\
  \mathbf q_A &= \sum_{i\in A,j\neq i}\mathbf q_A^{ij} \qquad
  \mathbf q_B = \sum_{i\in B,j\neq i}\mathbf q_B^{ij} .
\end{align}
When summing the stresses over like pairs, the midpoint cancels and the stress has the same form as Eq.~\ref{eq:mixgk-ik}. Hence
\begin{align}
 \boldsymbol\sigma_A &= - \sum_{i \in A, j \in A, j\neq i}\mathbf F_{ij}\otimes(\mathbf r_i-\mathbf r_j)
 b_{\mathbf r_i\mathbf r_j}(\mathbf x) - \sum_{i \in A, j\in B} \mathbf F_{ij}\otimes(\mathbf r_i-\mathbf m_{ij})
 b_{\mathbf r_i\mathbf m_{ij}}(\mathbf x) \\
 \boldsymbol\sigma_B &= - \sum_{i \in B, j \in B, j\neq i}\mathbf F_{ij}\otimes(\mathbf r_i-\mathbf r_j)
 b_{\mathbf r_i\mathbf r_j}(\mathbf x) - \sum_{i \in A, j\in B} \mathbf F_{ij}\otimes(\mathbf m_{ij}-\mathbf r_j)
 b_{\mathbf m_{ij}\mathbf r_j}(\mathbf x) .
\end{align}
When summing the exchange forces over like pairs, the exchange forces cancel. The exchange forces only appear in the unlike pairs. Hence
\begin{equation} \label{eq:mixgk-force-density}
  \mathbf q_A=-\mathbf q_B
 =\sum_{i\in A,j\in B}\mathbf F_{ij}\delta_{\mathbf m_{ij}}.
\end{equation}
The force density on A and B can be written as
\begin{equation}
 \mathbf f_\alpha
 =\nabla\cdot\boldsymbol\sigma_\alpha+\mathbf q_\alpha .
 \label{eq:mixgk-balance}
\end{equation}
Now we integrate the stress in space, $\overline{\boldsymbol\sigma^\alpha}= V^{-1}\int{\boldsymbol\sigma^\alpha dV}$. Note that when $i\in A$ and $j\in B$, $\mathbf r_i-\mathbf m_{ij}=\mathbf m_{ij}-\mathbf r_j=\mathbf r_{ij}/2$, therefore 
\begin{align}\label{eq:mixgk-virial}
 \overline{\boldsymbol\sigma}_{A} &=\overline{\boldsymbol\sigma}_{(AA)}+\tfrac12\overline{\boldsymbol\sigma}_{(AB)},\qquad \overline{\boldsymbol\sigma}_{B}=\overline{\boldsymbol\sigma}_{(BB)}+\tfrac12\overline{\boldsymbol\sigma}_{(AB)},\\
 \overline{\boldsymbol\sigma}^{(\alpha\beta)}
 &=-\frac1V\sum_{i \in A, j \in B, j\neq i}
 \mathbf F_{ij}\otimes\mathbf r_{ij}.
\end{align}
In the following text, we omit the overline when referring to the the virial stress.

Next, we derive the Green--Kubo formula for the two-fluid viscosities.
Let $X$ be the microscopic state, $P(X)$ is the probability of the state $X$. Let $L_0$ be the operator that governs the time evolution of probability densities, and $L_0P_{\rm ss}=0$ where $P_{\rm ss}$ is the stationary probability density.
A small perturbation with amplitudes $h_b^\beta(t)$ changes the operator to
$L=L_0+\sum_{\beta,b}h_b^\beta(t)L_{\beta b}+O(h^2)$.
In our case, we impose strain rate $b$ on species $\beta$, $h_b^\beta=\dot e_b^\beta$. 
Expanding the evolution of the probability density $\partial_tP=LP$ to first order around the stationary state gives
\begin{align}
 (\partial_t-L_0)\delta P(t)&=\sum_{\beta,b}h_b^\beta(t)L_{\beta b}P_{\rm ss},\\
 \delta P(t)&=\sum_{\beta,b}\int_{-\infty}^t dt'\,h_b^\beta(t')
 e^{(t-t')L_0}L_{\beta b}P_{\rm ss},
 \label{eq:mixgk-duhamel}
\end{align}
where $\delta P$ is the perturbed probability density.
The observable is the stress $a$ of species $\alpha$, $\langle s_a^\alpha\rangle_h(t)=\int dX\,s_a^\alpha(X)P(X,t)$.
Note that in our system, the stress is a function of the microscopic forces and does not explicitly depend on the particle velocities and hence the strain rate.
The perturbed stress is 
\begin{equation}
 \delta\langle s_a^\alpha\rangle(t)
 =\int dX\,s_a^\alpha(X)\delta P(X,t).
 \label{eq:mixgk-mean-expansion}
\end{equation}
Substituting Eq.~\eqref{eq:mixgk-duhamel} into the first term of
Eq.~\eqref{eq:mixgk-mean-expansion}:
\begin{equation}
 \int dX\,s_a^\alpha\delta P(X,t)
 =\sum_{\beta,b}\int_{-\infty}^{t}dt'\,
 h_b^\beta(t')R_{ab}^{\alpha\beta}(t-t')
\end{equation}
where the linear response kernel is
\begin{equation}
 R_{ab}^{\alpha\beta}(\tau)
 =\int dX\,s_a^\alpha(X)
       [e^{\tau L_0}L_{\beta b}P_{\rm ss}](X).
 \label{eq:mixgk-rate-kernel}
\end{equation}
We define the conjugate observable $B_b^\beta$ as $B_b^\beta(X)=P_{\rm ss}^{-1}(X)L_{\beta b}P_{\rm ss}(X)$.
Let the operator $e^{\tau L_0^\dagger}$ be the adjoint of $e^{\tau L_0}$, which advances observables in time. We get
\begin{equation}
 R_{ab}^{\alpha\beta}(\tau)
 =\langle\delta s_a^\alpha(\tau)B_b^\beta(0)\rangle_{\rm ss}.
 \label{eq:mixgk-response}
\end{equation}
where $\langle \rangle_{\rm ss}$ is the average over the stationary distribution and $\delta s_a^\alpha$ is the fluctuating stress.
When a constant strain rate is applied starting at $t=0$, $h_b^\beta(t>0)=h_0$,
for $t>0$, the stress response is
\begin{equation}
 \lim_{h_0\to0}\frac{\delta\langle s_a^\alpha\rangle(t)}{h_0}
 =\int_0^t R_{ab}^{\alpha\beta}(\tau)\,d\tau .
\end{equation}
The long-time limit is the zero-frequency viscosity,
\begin{equation}
 \eta_{ab}^{\alpha\beta}
 =\int_0^\infty d\tau\,
 \langle\delta s_a^\alpha(\tau)B_b^\beta(0)\rangle_{\rm ss}.
 \label{eq:mixgk-exact}
\end{equation}
Equation~\eqref{eq:mixgk-exact} is the stationary linear-response identity
for the specified shear perturbation.

We define the two independent shear as $\mathbf{E}_1 = \mathbf{e}_x\mathbf{e}_x - \mathbf{e}_y\mathbf{e}_y$ and $\mathbf{E}_2 = \mathbf{e}_x\mathbf{e}_y + \mathbf{e}_y\mathbf{e}_x$.
Imposing shear rate $b$ on species $\beta$ adds the affine drift $h_b^\beta \mathbf{E}_b\mathbf r_i$ to particles of species $\beta$. 
Therefore, the effect of the first-order perturbation to the operator $L$ on the probability distribution is 
\begin{equation}
 L_{\beta b}P=-\sum_{i\in\beta}\partial_{\mathbf{r}_i}\cdot[(\mathbf{E}_b\mathbf r_i)P],
\end{equation}
Hence, based on the definition of $B_b^\beta$, we have
\begin{equation}
 B_b^\beta=-\sum_{i\in\beta}(\mathbf{E}_b\mathbf r_i)\cdot\partial_{\mathbf{r}_i}\ln P_{\rm ss}.
 \label{eq:mixgk-source}
\end{equation}
Let's first consider the passive equilibrium and overdamped limit, $P_{\rm eq}\propto
\exp[-U/(k_BT)]$. Thus $k_BT\partial_{\mathbf{r}_i}\ln P_{\rm eq}=\mathbf F_i
=\sum_{j\ne i}\mathbf F_{ij}$. 
$B_b^\beta$ can be evaluated in a way similar to how we constructed the virial stress.
Specifically, by defining
\begin{equation} \label{eq:Mb_def}
 M_b=\sum_{i\in A,j\in B}\mathbf F_{ij}\cdot \mathbf{E}_b\mathbf m_{ij}.
\end{equation}
Summation of $\mathbf{E}_b\mathbf{r}_i \cdot \mathbf{F}_{ij}$ over a pair of particles of the same component gives $\mathbf{E}_b\mathbf{r}_{ij} \cdot \mathbf{F}_{ij}$. For an unlike pair the two contributions are $-\mathbf F_{ij}\cdot \mathbf{E}_b\mathbf r_i =-\tfrac12\mathbf F_{ij}\cdot \mathbf{E}_b\mathbf r_{ij} -\mathbf F_{ij}\cdot \mathbf{E}_b\mathbf m_{ij}$ and $\mathbf F_{ij}\cdot \mathbf{E}_b\mathbf r_j =-\tfrac12\mathbf F_{ij}\cdot \mathbf{E}_b\mathbf r_{ij} +\mathbf F_{ij}\cdot \mathbf{E}_b\mathbf m_{ij}$.
Therefore, we can show that 
\begin{align}
 k_BT B_b^A=Vs_b^A-M_b,\qquad k_BT B_b^B=Vs_b^B+M_b.
 \label{eq:mixgk-eqsource}
\end{align}
where $s_b^\alpha = \text{tr}(\boldsymbol{\sigma}_\alpha\mathbf{E}_b)$ is the shear stress $b$ of species $\alpha$.
For an individual entry $\eta_{ab}^{\alpha\beta}$, the exchange term
drops out if $\int_0^\infty dt\,\langle\delta s_a^\alpha(t)
\delta M_b(0)\rangle_{\rm eq}=0$. 
In a homogeneous bulk, $M_b$ in Eq.~\ref{eq:Mb_def} can also be written equivalently as
\begin{equation}
 M_b=\int_V{(\mathbf{E}_b\mathbf x)_kq_{A,k}(\mathbf x) dV}.
\end{equation}
where $\mathbf q_A$ is the force density defined in Eq.~\eqref{eq:mixgk-force-density}.
Define $G_{ak}^{\alpha}(\mathbf x,t)=
\langle\delta s_a^\alpha(t)\delta q_{A,k}(\mathbf x,0)\rangle_{\rm eq}$,
where $s_a^\alpha$ is averaged over the volume and $\langle \rangle_{\rm eq}$ denotes the equilibrium ensemble average. With a periodic boundary condition, translating all particles leaves it invariant and translates $\mathbf q_A$, so $G$ is independent of $\mathbf x$. Under a rotation by $\pi$ the shear stress is invariant, while the exchange force transforms as a vector. Invariance of the equilibrium ensemble and dynamics under this rotation gives $G_{ak}^{\alpha}(\mathbf x,t)=-G_{ak}^{\alpha}(-\mathbf x,t)$.
Combining the two identities yields $G=0$ and hence
\begin{equation}
 \langle\delta s_a^\alpha(t)\delta M_b(0)\rangle_{\rm eq}
 =\int_V{(\mathbf{E}_b\mathbf x)_kG_{ak}^{\alpha}(\mathbf x,t)dV}=0 .
 \label{eq:mixgk-exchange-cancel}
\end{equation}
This argument applies to either species and requires no exchange symmetry.
Therefore, the viscosity coefficient is
\begin{equation}
 \eta_{ab}^{\alpha\beta}
 =\frac{V}{k_BT}\int_0^\infty dt\,
 \langle\delta s_a^\alpha(t)\delta s_b^\beta(0)\rangle_{\rm eq}.
 \label{eq:mixgk-partial-equilibrium}
\end{equation}

For an active mixture, we postulate that the particles can be effectively modeled by a drift and thermal noise due to the transverse forces as a result of spinning. The thermal diffusivity is $D=k_BT_\textrm{eff} \Gamma^{-1}>0$.
The continuity equation of the probability in configuration space is
$\partial_t P=-\sum_i\partial_{\mathbf{r}_i}\cdot\mathbf J_i$, where $\mathbf{J}_i = \Gamma^{-1} \mathbf F_i P - D \partial_{\mathbf{r}_i}P$.
At steady state $\sum_i\partial_{\mathbf{r}_i}\cdot\mathbf J_i^{\rm ss}=0$, where $J_i^{\rm ss}$ is the steady-state probability flux, and $k_BT_\textrm{eff} \partial_{\mathbf{r}_i}\ln P_{\rm ss}=\mathbf F_i-\Gamma \frac{\mathbf J_i^{\rm ss}}{P_{\rm ss}}$.
Substituting into Eq.~\eqref{eq:mixgk-source} gives
\begin{equation}
 k_BT_\textrm{eff} B_b^\beta
 =-\sum_{i\in\beta}(\mathbf{E}_b\mathbf r_i)\cdot\mathbf F_i
 +\Gamma \sum_{i\in\beta}(\mathbf{E}_b\mathbf r_i)\cdot
 \frac{\mathbf J_i^{\rm ss}}{P_{\rm ss}}.
 \label{eq:mixgk-current}
\end{equation}
The first term generates the same result as the equilibrium case, while the second term is a correction due to the steady-state probability flux.
We approximate the zero-frequency viscosity matrix by retaining the contribution from correlations of the partial stresses, neglecting the contribution of the time-integrated stress-current correlation to the viscosity.
Therefore, the Green--Kubo formula for the active chiral mixture becomes
\begin{equation}
 \eta_{ab}^{\alpha\beta}
 =\frac{V}{k_BT_\textrm{eff}}\int_0^\infty dt\,
 \langle\delta s_a^\alpha(t)\delta s_b^\beta(0)\rangle.
 \label{eq:mixgk-partial-ss}
\end{equation}
We assess the closure by comparing the resulting viscosities with those inferred from the velocity fluctuation spectra in Sec.~\ref{sec:velocity_correlation}.

As explained in the main text, the stress--strain relation in an equal-composition mixture with parity-exchange symmetry is 
\begin{align}
    \begin{pmatrix}
        s_1^{A}\\
        s_2^{A}\\
        s_1^{B}\\
        s_2^{B}  
    \end{pmatrix} =
    \begin{pmatrix}
        \eta & \eta_{\mathrm{o}} & \eta^{\prime} & -\eta_{\mathrm{o}}^{\prime}\\
        -\eta_{\mathrm{o}} & \eta & \eta_{\mathrm{o}}^{\prime} & \eta^{\prime}\\
        \eta^{\prime} & \eta_{\mathrm{o}}^{\prime} & \eta & -\eta_{\mathrm{o}}\\
        -\eta_{\mathrm{o}}^{\prime} & \eta^{\prime} & \eta_{\mathrm{o}} & \eta\\
    \end{pmatrix}
    \begin{pmatrix}
        \dot{e}_1^{A} \\
        \dot{e}_2^{A}\\
        \dot{e}_1^{B} \\
        \dot{e}_2^{B} 
    \end{pmatrix}.~\label{visc_matrix}
\end{align}
This matrix relates $\eta_{ab}^{\alpha\beta}$ to $\eta_\mathrm{o}$ and $\eta_\mathrm{o}'$.

Here, we elaborate on the symmetry satisfied by the stress correlations. Denote $C_{ab}^{\alpha\beta} = \langle s_a^\alpha(t) s_b^\beta(0)\rangle$, where we use $s_a^\alpha$ to denote the fluctuating stress for simplicity of notation.
When rotating by $45^o$, $s_1^\alpha \to -s_2^\alpha$, $s_2^\alpha \to s_1^\alpha$, and because of rotational invariance, we then have
\begin{equation}
  \langle s_1^\alpha(t) s_2^\beta(0) \rangle = - \langle s_2^\alpha(t) s_1^\beta(0) \rangle = - \langle s_2^\alpha(0) s_1^\beta(-t) \rangle
\end{equation}
where the last equality is due to time translation invariance. Therefore, we have $C_{12}^{\alpha\beta}(t) = -C_{21}^{\alpha\beta}(t) = - C_{12}^{\beta\alpha}(-t) = C_{21}^{\beta\alpha}(-t)$.
When the two components rotate in opposite directions with the same angular speed, we also have the symmetry of $\mathcal{P} \times A\to B$, where $\mathcal{P}$ is the mirror reflection (about the $x$ axis or $y$ axis), and $A\to B$ means that component A and B are exchanged. Upon mirror reflection, $s_1^\alpha \to s_1^\alpha$, $s_2^\alpha \to -s_2^\alpha$, so we have $\langle s_1^A(t) s_2^B(0) \rangle = - \langle s_1^B(t) s_2^A(0) \rangle$, or $C_{12}^{AB}(t) = - C_{12}^{BA}(t)$. For the same reason, $C_{21}^{AB}(t) = - C_{21}^{BA}(t)$.
So by combining the two symmetries we have $C_{12}^{AB}(t) = C_{12}^{AB}(-t)$, $C_{21}^{AB}(t) = C_{21}^{AB}(-t)$ which are even with respect to the reversal of time lag.
Similarly, we can show that for even correlations $C_{ii}^{\alpha\alpha}(t)=C_{ii}^{\alpha\alpha}(-t)$ the same for $i=1,2$ and $\alpha=A,B$, $C_{ii}^{AB}(t)=C_{ii}^{BA}(t)=C_{ii}^{AB}(-t)=C_{ii}^{BA}(-t)$ the same for $i=1,2$, 
$C_{12}^{AB}(t)=C_{21}^{BA}(t)=-C_{21}^{AB}(t)=-C_{12}^{BA}(t)=C_{12}^{AB}(-t)=C_{21}^{BA}(-t)=C_{21}^{AB}(-t)=C_{12}^{BA}(-t)$, and
for odd correlations, $C_{12}^{AA}(t)=-C_{21}^{AA}(t)=-C_{12}^{BB}(t)=C_{21}^{BB}(t)=-C_{12}^{AA}(-t)=C_{21}^{AA}(-t)=C_{12}^{BB}(-t)=-C_{21}^{BB}(-t)$.

We obtain entries of the odd viscosity tensor from the antisymmetric correlation function $C_{\mathrm{o}}^{\alpha\beta}(t)= \frac{1}{2}(C_{12}^{\alpha\beta}(t)-C_{21}^{\alpha\beta}(t))$,
\begin{equation}
    \eta_{\mathrm{o},\alpha\beta}
    =
    \frac{V}{k_B T_{\mathrm{eff}}}
    \int_0^\infty
    C_{\mathrm{o}}^{\alpha\beta}(t)\,dt=\frac{1}{2}\epsilon_{ab}\eta_{ab}^{\alpha\beta}.
    \label{eq:crossodd_gk}
\end{equation}
Therefore we can define $\eta_{\mathrm{o}}^{\prime}$ from the $\bm{\eta}_{\mathrm{o}}$ matrix (Eq.~\ref{eq:oddmatrix}) as $\eta_{\mathrm{o}}^{\prime}=\frac{1}{2}(\eta_{\mathrm{o}, BA}-\eta_{\mathrm{o},AB})$ for a mixture constrained by the parity-exchange symmetry. Additionally, the odd viscosity of each individual species $\alpha$ is given by its diagonal component $\eta_{\mathrm{o},\alpha\alpha}$ in species space, and $\eta_{\mathrm{o}}=\frac{1}{2}(\eta_{\mathrm{o}, AA}-\eta_{\mathrm{o},BB})$.
The diagonal and off-diagonal entries of the shear viscosity matrix are computed from the auto correlation function $C^{\alpha\beta}(t)= \frac{1}{2}(C_{11}^{\alpha\beta}+ C_{22}^{\alpha\beta}(t))$,
\begin{equation}
\eta_{\alpha\beta}=\frac{V}{k_BT_{\mathrm{eff}}}\int_0^{\infty} C^{\alpha\beta}(t)\, dt = \frac12 \delta_{ab}\eta_{ab}^{\alpha\beta}.
\end{equation}
For the shear viscosity matrix in species space, we therefore have $\eta=\frac12(\eta_{AA}+\eta_{BB})$ and cross viscosity $\eta^{\prime}=\frac12(\eta_{BA}+\eta_{AB}$) for measurements.

\begin{figure}
    \centering
    \includegraphics[width=1.0\linewidth]{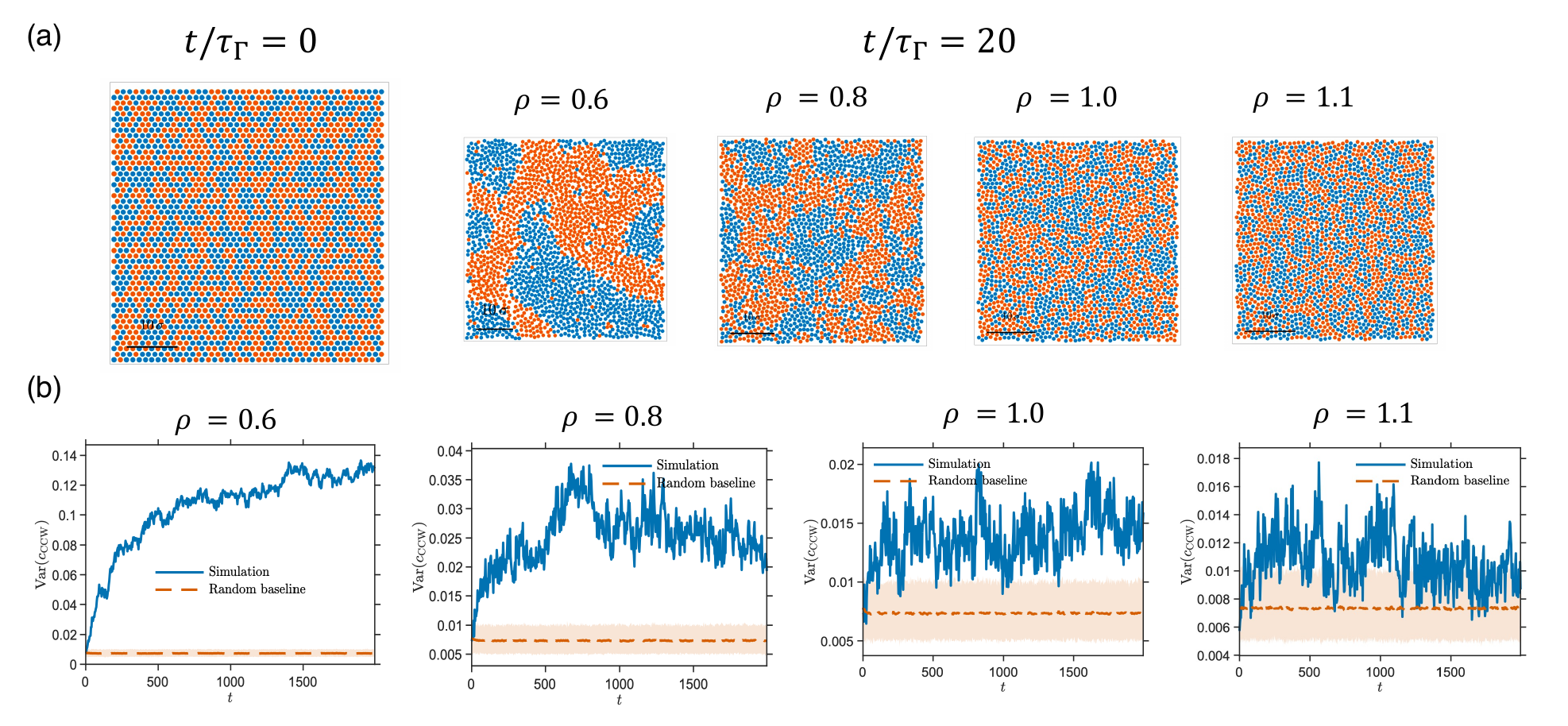}
    \caption{\textbf{Selection of the homogeneous regime for Green--Kubo analysis.} \textbf{(a)}~Particle configurations at $t/\tau_\Gamma=20$ for different packing fractions $\rho$, starting from the same random initial configuration. The mixture remains approximately homogeneous at $\rho=1.1$ throughout the evolution. \textbf{(b)}~Fraction of late time configurations at each packing fraction for which the variance of the local counterclockwise concentration $c_{\text{CCW}}=n_A/(n_A+n_B)$  falls within the random mixture baseline. At $\rho=1.1$, a substantial fraction of configurations have variances consistent with the random baseline, whereas at other packing fractions, most parts of the trajectories exhibit variances exceeding its upper bound. We therefore choose $\rho=1.1$ for the Green--Kubo analysis.}
    \label{fig:S5}
\end{figure}

In order to obtain viscosities of a homogeneous mixture, we sweep the packing fraction $\rho$ to a homogeneous regime. Homogeneity is assessed from the local composition variance across $64$ spatial blocks compared with a baseline by randomly shuffling the particles identified from $500$ independent runs. We report the fraction of late-time configurations for which the variance of the locally measured $c_{\text{CCW}}=n_A/(n_A+n_B)$ exceeds the upper bound of this baseline. The fraction is close to $1$ for $\rho=0.6, 0.8, 1.0$, whereas at $\rho=1.1$ it decreases to $<50\%$, consistent with a homogeneous mixture, as shown in Fig.~\ref{fig:S5}. Green--Kubo measurements are therefore performed at $\rho=1.1$. Packing fractions above $1$ are permitted here because the particles interact through soft repulsion. 

All other parameters are identical to those used in main text Fig.~2(a). The simulations are sampled at $\Delta t =0.05$ over $t\in [0,1000]$, with the first $20\%$ of each trajectory discarded as transient dynamics, such that correlation functions are evaluated over the remaining steady-state data. The Green--Kubo integrals are evaluated separately for each trajectory using trapezoidal integration and subsequently averaged over $80$ independent trajectories, with $95\%$ confidence intervals computed across trajectories. The integration cutoff $\tau_c$ is taken as the first nonzero time delay at which the $95\%$ confidence interval of the odd stress correlations includes zero.

\begin{figure}
    \centering
    \includegraphics[width=1.0\linewidth]{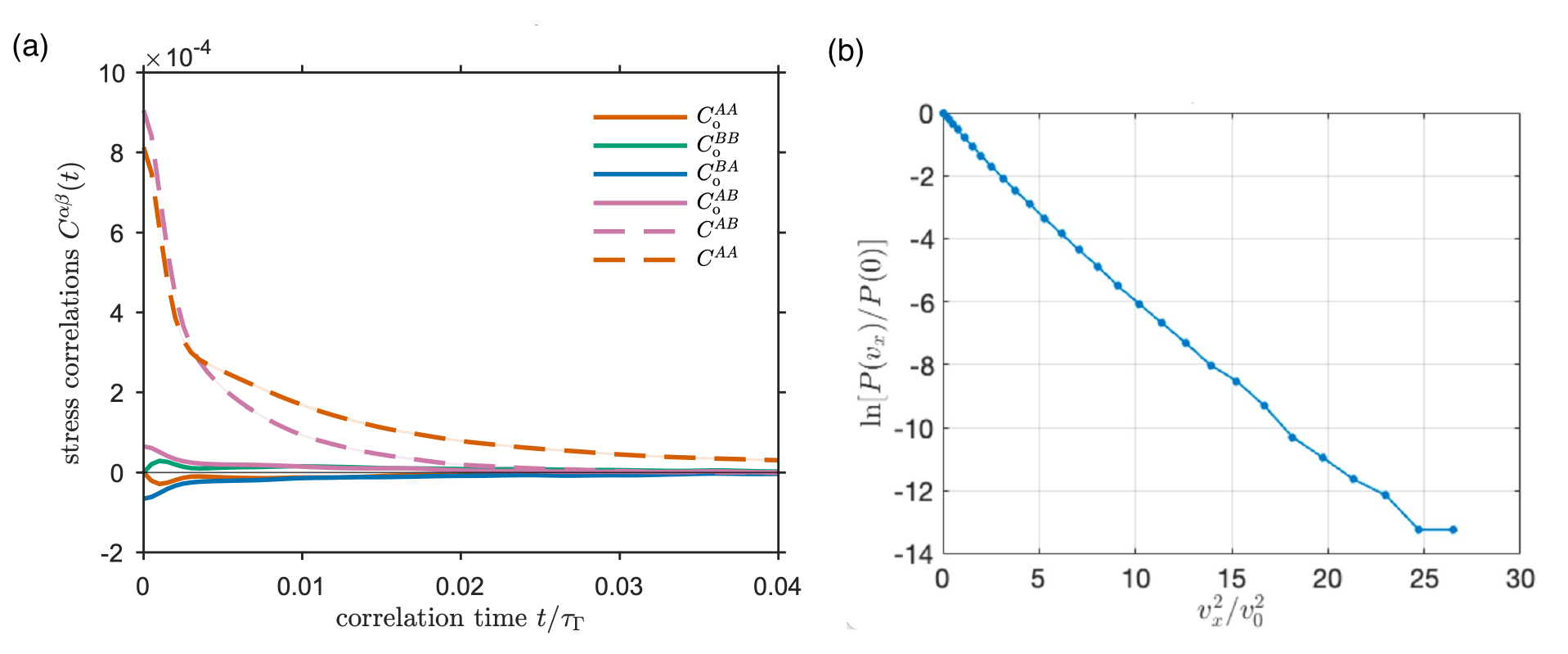}

    \caption{\textbf{Green--Kubo stress correlations and velocity statistics of the binary chiral active mixture.}\textbf{(a)} Stress correlations averaged over $80$ independent particle trajectories, including both the auto correlation functions $C^{\alpha\beta}(\tau)$ determining the generalized shear viscosity matrix and the antisymmetric correlation functions $C_\mathrm{o}^{\alpha\beta}(\tau)$ determining the generalized odd viscosity matrix. Their time integrals give the shear and transverse viscosity coefficients $\eta$, $\eta^\prime$, $\eta_\mathrm{o}$, and $\eta_\mathrm{o}^\prime$ entering the multicomponent viscosity matrix. The correlations exhibit the expected symmetry relations. \textbf{(b)} Semilog plot of the normalized velocity distribution $P(v_x)/P(0)$ against $v_x^2/v_0^2$, obtained from the same $80$ trajectories, where $v_0^2=\langle v_x^2\rangle$ is used for normalization. The linear behavior over the central range of the distribution is consistent with Gaussian velocity statistics, with deviations appearing in the long-tail limit.}
    \label{fig:S6}
\end{figure}

The normalized cross-species shear viscosity obtained from the Green--Kubo analysis is $\eta^{\prime}/\eta=0.686\pm 0.021$, satisfying $\eta>\eta^{\prime}>0$ as required for positive definiteness of the symmetric shear viscosity matrix. Both the diagonal and off-diagonal parity-violating viscosity responses exhibit the expected symmetry within statistical uncertainty: 
\begin{align}
(\eta_{\mathrm{o},AA}+\eta_{\mathrm{o},BB})/(2\eta) &= 0.010 \pm0.011, \\
(\eta_{\mathrm{o},BA}+\eta_{\mathrm{o},AB})/(2\eta) &= -0.007 \pm0.011.
\end{align}
For Green--Kubo measurements of transport values reported in the main text, we define and measure $\eta_{\mathrm{o}}^{\prime}/\eta=(\eta_{\mathrm{o}, BA}-\eta_{\mathrm{o},AB})/2\eta$ and $\eta_{\mathrm{o}}/\eta=(\eta_{\mathrm{o},AA}-\eta_{\mathrm{o},BB})/2\eta$ for the binary mixture. The stress correlation functions used to determine the viscosities are shown in Fig.~\ref{fig:S6}. 
While $\eta_{\mathrm{o}}$ shows up in the antisymmetric entries of the viscosity tensor and is therefore nondissipative \cite{Fruchart2023,Khain2022}, $\eta_{\mathrm{o}}^{\prime}$ is in the symmetric entries of the viscosity tensor and can contribute to dissipation. Correspondingly, as mentioned earlier, its stress correlations are even under time reversal and can remain finite at $\tau=0$. Therefore, $\eta_\mathrm{o}'$ is called cross parity-violating viscosity instead of odd viscosity, which by definition is nondissipative. 

As a further consistency check of the statistical properties of the nonequilibrium steady state, we examine the velocity distribution of single particles in Fig.~\ref{fig:S6}(b). The linear dependence of $\ln [P(v_x)/P(0)]$ on $v_x^2/v_0^2$ over the central range is consistent with a Gaussian velocity distribution, with deviations in the long tail. This confirms that the particle velocities in our system mostly follow a Maxwell-Boltzmann distribution, similar to the observation of chiral active fluids exhibiting effective thermodynamic behavior and Green--Kubo relations in Ref.~\cite{Han2021}. This observation supports the presence of equilibrium-like velocity statistics over a finite range.

\subsection{Fluctuating hydrodynamics and velocity correlations}
\label{sec:velocity_correlation}
In this section, we develop a theory of fluctuating hydrodynamics for active chiral mixtures.
The goal is to provide a test for the validity of the Green-Kubo relation for the nonequilibrium multicomponent system by comparing with the velocity correlation from theory and simulations (Eq.~\ref{eq:mixgk-partial-ss}).
The fluctuating hydrodynamic theory is constructed by adding random stresses and forces to the deterministic momentum balance (Eq.~\ref{eq:binary2}).
To study fluctuations, we linearize the continuity and momentum balance equations.
The linearized continuity equation is
\begin{equation}
  \partial_t \delta  n_\alpha +  n_\alpha^0 \nabla \cdot \mathbf{u}_\alpha = 0,
\end{equation}
where $\delta n_\alpha$ is the fluctuating density field around a state of uniform density $n_\alpha^0$ and $\mathbf{u}_\alpha$ is the fluctuating velocity around the stationary base state.
With the constitutive laws of stresses (Eqs.~\ref{eq:multi_viscous_stress}-\ref{eq:multi_torque}), the linearized momentum balance equation with the addition of noise is
\begin{multline}
  \eta_{\alpha\beta} \nabla^2 \mathbf{u}_\beta + \zeta_{\alpha\beta} \nabla\left(\nabla\cdot \mathbf{u}_\beta \right) + \eta_{\mathrm{o},\alpha\beta} \epsilon \nabla^2 \mathbf{u}_\beta - K_{\alpha\beta} \nabla \delta n_\beta + \epsilon \eta_R \Omega_\alpha \nabla \delta  n_\alpha - \Lambda_{\alpha\beta} \mathbf{u}_\beta + \nabla\cdot\boldsymbol{\sigma}^{R}_{\alpha} + \boldsymbol{\xi}_\alpha = 0
\end{multline}
where $K_{\alpha\beta} =  n_\alpha^0 \frac{\partial \mu_\alpha}{\partial  n_\beta}$, $\bm\sigma_\alpha^R$ and $\bm\xi_\alpha$ are the randomly fluctuating stress and forces, respectively. For the drag force $-\Gamma_\alpha\mathbf u_\alpha-\Gamma_f(\mathbf u_\alpha-\mathbf u_{\beta\ne\alpha})$, the positive semidefinite damping matrix is $\Lambda_{\alpha\beta}=(\Gamma_\alpha+\Gamma_f)\delta_{\alpha\beta}-\Gamma_f(1-\delta_{\alpha\beta})$, so that the friction term in the momentum equation is $-\Lambda_{\alpha\beta}\mathbf u_\beta$. 

For an equal composition mixture of counterrotating particles ($n_A^0 = n_B^0 = n_0$), as discussed in the previous section, spatial reflection combined with species exchange constrains the even transport matrices and odd viscosity tensor in the species space to
\begin{equation}
    \mathbf{X}=\begin{pmatrix}X_d&X_x\\X_x&X_d\end{pmatrix},\quad
  X\in\{K,\eta,\zeta,\Lambda\},\qquad
  \bm\eta_\mathrm{o}=\begin{pmatrix}\eta_\mathrm{o}&-\eta_\mathrm{o}^{\prime}\\\eta_\mathrm{o}^{\prime}&-\eta_\mathrm{o}\end{pmatrix}.
  \label{eq:species-symmetry}
\end{equation}
We define $X_\pm=X_d\pm X_x$ for even species matrices and
$\bm\eta_{\mathrm{o},s}=(\bm\eta_\mathrm{o}+\bm\eta_\mathrm{o}^T)/2$ and
$\bm\eta_{\mathrm{o},a}=(\bm\eta_\mathrm{o}-\bm\eta_\mathrm{o}^T)/2$ for the symmetric and antisymmetric parts of odd viscosity in species space. 

We use a minimal model for the random stresses that satisfies the symmetry requirement, and following Ref.~\cite{Han2021}, generalize the random stress correlations to a two-component system using the viscosity matrix introduced in Eq.~\ref{visc_matrix},
\begin{equation}
  \langle \sigma_i^{R,\alpha}(\mathbf{r},t) \sigma_j^{R,\beta}(\mathbf{r}',t') \rangle = 2 k_B T_\textrm{eff}  \delta(\mathbf{r}-\mathbf{r}') \left( \eta_{ij,\text{sym}}^{\alpha\beta} \delta(t-t') + \eta_{ij,\text{anti}}^{\alpha\beta} \psi(t-t') \right)
\end{equation}
where $\eta_{ij,\text{sym}}^{\alpha\beta} = \eta_{ji,\text{sym}}^{\beta\alpha}$, $\eta_{ij,\text{anti}}^{\alpha\beta} = - \eta_{ji,\text{anti}}^{\beta\alpha}$.
We assume spatially uncorrelated white noise $\delta(\mathbf{r}-\mathbf{r}')$, which is valid when the microscopic stress correlation length is small compared with the hydrodynamic length scales of interest.
For the time correlation, for correlations that are even with respect to the reversal of time lag, we assume temporally uncorrelated white noise $\delta(t-t')$, which is valid when the microscopic stress correlation time is small compared with the hydrodynamic time scales of interest. For correlations that are odd with respect to the reversal of time lag, we choose $\psi(t)=-\delta'(t)/\omega_0$ as the leading
nonvanishing term in a short-memory expansion of the odd correlations, since the first derivative of the delta function is the lowest-order odd correlation. Here, $\omega_0$ is a characteristic correlation frequency parameterizing the first temporal moment of these correlations. We define $\tilde{\omega}=\omega/\omega_0$. 
The amplitudes of the correlations $2k_B T_\textrm{eff} \eta_{ij}^{\alpha\beta}$ are determined by consistency with the proposed Green-Kubo relation (Eq.~\ref{eq:mixgk-partial-ss}).
Similarly, the random force associated with this damping obeys
\begin{equation}
  \left\langle \xi_{i,\alpha}(\mathbf r,t)\xi_{j,\beta}(\mathbf r',t')\right\rangle
  =2k_BT_\textrm{eff}\Lambda_{\alpha\beta}\delta_{ij}
  \delta(\mathbf r-\mathbf r')\delta(t-t'),
\end{equation}
where we assume that all random stresses and forces share the same effective temperature $T_\textrm{eff}$.

We use the following convention to define inverse Fourier transform $f(\mathbf{r},t) = \int \frac{d^d\mathbf{k}}{(2\pi)^d} \int \frac{d\omega}{2\pi} \hat f(\mathbf{k},\omega) e^{i(\mathbf{k}\cdot \mathbf{r}+\omega t)}$, and use hat to denote Fourier amplitudes, the random stress covariance in the Fourier space becomes
\begin{multline}
  \langle 
  \hat\sigma_{ij,\alpha}^{R}(\mathbf{k},\omega) \hat\sigma_{kl,\beta}^{R,*}(\mathbf{k}',\omega') \rangle = 2 k_B T_\textrm{eff} (2\pi)^d \delta(\mathbf{k}-\mathbf{k}') \delta(\omega-\omega') \left( \eta_{\alpha\beta} (\delta_{ik}\delta_{jl} + \delta_{il}\delta_{jk}) + (\zeta_{\alpha\beta} - \eta_{\alpha\beta}) \delta_{ij}\delta_{kl}  \right. \\ \left. +   \left(\epsilon_{il}\delta_{jk}+\epsilon_{jk}\delta_{il}\right) ( -i \tilde{\omega} \eta_{\mathrm{o},s,\alpha\beta} + \eta_{\mathrm{o},a,\alpha\beta}  )  \right)
\end{multline}
where for our system $\eta_{\mathrm{o},s,\alpha\beta}=\eta_{\mathrm{o},\alpha\beta}\delta_{\alpha\beta}$, and $\eta_{\mathrm{o},a,\alpha\beta}=-\epsilon^{\alpha\beta}\eta_\mathrm{o}'$.

We now calculate the velocity correlations using a consistent Fourier convention. Introducing the longitudinal and transverse velocities $\hat{u}_L^\alpha = \hat{\mathbf k}\cdot \hat{\mathbf{u}}_\alpha$, $\hat{u}_{T,\alpha}=(\bm\epsilon\hat{\mathbf k})\cdot \hat{\mathbf{u}}_\alpha$, we define $\mathbf{u}_L=(u_L^\alpha, u_L^\beta)^T$, $\mathbf{u}_T=(u_T^\alpha, u_T^\beta)^T$,
$\delta \mathbf{n}=(\delta n_A,\delta n_B)^T$,
$N_{\alpha\beta}=n_0 \delta_{\alpha\beta}$, and
$\tau_{\alpha\beta}=\eta_R\Omega_\alpha\delta_{\alpha\beta}$.
For opposite torques, $\bm\tau=\operatorname{diag}(\tau,-\tau)$,
where $\tau=\eta_R\Omega$. The projected random stresses are
$\hat\sigma_{L,\alpha}=\hat{\mathbf k}\cdot(\hat{\boldsymbol\sigma}_\alpha^{R} \hat{\mathbf k})$
and $\hat\sigma_{T,\alpha}=(\bm\epsilon\hat{\mathbf k})\cdot(\hat{\boldsymbol\sigma}^{R}_{\alpha}\hat{\mathbf k})$,
with $\hat\xi_{L,\alpha} = \hat{\mathbf{k}} \cdot \hat{\boldsymbol{\xi}}_\alpha$ and $\hat\xi_{T,\alpha} = \hat{\mathbf{k}}_\perp \cdot \hat{\boldsymbol{\xi}}_\alpha$ .
We define $\hat{\bm\sigma}_L=(\hat{\sigma}_{L,A},\hat{\sigma}_{L,B})^T$, $\hat{\bm\sigma}_T=(\hat{\sigma}_{T,A},\hat{\sigma}_{T,B})^T$, $\hat{\bm\xi}_L=(\hat{\xi}_{L,A}, \hat{\xi}_{L,B})^T$, and $\hat{\bm\xi}_T=(\hat{\xi}_{T,A}, \hat{\xi}_{T,B})^T$.
The linearized continuity and momentum equations in Fourier space are then
\begin{equation}
  \begin{pmatrix}
    i \omega \mathbf{I} & i k \mathbf{N} & 0 \\
    -i k \mathbf{K} & -(\bm\eta+\bm\zeta) k^2 - \bm\Lambda & k^2 \bm\eta_\mathrm{o} \\
    i k \bm\tau & -\bm\eta k^2 - \bm\Lambda & - k^2 \bm\eta_\mathrm{o} 
  \end{pmatrix}
  \begin{pmatrix} \delta \mathbf{n} \\ \mathbf{\hat{u}}_L \\ \mathbf{\hat{u}}_T \end{pmatrix}
  + i k
  \begin{pmatrix} 0 \\ \hat{\bm\sigma}_L \\ \hat{\bm\sigma}_T \end{pmatrix}
  + \begin{pmatrix}
    0 \\ \hat{\bm\xi}_L \\ \hat{\bm\xi}_T
  \end{pmatrix} = 0.~\label{eq:full-response}
\end{equation}
We denote the 6-by-6 matrix in the linearized equations by $\mathbf{M}$.

We can eliminate the density fluctuations to get a 4-by-4 matrix equation for the velocities only.  The result is
\begin{equation}
  \begin{pmatrix}
    -(\bm\eta+\bm\zeta) k^2 - \bm\Lambda + ik^2 \omega^{-1} \mathbf{K} \mathbf{N} & \bm\eta_\mathrm{o} k^2 \\
    - k^2 \bm\eta_\mathrm{o} - i k^2 \omega^{-1} \bm\tau \mathbf{N} &  -\bm\eta k^2 - \bm\Lambda
  \end{pmatrix}
  \begin{pmatrix} \hat{\bm\sigma}_L \\ \hat{\bm\sigma}_T \end{pmatrix}
  + i k
  \begin{pmatrix} \hat{\bm\sigma}_L \\ \hat{\bm\sigma}_T \end{pmatrix}
  + \begin{pmatrix}
    \hat{\bm\xi}_L \\ \hat{\bm\xi}_T
  \end{pmatrix} = 0 ~\label{vel_1}
\end{equation}
The matrix in the above equation is denoted by $\mathbf{M}_u$.

To calculate the covariance between $\hat\sigma_L$ and $\hat\sigma_T$, we take the following as an example, 
\begin{multline}
  \langle \hat\sigma_{L,\alpha}(\mathbf{k},\omega) \hat\sigma_{T,\beta}^{*}(\mathbf{k}',\omega') \rangle = \hat{k}_i \hat{k}_j \hat{k}^\perp_k \hat{k}_l \langle \hat\sigma_{ij,\alpha}^{R}(\mathbf{k},\omega) \hat\sigma_{kl,\beta}^{R,*}(\mathbf{k}',\omega') \rangle \\ = (-i\tilde{\omega} \eta_{\mathrm{o},\alpha\beta}\delta_{\alpha\beta} + \epsilon_{\alpha\beta} \eta_\mathrm{o}') \hat{k}_j \hat{k}^\perp_k \epsilon_{jk} = i\tilde{\omega} \eta_{\mathrm{o},\alpha\beta}\delta_{\alpha\beta} - \epsilon_{\alpha\beta} \eta_\mathrm{o}'
\end{multline}
\begin{multline}
  \langle \hat\sigma_{T,\alpha}(\mathbf{k},\omega) \hat\sigma_{L,\beta}^{*}(\mathbf{k}',\omega') \rangle = \hat{k}^\perp_i \hat{k}_j \hat{k}_k \hat{k}_l \langle \hat\sigma_{ij,\alpha}^{R}(\mathbf{k},\omega) \hat\sigma_{kl,\beta}^{R,*}(\mathbf{k}',\omega') \rangle \\  = (-i\tilde{\omega} \eta_{\mathrm{o},\alpha\beta}\delta_{\alpha\beta} + \epsilon_{\alpha\beta} \eta_\mathrm{o}') \hat{k}^\perp_i  \hat{k}_l \epsilon_{il} = -i\tilde{\omega} \eta_{\mathrm{o},\alpha\beta}\delta_{\alpha\beta} + \epsilon_{\alpha\beta} \eta_\mathrm{o}'
\end{multline}
where $\hat{k}_i^\perp=\epsilon_{ij}\hat{k}_j$.
The cross species covariance when $\alpha \neq \beta$ satisfies $\langle \hat\sigma_{T,\alpha}(\mathbf{k},\omega) \hat\sigma_{L,\beta}^{*}(\mathbf{k}',\omega') \rangle = \langle \hat\sigma_{L,\beta}(\mathbf{k},\omega) \hat\sigma_{T,\alpha}^{*}(\mathbf{k}',\omega') \rangle^*$. 
In summary, defining $\hat{\bm\sigma}=(\hat{\bm\sigma}_L,\hat{\bm\sigma}_R)^T$, $\hat{\bm\xi}=(\hat{\bm\xi}_L,\hat{\bm\xi}_R)$, the covariance of the noise is
\begin{align}
  \langle \hat{\bm\sigma} \hat{\bm\sigma}^\dagger\rangle
  &=2k_B T_\textrm{eff}(2\pi)^{d+1}\delta(\mathbf k-\mathbf k')\delta(\omega-\omega')\begin{pmatrix}
    \bm\eta+\bm\zeta & i\tilde\omega\bm\eta_{\mathrm{o},s}-\bm\eta_{\mathrm{o},a}\\
    -i\tilde\omega\bm\eta_{\mathrm{o},s}+\bm\eta_{\mathrm{o},a} & \bm\eta
  \end{pmatrix},\nonumber\\
  \langle\hat{\bm\xi}\hat{\bm\xi}^\dagger\rangle
  &=2k_B T_\textrm{eff}(2\pi)^{d+1}\delta(\mathbf k-\mathbf k')\delta(\omega-\omega')\begin{pmatrix}\bm\Lambda&0\\0&\bm\Lambda\end{pmatrix},
\end{align}

The covariance of the density and velocity components can be obtained from the inverse of the matrix $\mathbf{M}$ and the noise covariance:
\begin{equation}
  \langle 
  \begin{pmatrix}
    \delta  \mathbf{n} \\ \hat{\mathbf{u}}_L \\ \hat{\mathbf{u}}_T
  \end{pmatrix}
  \begin{pmatrix}
    \delta  \mathbf{n} \\ \hat{\mathbf{u}}_L \\ \hat{\mathbf{u}}_T
  \end{pmatrix}^\dagger
  \rangle
  = \mathcal{D} \mathbf{M}^{-1} 
  \begin{pmatrix}
    0 & 0 & 0 \\
    0 & (\bm\eta+\bm\zeta)k^2 + \bm\Lambda & (i \tilde{\omega} \bm\eta_{\mathrm{o},s} - \bm\eta_{\mathrm{o},a}) k^2 \\
    0 & (-i \tilde{\omega} \bm\eta_{\mathrm{o},s} + \bm\eta_{\mathrm{o},a}) k^2 & \bm\eta k^2 + \bm\Lambda
  \end{pmatrix}
  (\mathbf{M}^{-1})^\dagger
\end{equation}
where $\mathcal{D}=2k_B T_\textrm{eff}(2\pi)^{d+1}$.
Alternatively if we are more in the velocity correlation only, we can use the inverse of $\mathbf{M}_u$ and the noise covariance:
\begin{equation}
  \langle 
  \begin{pmatrix}
   \hat{\mathbf{u}}_L \\ \hat{\mathbf{u}}_T
  \end{pmatrix}
  \begin{pmatrix}
   \hat{\mathbf{u}}_L \\ \hat{\mathbf{u}}_T
  \end{pmatrix}^\dagger
  \rangle
  = \mathbf{M}_u^{-1} 
  \begin{pmatrix}
    (\bm\eta+\bm\zeta)k^2 + \bm\Lambda & (i \tilde{\omega} \bm\eta_{\mathrm{o},s} - \bm\eta_{\mathrm{o},a}) k^2 \\
    (-i \tilde{\omega} \bm\eta_{\mathrm{o},s} + \bm\eta_{\mathrm{o},a}) k^2 & \bm\eta k^2 + \bm\Lambda
  \end{pmatrix}
  (\mathbf{M}_u^{-1})^\dagger ~\label{vel_2}
\end{equation}

\begin{figure}
    \centering   \includegraphics[width=1.0\linewidth]{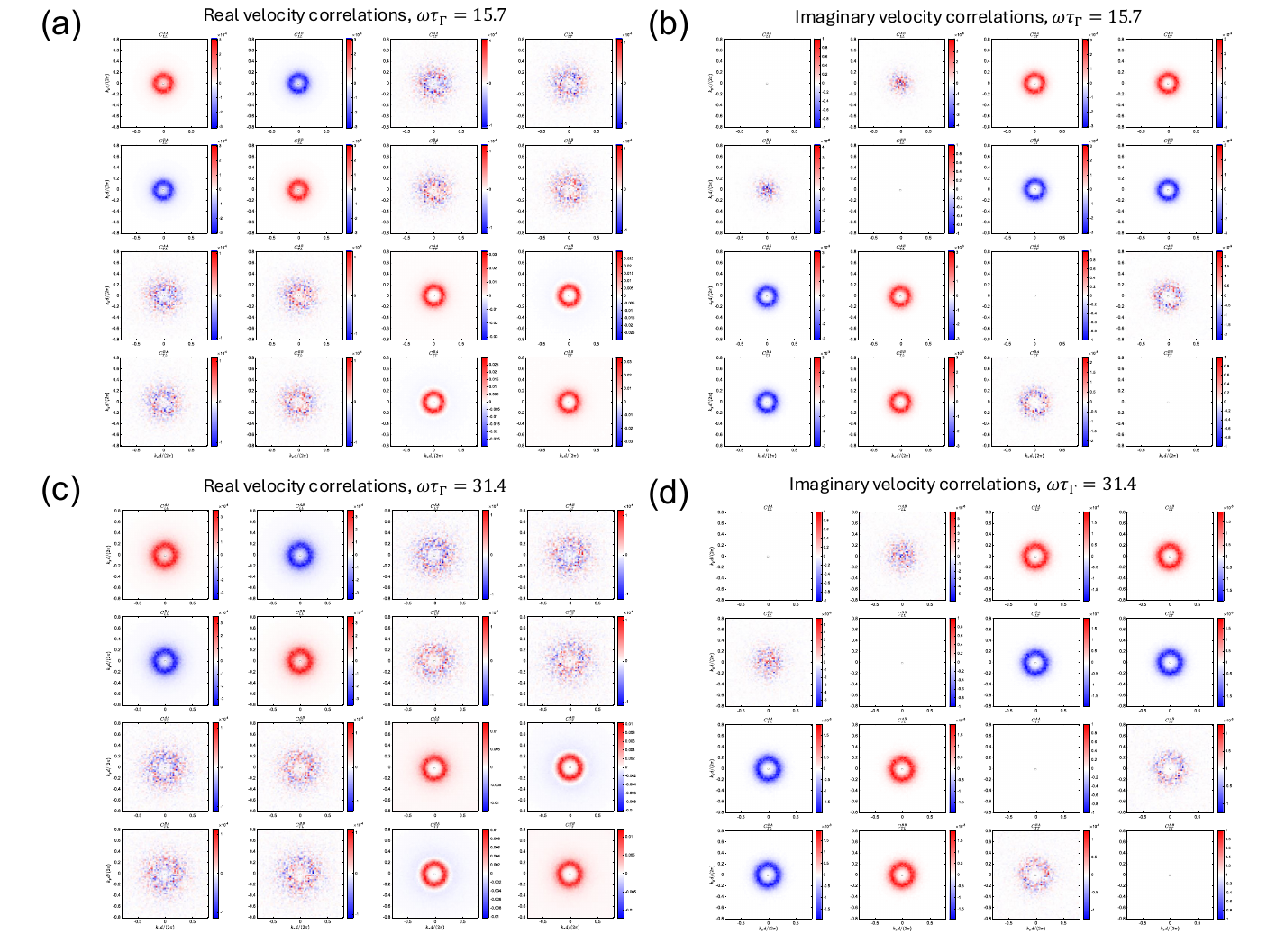}
    \caption{\textbf{Measured velocity correlations from microscopic simulations.} \textbf{(a)} The real part of velocity correlations at $\omega\tau_\Gamma=15.7$. Rows and columns label the first velocity and the conjugated second velocity in the order of $\hat{u}_L^A,\hat{u}_L^B,\hat{u}_T^A,\hat{u}_T^B$. The displayed wavevector coordinates are $k_x\sigma/(2\pi)$ and $k_y\sigma/(2\pi)$, and each heatmap colorbar has its own scale. \textbf{(b)} The imaginary part of velocity correlations at $\omega\tau_\Gamma=15.7$. \textbf{(c)} The real part of velocity correlations at $\omega\tau_\Gamma=31.4$. \textbf{(d)} The imaginary part of velocity correlations at $\omega\tau_\Gamma=31.4$.}
    \label{fig:S09}
\end{figure}

From the particle simulation data, we project the Fourier space velocities onto the longitudinal and transverse directions and calculate their correlations by averaging over 80 independent simulation trajectories. For each wavevector $\mathbf{k}$ and frequency $\omega$, the correlation matrix is computed as
\begin{equation}
C_{\gamma\delta}^{\alpha\beta}(\mathbf{k},\omega)=\frac{1}{N_{\text{traj}}}\sum_{r=1}^{N_{\text{traj}}}\hat{u}_\gamma^{\alpha,(r)}(\mathbf{k},\omega)\hat{u}_\delta^{\beta,(r)*}(\mathbf{k},\omega),
\end{equation}
where $\gamma,\delta=\{L,T\}$ label the velocity components, $r$ labels the different trajectories. Fig.~\ref{fig:S09} shows the real and imaginary parts of the velocity correlations at $\omega\tau_\Gamma=15.7$ and $31.4$.
The results confirm the isotropy and the parity-exchange symmetry, which requires that that the velocity correlations $C_{\gamma\delta}^{\alpha\beta}$ is rotationally invariant and depends only on $|\mathbf{k}|$, $C_{LL}^{AA}=C_{LL}^{BB}$, $C_{TT}^{AA}=C_{TT}^{BB}$, $C_{LL}^{AB}=C_{LL}^{BA}$ and $C_{TT}^{AB}=C_{TT}^{BA}$ are real, while $C_{LT}^{AA}=-C_{LT}^{BB}$ and $C_{LT}^{AB}=-C_{LT}^{BA}$ are complex.

\begin{figure}
    \centering 
    \includegraphics[width=1.0\linewidth]{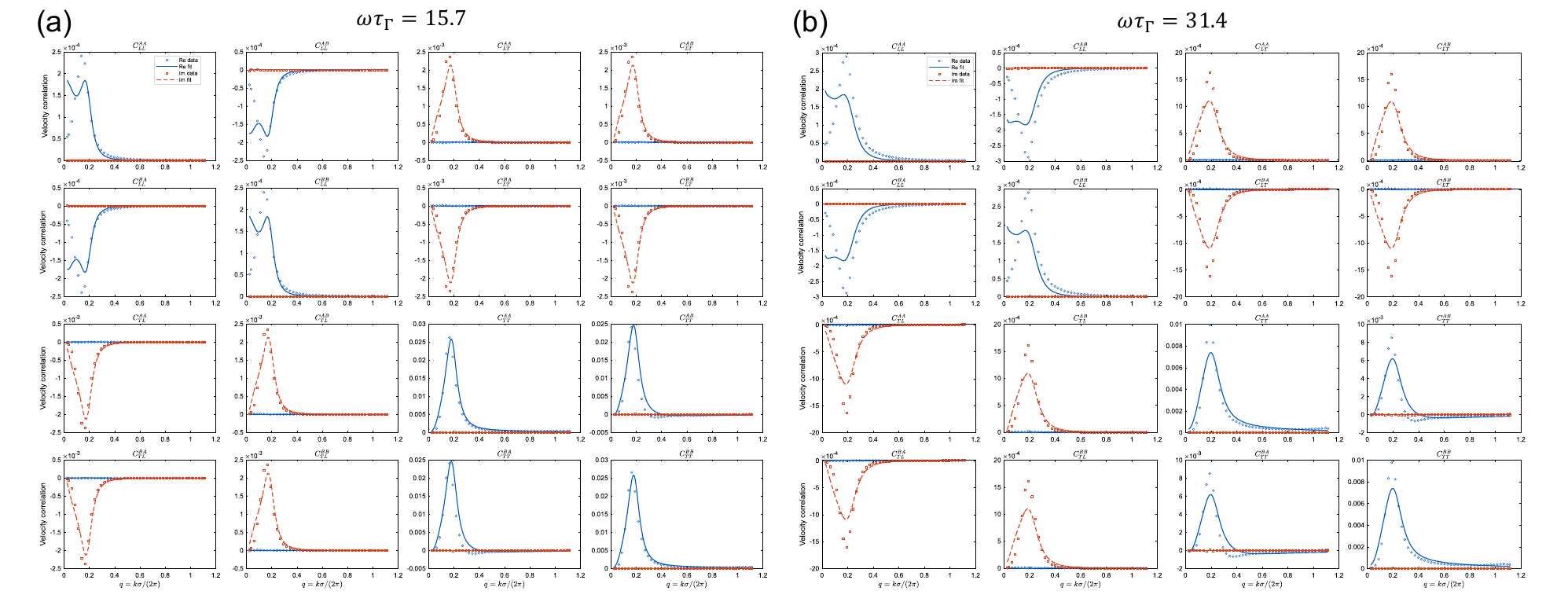}%
    \caption{\textbf{Fitting the microscopic to the theory predicted velocity correlations.} \textbf{(a)} Radially averaged velocity correlations at $\omega\tau_\Gamma=15.7$. Blue circles/solid curves show real parts; orange squares/dashed curves show imaginary parts. Circles and squares are from simulations. Solid and dashed curves are from fitted theory. \textbf{(b)} The corresponding radial comparison at $\omega\tau_\Gamma=31.4$, using the same fitted parameters.}
    \label{fig:S011}
\end{figure}
\begin{figure}
    \centering 
    \includegraphics[width=1.0\linewidth]{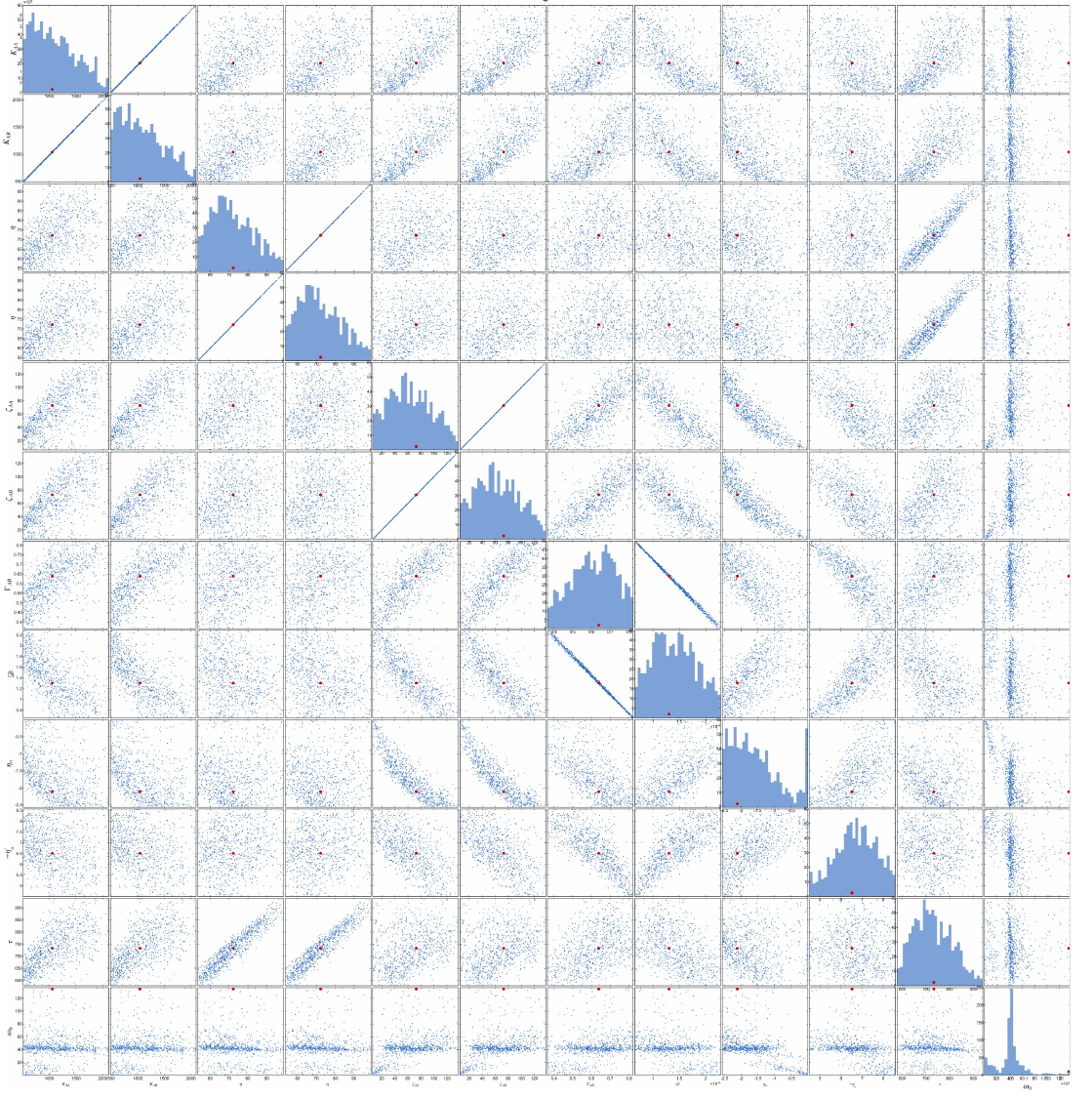}%
    \caption{\textbf{Bootstrap distribution of fitted parameters.} The displayed distributions use the 997 converged optimization results. Red dots mark the full data best-fit estimate. Each axis is restricted to its own marginal 5th--95th percentile.}
    \label{fig:S011B}
\end{figure}
We fit the simulation velocity correlations to the fluctuating hydrodynamic model using Eqs.~\eqref{vel_1} and \eqref{vel_2}, subject to the symmetry constraints in Eq.~\eqref{eq:species-symmetry}. The resulting velocity correlations in species space at the same frequencies as Fig.~\ref{fig:S09} are shown in Fig.~\ref{fig:S011}. Since the spectra are invariant under a common rescaling of the response coefficients and noise amplitude, we fix the overall scale by setting $\Lambda_{AA}=\Lambda_{BB}=1$, while fitting $\Lambda_{AB}=\Lambda_{BA}$. To ensure the positive definiteness of $\mathbf{K}$, $\bm\eta$, $\bm\zeta$, and $\bm\Lambda$, we require $K_\pm,\eta_\pm,\zeta_\pm>0$, $|\Lambda_{AB}|<1$. We also require that $\tau,\omega_0,\mathcal D>0$. We impose positive semidefiniteness of the noise covariance over the fitted wavevector and frequency range. We restrict the fit to the lowest temporal frequencies satisfying $\omega\tau_{\Gamma}<100$, to focus on the lower frequency regime of the fluctuating hydrodynamic model. The theoretical and measured spectra are averaged over the sampled wavenumbers in each of the radial bins $k_l<|\mathbf{k}|<k_{l+1}$. All 16 independent components of the Hermitian correlation matrix are included in a weighted least-squares fit, minimizing $J(\theta)=\sum_{i,j,\ell}
w_j[\widehat y_{ij\ell}(\theta)-y_{ij\ell}]^2$, where $y_{ij\ell}$ and $\widehat y_{ij\ell}(\theta)$ are the measured and predicted correlations for wavenumber $i$, correlation channel $j$, and temporal frequency $\ell$, with $\theta$ denoting the fitting parameters. The channel weights $w_j$ are held fixed and normalized by the corresponding diagonal correlation scales.

The best fit, obtained from 100 independent initial guesses for the optimization, gives an unweighted full matrix relative error $E=\|\hat{y} -y\|/\|y\|=0.259$, where the norm includes all fitted wavenumbers, frequencies, and correlation components. We estimate parameter uncertainties from repeating the optimization over 1000 bootstrap samples by randomly resampling temporal frequencies and correlation channels, while keeping all wavenumbers. The reported $95\%$ percentile intervals of the fitted parameters are obtained from 997 converged fits and characterize sensitivity to the selected spectra. The full data best-fit values are $\eta_\mathrm{o}/\eta=-0.029$, $\eta_\mathrm{o}^{\prime}/\eta=-0.090$, and $\eta_\mathrm{o}/\eta_\mathrm{o}^{\prime} = 0.324$.
The mean parameter values fitted over the converged bootstrap samples are $\eta_\mathrm{o}/\eta=-0.022$, $\eta_\mathrm{o}'/\eta=-0.094$, $\eta_\mathrm{o}/\eta_\mathrm{o}'=0.25$.
The $95\%$ bootstrap percentile intervals
\begin{equation}
 \frac{\eta_\mathrm{o}}{\eta}\equiv\frac{\eta^o_{AA}}{\eta_{AA}}\in[-0.039,\;0],\qquad
 \frac{\eta_\mathrm{o}^{\prime}}{\eta}\equiv-\frac{\eta^o_{AB}}{\eta_{11}}\in[-0.143,\;-0.057], \qquad
 \frac{\eta_\mathrm{o}}{\eta} = -\frac{\eta^o_{AA}}{\eta^o_{AB}} \in [0,0.570]
\end{equation}

Fig.~\ref{fig:S011B} shows strong correlations between $K_{AA}$ and $K_{AB}$, $\eta_{AA}$ and $\eta_{AB}$, and $\zeta_{AA}$ and $\zeta_{AB}$, and that $K_{AA}\approx K_{AB}$, $\eta_{AA}\approx \eta_{AB}$, $\zeta_{AA}\approx\zeta_{AB}$. $K_{AA}\approx K_{AB}$ indicates that the passive free energy does not promote or suppress the mixing of A and B, which is expected since passive A and B are identical and there is no thermal noise that promotes the mixing of A and B. $\eta_{AA}\approx \eta_{AB}$ ($\eta\approx \eta'$) and $\zeta_{AA}\approx\zeta_{AB}$ indicate that passive stress responses are governed by collective modes of the two components. Recall that the Green-Kubo analysis also shows a consistently large self and cross shear viscosity ratio $\eta'/\eta=0.686$.

\subsection{Continuum simulations of the binary mixture}
\label{sec:binary_compressible_simulation}
To examine the onset of bulk demixing, we simulate the full two-fluid model with the generalized viscosity tensor. In contrast to the incompressible continuum simulations in Sec.~\ref{sec:incompressible_two_fluid} that start from a segregated state with preexisting interfaces, here, the full simulations, where we remove the incompressible assumption, allow density variations to develop from an initially homogeneous mixture.

\begin{figure}
    \centering
    \includegraphics[width=0.95\linewidth]{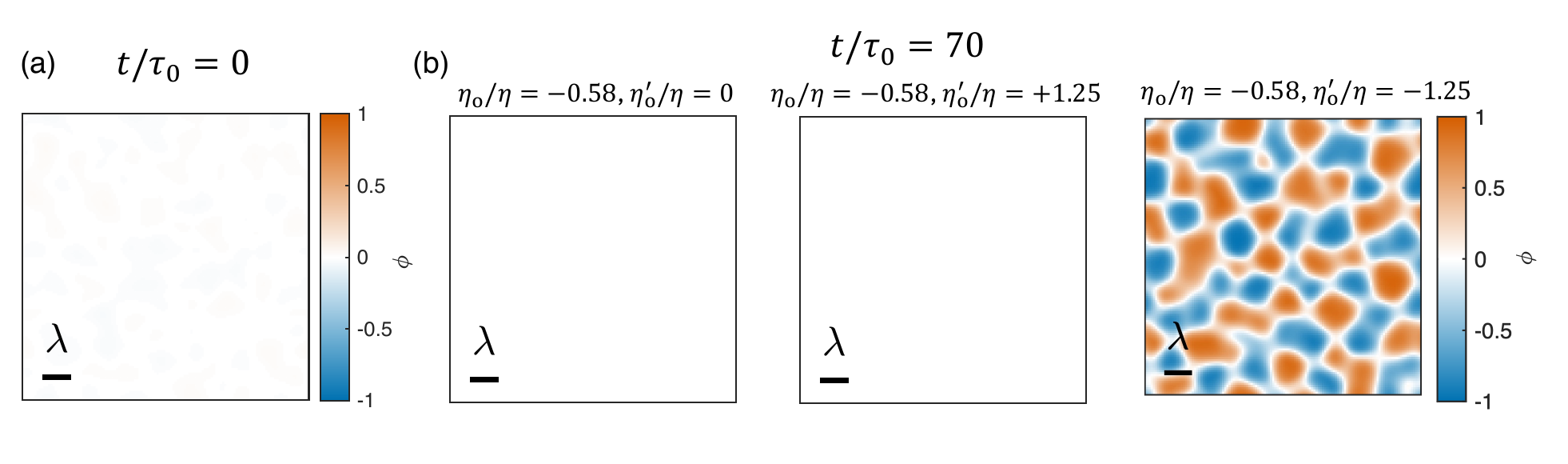}
    \caption{\textbf{Compressible continuum simulations of binary bulk dynamics.} \textbf{(a)}~ Initial configuration at $t/\tau_0=0$, generated from a weakly perturbed homogeneous binary state with $n_A=0.4+\delta n$ and $n_B=0.4-\delta n$. \textbf{(b)}~Composition fields at $t/\tau_0=70$ for $\eta_\mathrm{o}^{\prime}/\eta = 0,+1.25,-1.25$ with $\eta_\mathrm{o}/\eta = -0.583$ held fixed for the $\Omega>0$ component. Demixing occurs for $|\eta_\mathrm{o}^{\prime}|/\eta = -1.25$, for which $\eta_{\mathrm{o}},\eta_{\mathrm{o}}'<0$ and $|\eta_{\mathrm{o}}'|>|\eta_{\mathrm{o}}|$, consistent with the Green--Kubo measurements and the linear instability condition. The color bar denotes the composition $\phi=(n_A-n_B)/(n_A+n_B)$, with positive and negative values corresponding to enrichment of the clockwise and counterclockwise components.}
    \label{fig:S102}
\end{figure}

We use a simple free energy model based on the ideal entropy of mixing. Defining the dimensionless densities $\tilde{n}_\alpha=n_\alpha/\bar{n}$, where $\bar{n}$ is the maximum density, the free energy density is $f(n) = A \left( \tilde{n}_A \ln{\tilde{n}_A} + \tilde n_B \ln{\tilde n_B} +\left(1 - \tilde n_A-\tilde n_B\right) \ln{\left(1-\tilde n_A-\tilde n_B\right)} \right)+\frac{\kappa}{2}\left(|\nabla \tilde n_A|^2+|\nabla \tilde n_B|^2\right)$. The self viscosities of each species $\alpha$ depend linearly on the local densities with $\eta_{\alpha\alpha}=\bar{\eta}_{\alpha\alpha} \tilde{n}_\alpha$, $\eta_{\mathrm{o},\alpha\alpha}=\bar{\eta}_{\mathrm{o},\alpha\alpha} \tilde{n}_\alpha$, where $\bar{\eta}_{AA}=\bar{\eta}_{BB}\bar{\eta}$, $\bar{\eta}_{\mathrm{o},AA}=-\bar{\eta}_{\mathrm{o},BB}=\bar{\eta}_{\mathrm{o}}$. We assume the cross viscosities depend on the local overlap of the two components, 
\begin{equation}
\eta^{\prime}=\bar\eta^{\prime}\frac{\tilde n_A \tilde n_B}{\tilde n_A+\tilde n_B},\quad\eta_{\mathrm{o}}^{\prime}=\bar\eta_{\mathrm{o}}^{\prime}\frac{\tilde n_A \tilde n_B}{\tilde n_A+\tilde n_B}.
\end{equation}

We use the same nondimensionalization as in Sec.~\ref{sec:one_cont}, with characteristic length scale $\lambda$, velocity $u_0$, and the reference active torque coefficient $\Pi_0 = \eta_R\Omega_0$. The relevant time, viscosity, and friction scales are then $\tau_0=\lambda/u_0$, $\eta_0=\Pi_0 \bar n\lambda/u_0$, and $\Gamma_0=\eta_0/\lambda^2$. We set $\lambda=0.1, u_0=1$, and $\Pi_0=1$, giving $\tau_0=0.1, \eta_0=0.1$, and $\Gamma_0=10$. These reference scales remain fixed when the material coefficients are varied.

For the simulations in Fig.~\ref{fig:S102}, the two components have equal self shear viscosity coefficients and substrate friction coefficients, with $\bar\eta/\eta_0\equiv\bar \eta_{AA}/\eta_0=\bar\eta_{BB}/\eta_0=1.5$ and $\Gamma \lambda^2/\eta_0=5$. The bulk viscosities and explicit interspecies drag are zero with $\Gamma_{f0}\bar n/\Gamma = 0$. The active torques are equal and opposite, with dimensionless strengths $\Omega_A/\Omega_0=+10$ and $\Omega_B/\Omega_0=-10$. The free energy parameters are $A/\Pi_0=0.8$ and $\kappa \bar n/(A\lambda^2)=0.3125$.

The simulations are initialized from a weakly perturbed homogeneous mixture, with $\tilde{n}_A=0.4+\delta n_A$ and $\tilde{n}_B=0.4-\delta n_B$, where $\delta n_A$ and $\delta n_B$ are uncorrelated random spatial fields with zero mean and maximum magnitude $0.01$. At the homogeneous state where $\tilde{n}_A=0.4$ and $\tilde{n}_B=0.4$, the odd viscosities are fixed at $\eta_\mathrm{o}/\eta\equiv\eta_\mathrm{o,AA}/\eta_{AA} =-\eta_\mathrm{o,BB}/\eta_{BB}=-0.583$, 
while $\eta^\prime/\eta=0.333$. We vary the dimensionless cross parity-violating coefficient $\bar\eta_\mathrm{o}^{\prime}/\bar{\eta}_0=0, -3.75, +3.75$, corresponding to $\eta_\mathrm{o}^{\prime}/\eta=0, -1.25, +1.25$ at the initial homogeneous state, keeping all other parameters fixed. 

As shown in Fig.~\ref{fig:S102}, the homogeneous state remains stable for $\eta_\mathrm{o}^{\prime}/\eta=0$ and $+1.25$, whereas composition fluctuations grow and phase separation develops for $\eta_\mathrm{o}^{\prime}/\eta=-1.25$. In the unstable case,
\begin{equation}
    \eta_\mathrm{o}-\eta_\mathrm{o}^{\prime}>0,
\end{equation}
as required by the demixing condition obtained from the linear stability analysis. Importantly, this is also the relationship obtained independently from the Green--Kubo measurements of the particle system, for which both $\eta_\mathrm{o}$ and $\eta_\mathrm{o}^{\prime}$ have the same sign while  $|\eta_\mathrm{o}^{\prime}|>|\eta_\mathrm{o}|$. The resulting growth of composition fluctuations into phase separated domains is qualitatively consistent with the onset of demixing observed in the agent-based simulations. Together, these results show that the parity-violating viscosities $\eta_{\mathrm{o},\alpha\beta}$ inferred from the microscopic dynamics produce the relationship required for the onset of bulk demixing instability in the continuum model, complementary to the incompressible simulation results in Sec.~\ref{sec:4}, which characterize the late-stage dynamics.


\begin{thebibliography}{49}%
\makeatletter
\providecommand \@ifxundefined [1]{%
 \@ifx{#1\undefined}
}%
\providecommand \@ifnum [1]{%
 \ifnum #1\expandafter \@firstoftwo
 \else \expandafter \@secondoftwo
 \fi
}%
\providecommand \@ifx [1]{%
 \ifx #1\expandafter \@firstoftwo
 \else \expandafter \@secondoftwo
 \fi
}%
\providecommand \natexlab [1]{#1}%
\providecommand \enquote  [1]{``#1''}%
\providecommand \bibnamefont  [1]{#1}%
\providecommand \bibfnamefont [1]{#1}%
\providecommand \citenamefont [1]{#1}%
\providecommand \href@noop [0]{\@secondoftwo}%
\providecommand \href [0]{\begingroup \@sanitize@url \@href}%
\providecommand \@href[1]{\@@startlink{#1}\@@href}%
\providecommand \@@href[1]{\endgroup#1\@@endlink}%
\providecommand \@sanitize@url [0]{\catcode `\\12\catcode `\$12\catcode `\&12\catcode `\#12\catcode `\^12\catcode `\_12\catcode `\%12\relax}%
\providecommand \@@startlink[1]{}%
\providecommand \@@endlink[0]{}%
\providecommand \url  [0]{\begingroup\@sanitize@url \@url }%
\providecommand \@url [1]{\endgroup\@href {#1}{\urlprefix }}%
\providecommand \urlprefix  [0]{URL }%
\providecommand \Eprint [0]{\href }%
\providecommand \doibase [0]{https://doi.org/}%
\providecommand \selectlanguage [0]{\@gobble}%
\providecommand \bibinfo  [0]{\@secondoftwo}%
\providecommand \bibfield  [0]{\@secondoftwo}%
\providecommand \translation [1]{[#1]}%
\providecommand \BibitemOpen [0]{}%
\providecommand \bibitemStop [0]{}%
\providecommand \bibitemNoStop [0]{.\EOS\space}%
\providecommand \EOS [0]{\spacefactor3000\relax}%
\providecommand \BibitemShut  [1]{\csname bibitem#1\endcsname}%
\let\auto@bib@innerbib\@empty
\bibitem [{\citenamefont {Cates}\ and\ \citenamefont {Tailleur}(2015)}]{Cates2015}%
  \BibitemOpen
  \bibfield  {author} {\bibinfo {author} {\bibfnamefont {M.~E.}\ \bibnamefont {Cates}}\ and\ \bibinfo {author} {\bibfnamefont {J.}~\bibnamefont {Tailleur}},\ }\href {https://doi.org/10.1146/annurev-conmatphys-031214-014710} {\bibfield  {journal} {\bibinfo  {journal} {Annual Review of Condensed Matter Physics}\ }\textbf {\bibinfo {volume} {6}},\ \bibinfo {pages} {219} (\bibinfo {year} {2015})}\BibitemShut {NoStop}%
\bibitem [{\citenamefont {Wysocki}\ \emph {et~al.}(2016)\citenamefont {Wysocki}, \citenamefont {Winkler},\ and\ \citenamefont {Gompper}}]{Wysocki2016}%
  \BibitemOpen
  \bibfield  {author} {\bibinfo {author} {\bibfnamefont {A.}~\bibnamefont {Wysocki}}, \bibinfo {author} {\bibfnamefont {R.~G.}\ \bibnamefont {Winkler}},\ and\ \bibinfo {author} {\bibfnamefont {G.}~\bibnamefont {Gompper}},\ }\href {https://doi.org/10.1088/1367-2630/aa529d} {\bibfield  {journal} {\bibinfo  {journal} {New Journal of Physics}\ }\textbf {\bibinfo {volume} {18}},\ \bibinfo {pages} {123030} (\bibinfo {year} {2016})}\BibitemShut {NoStop}%
\bibitem [{\citenamefont {Stenhammar}\ \emph {et~al.}(2015)\citenamefont {Stenhammar}, \citenamefont {Wittkowski}, \citenamefont {Marenduzzo},\ and\ \citenamefont {Cates}}]{Stenhammar2015}%
  \BibitemOpen
  \bibfield  {author} {\bibinfo {author} {\bibfnamefont {J.}~\bibnamefont {Stenhammar}}, \bibinfo {author} {\bibfnamefont {R.}~\bibnamefont {Wittkowski}}, \bibinfo {author} {\bibfnamefont {D.}~\bibnamefont {Marenduzzo}},\ and\ \bibinfo {author} {\bibfnamefont {M.~E.}\ \bibnamefont {Cates}},\ }\href {https://doi.org/10.1103/PhysRevLett.114.018301} {\bibfield  {journal} {\bibinfo  {journal} {Phys. Rev. Lett.}\ }\textbf {\bibinfo {volume} {114}},\ \bibinfo {pages} {018301} (\bibinfo {year} {2015})}\BibitemShut {NoStop}%
\bibitem [{\citenamefont {Bhattacharyya}\ and\ \citenamefont {Yeomans}(2023)}]{Bhattacharyya2023}%
  \BibitemOpen
  \bibfield  {author} {\bibinfo {author} {\bibfnamefont {S.}~\bibnamefont {Bhattacharyya}}\ and\ \bibinfo {author} {\bibfnamefont {J.~M.}\ \bibnamefont {Yeomans}},\ }\href {https://doi.org/10.1103/PhysRevLett.130.238201} {\bibfield  {journal} {\bibinfo  {journal} {Phys. Rev. Lett.}\ }\textbf {\bibinfo {volume} {130}},\ \bibinfo {pages} {238201} (\bibinfo {year} {2023})}\BibitemShut {NoStop}%
\bibitem [{\citenamefont {Bhattacharyya}\ and\ \citenamefont {Yeomans}(2024)}]{Bhattacharyya2024}%
  \BibitemOpen
  \bibfield  {author} {\bibinfo {author} {\bibfnamefont {S.}~\bibnamefont {Bhattacharyya}}\ and\ \bibinfo {author} {\bibfnamefont {J.~M.}\ \bibnamefont {Yeomans}},\ }\href {https://doi.org/10.1103/PhysRevE.110.024607} {\bibfield  {journal} {\bibinfo  {journal} {Phys. Rev. E}\ }\textbf {\bibinfo {volume} {110}},\ \bibinfo {pages} {024607} (\bibinfo {year} {2024})}\BibitemShut {NoStop}%
\bibitem [{\citenamefont {Liebchen}\ and\ \citenamefont {Levis}(2022)}]{Liebchen2022}%
  \BibitemOpen
  \bibfield  {author} {\bibinfo {author} {\bibfnamefont {B.}~\bibnamefont {Liebchen}}\ and\ \bibinfo {author} {\bibfnamefont {D.}~\bibnamefont {Levis}},\ }\href {https://doi.org/10.1209/0295-5075/ac8f69} {\bibfield  {journal} {\bibinfo  {journal} {Europhysics Letters}\ }\textbf {\bibinfo {volume} {139}},\ \bibinfo {pages} {67001} (\bibinfo {year} {2022})}\BibitemShut {NoStop}%
\bibitem [{\citenamefont {Levis}\ and\ \citenamefont {Liebchen}(2019)}]{Levis2019}%
  \BibitemOpen
  \bibfield  {author} {\bibinfo {author} {\bibfnamefont {D.}~\bibnamefont {Levis}}\ and\ \bibinfo {author} {\bibfnamefont {B.}~\bibnamefont {Liebchen}},\ }\href {https://doi.org/10.1103/PhysRevE.100.012406} {\bibfield  {journal} {\bibinfo  {journal} {Physical Review E}\ }\textbf {\bibinfo {volume} {100}},\ \bibinfo {pages} {012406} (\bibinfo {year} {2019})}\BibitemShut {NoStop}%
\bibitem [{\citenamefont {Ventejou}\ \emph {et~al.}(2021)\citenamefont {Ventejou}, \citenamefont {Chat{\'e}}, \citenamefont {Montagne},\ and\ \citenamefont {Shi}}]{Ventejou2021}%
  \BibitemOpen
  \bibfield  {author} {\bibinfo {author} {\bibfnamefont {B.}~\bibnamefont {Ventejou}}, \bibinfo {author} {\bibfnamefont {H.}~\bibnamefont {Chat{\'e}}}, \bibinfo {author} {\bibfnamefont {R.}~\bibnamefont {Montagne}},\ and\ \bibinfo {author} {\bibfnamefont {X.-q.}\ \bibnamefont {Shi}},\ }\href {https://doi.org/10.1103/PhysRevLett.127.238001} {\bibfield  {journal} {\bibinfo  {journal} {Physical Review Letters}\ }\textbf {\bibinfo {volume} {127}},\ \bibinfo {pages} {238001} (\bibinfo {year} {2021})}\BibitemShut {NoStop}%
\bibitem [{\citenamefont {Ai}\ \emph {et~al.}(2023)\citenamefont {Ai}, \citenamefont {Quan},\ and\ \citenamefont {Li}}]{Ai2023}%
  \BibitemOpen
  \bibfield  {author} {\bibinfo {author} {\bibfnamefont {B.-Q.}\ \bibnamefont {Ai}}, \bibinfo {author} {\bibfnamefont {S.}~\bibnamefont {Quan}},\ and\ \bibinfo {author} {\bibfnamefont {F.-g.}\ \bibnamefont {Li}},\ }\href {https://doi.org/10.1088/1367-2630/acdf79} {\bibfield  {journal} {\bibinfo  {journal} {New Journal of Physics}\ }\textbf {\bibinfo {volume} {25}},\ \bibinfo {pages} {063025} (\bibinfo {year} {2023})}\BibitemShut {NoStop}%
\bibitem [{\citenamefont {Nguyen}\ \emph {et~al.}(2014)\citenamefont {Nguyen}, \citenamefont {Klotsa}, \citenamefont {Engel},\ and\ \citenamefont {Glotzer}}]{Nguyen2014}%
  \BibitemOpen
  \bibfield  {author} {\bibinfo {author} {\bibfnamefont {N.~H.~P.}\ \bibnamefont {Nguyen}}, \bibinfo {author} {\bibfnamefont {D.}~\bibnamefont {Klotsa}}, \bibinfo {author} {\bibfnamefont {M.}~\bibnamefont {Engel}},\ and\ \bibinfo {author} {\bibfnamefont {S.~C.}\ \bibnamefont {Glotzer}},\ }\href {https://doi.org/10.1103/PhysRevLett.112.075701} {\bibfield  {journal} {\bibinfo  {journal} {Phys. Rev. Lett.}\ }\textbf {\bibinfo {volume} {112}},\ \bibinfo {pages} {075701} (\bibinfo {year} {2014})}\BibitemShut {NoStop}%
\bibitem [{\citenamefont {Guo}\ \emph {et~al.}(2025)\citenamefont {Guo}, \citenamefont {Li},\ and\ \citenamefont {Ai}}]{Guo2025}%
  \BibitemOpen
  \bibfield  {author} {\bibinfo {author} {\bibfnamefont {R.-x.}\ \bibnamefont {Guo}}, \bibinfo {author} {\bibfnamefont {J.-j.}\ \bibnamefont {Li}},\ and\ \bibinfo {author} {\bibfnamefont {B.-q.}\ \bibnamefont {Ai}},\ }\href {https://doi.org/10.1103/PhysRevE.111.045423} {\bibfield  {journal} {\bibinfo  {journal} {Phys. Rev. E}\ }\textbf {\bibinfo {volume} {111}},\ \bibinfo {pages} {045423} (\bibinfo {year} {2025})}\BibitemShut {NoStop}%
\bibitem [{\citenamefont {Yeo}\ \emph {et~al.}(2015)\citenamefont {Yeo}, \citenamefont {Lushi},\ and\ \citenamefont {Vlahovska}}]{Yeo2015}%
  \BibitemOpen
  \bibfield  {author} {\bibinfo {author} {\bibfnamefont {K.}~\bibnamefont {Yeo}}, \bibinfo {author} {\bibfnamefont {E.}~\bibnamefont {Lushi}},\ and\ \bibinfo {author} {\bibfnamefont {P.~M.}\ \bibnamefont {Vlahovska}},\ }\href {https://doi.org/10.1103/PhysRevLett.114.188301} {\bibfield  {journal} {\bibinfo  {journal} {Phys. Rev. Lett.}\ }\textbf {\bibinfo {volume} {114}},\ \bibinfo {pages} {188301} (\bibinfo {year} {2015})}\BibitemShut {NoStop}%
\bibitem [{\citenamefont {Scholz}\ \emph {et~al.}(2018)\citenamefont {Scholz}, \citenamefont {Engel},\ and\ \citenamefont {P{\"o}schel}}]{Scholz2018}%
  \BibitemOpen
  \bibfield  {author} {\bibinfo {author} {\bibfnamefont {C.}~\bibnamefont {Scholz}}, \bibinfo {author} {\bibfnamefont {M.}~\bibnamefont {Engel}},\ and\ \bibinfo {author} {\bibfnamefont {T.}~\bibnamefont {P{\"o}schel}},\ }\href {https://doi.org/10.1038/s41467-018-03154-7} {\bibfield  {journal} {\bibinfo  {journal} {Nature Communications}\ }\textbf {\bibinfo {volume} {9}},\ \bibinfo {pages} {931} (\bibinfo {year} {2018})}\BibitemShut {NoStop}%
\bibitem [{\citenamefont {Deb}\ and\ \citenamefont {Singh}(2026)}]{Deb2026}%
  \BibitemOpen
  \bibfield  {author} {\bibinfo {author} {\bibfnamefont {M.}~\bibnamefont {Deb}}\ and\ \bibinfo {author} {\bibfnamefont {R.}~\bibnamefont {Singh}},\ }\href {https://doi.org/10.1039/d6sm00188b} {\bibfield  {journal} {\bibinfo  {journal} {Soft Matter}\ }\textbf {\bibinfo {volume} {22}},\ \bibinfo {pages} {4846} (\bibinfo {year} {2026})}\BibitemShut {NoStop}%
\bibitem [{\citenamefont {Cates}\ and\ \citenamefont {Nardini}(2025)}]{Cates2025}%
  \BibitemOpen
  \bibfield  {author} {\bibinfo {author} {\bibfnamefont {M.~E.}\ \bibnamefont {Cates}}\ and\ \bibinfo {author} {\bibfnamefont {C.}~\bibnamefont {Nardini}},\ }\href {https://doi.org/10.1088/1361-6633/add278} {\bibfield  {journal} {\bibinfo  {journal} {Reports on Progress in Physics}\ }\textbf {\bibinfo {volume} {88}},\ \bibinfo {pages} {056601} (\bibinfo {year} {2025})}\BibitemShut {NoStop}%
\bibitem [{\citenamefont {Khain}\ \emph {et~al.}(2022)\citenamefont {Khain}, \citenamefont {Scheibner}, \citenamefont {Fruchart},\ and\ \citenamefont {Vitelli}}]{Khain2022}%
  \BibitemOpen
  \bibfield  {author} {\bibinfo {author} {\bibfnamefont {T.}~\bibnamefont {Khain}}, \bibinfo {author} {\bibfnamefont {C.}~\bibnamefont {Scheibner}}, \bibinfo {author} {\bibfnamefont {M.}~\bibnamefont {Fruchart}},\ and\ \bibinfo {author} {\bibfnamefont {V.}~\bibnamefont {Vitelli}},\ }\href {https://doi.org/10.1017/jfm.2021.1079} {\bibfield  {journal} {\bibinfo  {journal} {Journal of Fluid Mechanics}\ }\textbf {\bibinfo {volume} {934}},\ \bibinfo {eid} {A23} (\bibinfo {year} {2022})}\BibitemShut {NoStop}%
\bibitem [{\citenamefont {Fruchart}\ \emph {et~al.}(2023)\citenamefont {Fruchart}, \citenamefont {Scheibner},\ and\ \citenamefont {Vitelli}}]{Fruchart2023}%
  \BibitemOpen
  \bibfield  {author} {\bibinfo {author} {\bibfnamefont {M.}~\bibnamefont {Fruchart}}, \bibinfo {author} {\bibfnamefont {C.}~\bibnamefont {Scheibner}},\ and\ \bibinfo {author} {\bibfnamefont {V.}~\bibnamefont {Vitelli}},\ }\href {https://doi.org/https://doi.org/10.1146/annurev-conmatphys-040821-125506} {\bibfield  {journal} {\bibinfo  {journal} {Annual Review of Condensed Matter Physics}\ }\textbf {\bibinfo {volume} {14}},\ \bibinfo {pages} {471} (\bibinfo {year} {2023})}\BibitemShut {NoStop}%
\bibitem [{\citenamefont {Banerjee}\ \emph {et~al.}(2017)\citenamefont {Banerjee}, \citenamefont {Souslov}, \citenamefont {Abanov},\ and\ \citenamefont {Vitelli}}]{Banerjee2017}%
  \BibitemOpen
  \bibfield  {author} {\bibinfo {author} {\bibfnamefont {D.}~\bibnamefont {Banerjee}}, \bibinfo {author} {\bibfnamefont {A.}~\bibnamefont {Souslov}}, \bibinfo {author} {\bibfnamefont {A.~G.}\ \bibnamefont {Abanov}},\ and\ \bibinfo {author} {\bibfnamefont {V.}~\bibnamefont {Vitelli}},\ }\href {https://doi.org/10.1038/s41467-017-01378-7} {\bibfield  {journal} {\bibinfo  {journal} {Nature Communications}\ }\textbf {\bibinfo {volume} {8}},\ \bibinfo {pages} {1573} (\bibinfo {year} {2017})}\BibitemShut {NoStop}%
\bibitem [{\citenamefont {Markovich}\ and\ \citenamefont {Lubensky}(2024)}]{Markovich2024}%
  \BibitemOpen
  \bibfield  {author} {\bibinfo {author} {\bibfnamefont {T.}~\bibnamefont {Markovich}}\ and\ \bibinfo {author} {\bibfnamefont {T.~C.}\ \bibnamefont {Lubensky}},\ }\href {https://doi.org/10.1073/pnas.2219385121} {\bibfield  {journal} {\bibinfo  {journal} {Proceedings of the National Academy of Sciences}\ }\textbf {\bibinfo {volume} {121}},\ \bibinfo {pages} {e2219385121} (\bibinfo {year} {2024})}\BibitemShut {NoStop}%
\bibitem [{\citenamefont {Markovich}\ and\ \citenamefont {Lubensky}(2021)}]{Markovich2021}%
  \BibitemOpen
  \bibfield  {author} {\bibinfo {author} {\bibfnamefont {T.}~\bibnamefont {Markovich}}\ and\ \bibinfo {author} {\bibfnamefont {T.~C.}\ \bibnamefont {Lubensky}},\ }\href {https://doi.org/10.1103/PhysRevLett.127.048001} {\bibfield  {journal} {\bibinfo  {journal} {Phys. Rev. Lett.}\ }\textbf {\bibinfo {volume} {127}},\ \bibinfo {pages} {048001} (\bibinfo {year} {2021})}\BibitemShut {NoStop}%
\bibitem [{\citenamefont {Caprini}\ and\ \citenamefont {Marini Bettolo~Marconi}(2025)}]{Caprini2025}%
  \BibitemOpen
  \bibfield  {author} {\bibinfo {author} {\bibfnamefont {L.}~\bibnamefont {Caprini}}\ and\ \bibinfo {author} {\bibfnamefont {U.}~\bibnamefont {Marini Bettolo~Marconi}},\ }\href {https://doi.org/10.1063/5.0262594} {\bibfield  {journal} {\bibinfo  {journal} {The Journal of Chemical Physics}\ }\textbf {\bibinfo {volume} {162}},\ \bibinfo {pages} {161101} (\bibinfo {year} {2025})}\BibitemShut {NoStop}%
\bibitem [{\citenamefont {Marini Bettolo~Marconi}\ \emph {et~al.}(2026)\citenamefont {Marini Bettolo~Marconi}, \citenamefont {Petrini}, \citenamefont {Maire},\ and\ \citenamefont {Caprini}}]{Marconi2026}%
  \BibitemOpen
  \bibfield  {author} {\bibinfo {author} {\bibfnamefont {U.}~\bibnamefont {Marini Bettolo~Marconi}}, \bibinfo {author} {\bibfnamefont {A.}~\bibnamefont {Petrini}}, \bibinfo {author} {\bibfnamefont {R.}~\bibnamefont {Maire}},\ and\ \bibinfo {author} {\bibfnamefont {L.}~\bibnamefont {Caprini}},\ }\href {https://doi.org/10.1088/1367-2630/ae6c55} {\bibfield  {journal} {\bibinfo  {journal} {New Journal of Physics}\ }\textbf {\bibinfo {volume} {28}},\ \bibinfo {pages} {064401} (\bibinfo {year} {2026})}\BibitemShut {NoStop}%
\bibitem [{\citenamefont {Ding}\ \emph {et~al.}(2024)\citenamefont {Ding}, \citenamefont {Wang}, \citenamefont {Yang}, \citenamefont {Zhao}, \citenamefont {Komura}, \citenamefont {Seto}, \citenamefont {Yang},\ and\ \citenamefont {Ye}}]{Yu2024}%
  \BibitemOpen
  \bibfield  {author} {\bibinfo {author} {\bibfnamefont {Y.}~\bibnamefont {Ding}}, \bibinfo {author} {\bibfnamefont {B.}~\bibnamefont {Wang}}, \bibinfo {author} {\bibfnamefont {Q.}~\bibnamefont {Yang}}, \bibinfo {author} {\bibfnamefont {Z.}~\bibnamefont {Zhao}}, \bibinfo {author} {\bibfnamefont {S.}~\bibnamefont {Komura}}, \bibinfo {author} {\bibfnamefont {R.}~\bibnamefont {Seto}}, \bibinfo {author} {\bibfnamefont {M.}~\bibnamefont {Yang}},\ and\ \bibinfo {author} {\bibfnamefont {F.}~\bibnamefont {Ye}},\ }\href {https://doi.org/10.34133/research.0356} {\bibfield  {journal} {\bibinfo  {journal} {Research}\ }\textbf {\bibinfo {volume} {7}},\ \bibinfo {pages} {0356} (\bibinfo {year} {2024})}\BibitemShut {NoStop}%
\bibitem [{\citenamefont {{Soni}}\ \emph {et~al.}(2019)\citenamefont {{Soni}}, \citenamefont {{Bililign}}, \citenamefont {{Magkiriadou}}, \citenamefont {{Sacanna}}, \citenamefont {{Bartolo}}, \citenamefont {{Shelley}},\ and\ \citenamefont {{Irvine}}}]{Soni2019}%
  \BibitemOpen
  \bibfield  {author} {\bibinfo {author} {\bibfnamefont {V.}~\bibnamefont {{Soni}}}, \bibinfo {author} {\bibfnamefont {E.~S.}\ \bibnamefont {{Bililign}}}, \bibinfo {author} {\bibfnamefont {S.}~\bibnamefont {{Magkiriadou}}}, \bibinfo {author} {\bibfnamefont {S.}~\bibnamefont {{Sacanna}}}, \bibinfo {author} {\bibfnamefont {D.}~\bibnamefont {{Bartolo}}}, \bibinfo {author} {\bibfnamefont {M.~J.}\ \bibnamefont {{Shelley}}},\ and\ \bibinfo {author} {\bibfnamefont {W.~T.~M.}\ \bibnamefont {{Irvine}}},\ }\href {https://doi.org/10.1038/s41567-019-0603-8} {\bibfield  {journal} {\bibinfo  {journal} {Nature Physics}\ }\textbf {\bibinfo {volume} {15}},\ \bibinfo {pages} {1188} (\bibinfo {year} {2019})}\BibitemShut {NoStop}%
\bibitem [{\citenamefont {Caporusso}\ \emph {et~al.}(2024)\citenamefont {Caporusso}, \citenamefont {Gonnella},\ and\ \citenamefont {Levis}}]{Caporusso2024}%
  \BibitemOpen
  \bibfield  {author} {\bibinfo {author} {\bibfnamefont {C.~B.}\ \bibnamefont {Caporusso}}, \bibinfo {author} {\bibfnamefont {G.}~\bibnamefont {Gonnella}},\ and\ \bibinfo {author} {\bibfnamefont {D.}~\bibnamefont {Levis}},\ }\href {https://doi.org/10.1103/PhysRevLett.132.168201} {\bibfield  {journal} {\bibinfo  {journal} {Phys. Rev. Lett.}\ }\textbf {\bibinfo {volume} {132}},\ \bibinfo {pages} {168201} (\bibinfo {year} {2024})}\BibitemShut {NoStop}%
\bibitem [{\citenamefont {Neville}\ \emph {et~al.}(2026)\citenamefont {Neville}, \citenamefont {Eggers},\ and\ \citenamefont {Liverpool}}]{Neville2026}%
  \BibitemOpen
  \bibfield  {author} {\bibinfo {author} {\bibfnamefont {L.}~\bibnamefont {Neville}}, \bibinfo {author} {\bibfnamefont {J.}~\bibnamefont {Eggers}},\ and\ \bibinfo {author} {\bibfnamefont {T.~B.}\ \bibnamefont {Liverpool}},\ }\href {https://doi.org/10.1103/w2n2-zzkd} {\bibfield  {journal} {\bibinfo  {journal} {Phys. Rev. Lett.}\ }\textbf {\bibinfo {volume} {137}},\ \bibinfo {pages} {088302} (\bibinfo {year} {2026})}\BibitemShut {NoStop}%
\bibitem [{\citenamefont {Abanov}\ \emph {et~al.}(2018)\citenamefont {Abanov}, \citenamefont {Can},\ and\ \citenamefont {Ganeshan}}]{Abanov2018}%
  \BibitemOpen
  \bibfield  {author} {\bibinfo {author} {\bibfnamefont {A.~G.}\ \bibnamefont {Abanov}}, \bibinfo {author} {\bibfnamefont {T.}~\bibnamefont {Can}},\ and\ \bibinfo {author} {\bibfnamefont {S.}~\bibnamefont {Ganeshan}},\ }\bibfield  {journal} {\bibinfo  {journal} {SciPost Physics}\ }\textbf {\bibinfo {volume} {5}},\ \href {https://doi.org/10.21468/scipostphys.5.1.010} {10.21468/scipostphys.5.1.010} (\bibinfo {year} {2018})\BibitemShut {NoStop}%
\bibitem [{\citenamefont {Souslov}\ \emph {et~al.}(2019)\citenamefont {Souslov}, \citenamefont {Dasbiswas}, \citenamefont {Fruchart}, \citenamefont {Vaikuntanathan},\ and\ \citenamefont {Vitelli}}]{Souslov2019}%
  \BibitemOpen
  \bibfield  {author} {\bibinfo {author} {\bibfnamefont {A.}~\bibnamefont {Souslov}}, \bibinfo {author} {\bibfnamefont {K.}~\bibnamefont {Dasbiswas}}, \bibinfo {author} {\bibfnamefont {M.}~\bibnamefont {Fruchart}}, \bibinfo {author} {\bibfnamefont {S.}~\bibnamefont {Vaikuntanathan}},\ and\ \bibinfo {author} {\bibfnamefont {V.}~\bibnamefont {Vitelli}},\ }\href {https://doi.org/10.1103/PhysRevLett.122.128001} {\bibfield  {journal} {\bibinfo  {journal} {Phys. Rev. Lett.}\ }\textbf {\bibinfo {volume} {122}},\ \bibinfo {pages} {128001} (\bibinfo {year} {2019})}\BibitemShut {NoStop}%
\bibitem [{\citenamefont {Han}\ \emph {et~al.}(2021)\citenamefont {Han}, \citenamefont {Fruchart}, \citenamefont {Scheibner}, \citenamefont {Vaikuntanathan}, \citenamefont {de~Pablo},\ and\ \citenamefont {Vitelli}}]{Han2021}%
  \BibitemOpen
  \bibfield  {author} {\bibinfo {author} {\bibfnamefont {M.}~\bibnamefont {Han}}, \bibinfo {author} {\bibfnamefont {M.}~\bibnamefont {Fruchart}}, \bibinfo {author} {\bibfnamefont {C.}~\bibnamefont {Scheibner}}, \bibinfo {author} {\bibfnamefont {S.}~\bibnamefont {Vaikuntanathan}}, \bibinfo {author} {\bibfnamefont {J.~J.}\ \bibnamefont {de~Pablo}},\ and\ \bibinfo {author} {\bibfnamefont {V.}~\bibnamefont {Vitelli}},\ }\href {https://doi.org/10.1038/s41567-021-01360-7} {\bibfield  {journal} {\bibinfo  {journal} {Nature Physics}\ }\textbf {\bibinfo {volume} {17}},\ \bibinfo {pages} {1260} (\bibinfo {year} {2021})}\BibitemShut {NoStop}%
\bibitem [{\citenamefont {Tan}\ \emph {et~al.}(2022)\citenamefont {Tan}, \citenamefont {Mietke}, \citenamefont {Li}, \citenamefont {Chen}, \citenamefont {Higinbotham}, \citenamefont {Foster}, \citenamefont {Gokhale}, \citenamefont {Dunkel},\ and\ \citenamefont {Fakhri}}]{Tan2022}%
  \BibitemOpen
  \bibfield  {author} {\bibinfo {author} {\bibfnamefont {T.~H.}\ \bibnamefont {Tan}}, \bibinfo {author} {\bibfnamefont {A.}~\bibnamefont {Mietke}}, \bibinfo {author} {\bibfnamefont {J.}~\bibnamefont {Li}}, \bibinfo {author} {\bibfnamefont {Y.}~\bibnamefont {Chen}}, \bibinfo {author} {\bibfnamefont {H.}~\bibnamefont {Higinbotham}}, \bibinfo {author} {\bibfnamefont {P.~J.}\ \bibnamefont {Foster}}, \bibinfo {author} {\bibfnamefont {S.}~\bibnamefont {Gokhale}}, \bibinfo {author} {\bibfnamefont {J.}~\bibnamefont {Dunkel}},\ and\ \bibinfo {author} {\bibfnamefont {N.}~\bibnamefont {Fakhri}},\ }\href {https://doi.org/10.1038/s41586-022-04889-6} {\bibfield  {journal} {\bibinfo  {journal} {Nature}\ }\textbf {\bibinfo {volume} {607}},\ \bibinfo {pages} {287} (\bibinfo {year} {2022})}\BibitemShut {NoStop}%
\bibitem [{\citenamefont {Bhattacharyya}\ and\ \citenamefont {Yeomans}(2025)}]{Bhattacharyya2025}%
  \BibitemOpen
  \bibfield  {author} {\bibinfo {author} {\bibfnamefont {S.}~\bibnamefont {Bhattacharyya}}\ and\ \bibinfo {author} {\bibfnamefont {J.~M.}\ \bibnamefont {Yeomans}},\ }\href {https://doi.org/10.1039/D5SM00636H} {\bibfield  {journal} {\bibinfo  {journal} {Soft Matter}\ }\textbf {\bibinfo {volume} {21}},\ \bibinfo {pages} {7641} (\bibinfo {year} {2025})}\BibitemShut {NoStop}%
\bibitem [{\citenamefont {Bruinsma}\ \emph {et~al.}(2014)\citenamefont {Bruinsma}, \citenamefont {Grosberg}, \citenamefont {Rabin},\ and\ \citenamefont {Zidovska}}]{Bruinsma2014}%
  \BibitemOpen
  \bibfield  {author} {\bibinfo {author} {\bibfnamefont {R.}~\bibnamefont {Bruinsma}}, \bibinfo {author} {\bibfnamefont {A.~Y.}\ \bibnamefont {Grosberg}}, \bibinfo {author} {\bibfnamefont {Y.}~\bibnamefont {Rabin}},\ and\ \bibinfo {author} {\bibfnamefont {A.}~\bibnamefont {Zidovska}},\ }\href {https://doi.org/10.1016/j.bpj.2014.03.038} {\bibfield  {journal} {\bibinfo  {journal} {Biophysical Journal}\ }\textbf {\bibinfo {volume} {106}},\ \bibinfo {pages} {1871} (\bibinfo {year} {2014})}\BibitemShut {NoStop}%
\bibitem [{\citenamefont {Rautu}\ \emph {et~al.}(2026)\citenamefont {Rautu}, \citenamefont {Zidovska}, \citenamefont {Saintillan},\ and\ \citenamefont {Shelley}}]{Rautu2026}%
  \BibitemOpen
  \bibfield  {author} {\bibinfo {author} {\bibfnamefont {S.~A.}\ \bibnamefont {Rautu}}, \bibinfo {author} {\bibfnamefont {A.}~\bibnamefont {Zidovska}}, \bibinfo {author} {\bibfnamefont {D.}~\bibnamefont {Saintillan}},\ and\ \bibinfo {author} {\bibfnamefont {M.~J.}\ \bibnamefont {Shelley}},\ }\href {https://doi.org/10.1103/8n8h-7gx6} {\bibfield  {journal} {\bibinfo  {journal} {Physical Review X}\ }\textbf {\bibinfo {volume} {16}},\ \bibinfo {pages} {021009} (\bibinfo {year} {2026})}\BibitemShut {NoStop}%
\bibitem [{\citenamefont {Hargus}\ \emph {et~al.}(2020)\citenamefont {Hargus}, \citenamefont {Klymko}, \citenamefont {Epstein},\ and\ \citenamefont {Mandadapu}}]{Hargus2020}%
  \BibitemOpen
  \bibfield  {author} {\bibinfo {author} {\bibfnamefont {C.}~\bibnamefont {Hargus}}, \bibinfo {author} {\bibfnamefont {K.}~\bibnamefont {Klymko}}, \bibinfo {author} {\bibfnamefont {J.~M.}\ \bibnamefont {Epstein}},\ and\ \bibinfo {author} {\bibfnamefont {K.~K.}\ \bibnamefont {Mandadapu}},\ }\href {https://doi.org/10.1063/5.0006441} {\bibfield  {journal} {\bibinfo  {journal} {The Journal of Chemical Physics}\ }\textbf {\bibinfo {volume} {152}},\ \bibinfo {pages} {201102} (\bibinfo {year} {2020})}\BibitemShut {NoStop}%
\bibitem [{\citenamefont {Liao}\ \emph {et~al.}(2019)\citenamefont {Liao}, \citenamefont {Han}, \citenamefont {Fruchart}, \citenamefont {Vitelli},\ and\ \citenamefont {Vaikuntanathan}}]{Liao2019}%
  \BibitemOpen
  \bibfield  {author} {\bibinfo {author} {\bibfnamefont {Z.}~\bibnamefont {Liao}}, \bibinfo {author} {\bibfnamefont {M.}~\bibnamefont {Han}}, \bibinfo {author} {\bibfnamefont {M.}~\bibnamefont {Fruchart}}, \bibinfo {author} {\bibfnamefont {V.}~\bibnamefont {Vitelli}},\ and\ \bibinfo {author} {\bibfnamefont {S.}~\bibnamefont {Vaikuntanathan}},\ }\href {https://doi.org/10.1063/1.5126962} {\bibfield  {journal} {\bibinfo  {journal} {The Journal of Chemical Physics}\ }\textbf {\bibinfo {volume} {151}},\ \bibinfo {pages} {194108} (\bibinfo {year} {2019})}\BibitemShut {NoStop}%
\bibitem [{\citenamefont {Mecke}\ \emph {et~al.}(2023)\citenamefont {Mecke}, \citenamefont {Gao}, \citenamefont {Ram{\'i}rez~Medina}, \citenamefont {Aarts}, \citenamefont {Gompper},\ and\ \citenamefont {Ripoll}}]{Mecke2023}%
  \BibitemOpen
  \bibfield  {author} {\bibinfo {author} {\bibfnamefont {J.}~\bibnamefont {Mecke}}, \bibinfo {author} {\bibfnamefont {Y.}~\bibnamefont {Gao}}, \bibinfo {author} {\bibfnamefont {C.~A.}\ \bibnamefont {Ram{\'i}rez~Medina}}, \bibinfo {author} {\bibfnamefont {D.~G. A.~L.}\ \bibnamefont {Aarts}}, \bibinfo {author} {\bibfnamefont {G.}~\bibnamefont {Gompper}},\ and\ \bibinfo {author} {\bibfnamefont {M.}~\bibnamefont {Ripoll}},\ }\href {https://doi.org/10.1038/s42005-023-01442-3} {\bibfield  {journal} {\bibinfo  {journal} {Communications Physics}\ }\textbf {\bibinfo {volume} {6}},\ \bibinfo {pages} {324} (\bibinfo {year} {2023})}\BibitemShut {NoStop}%
\bibitem [{\citenamefont {Eren}\ \emph {et~al.}(2026)\citenamefont {Eren}, \citenamefont {Fruchart},\ and\ \citenamefont {Vitelli}}]{Eren2026}%
  \BibitemOpen
  \bibfield  {author} {\bibinfo {author} {\bibfnamefont {E.}~\bibnamefont {Eren}}, \bibinfo {author} {\bibfnamefont {M.}~\bibnamefont {Fruchart}},\ and\ \bibinfo {author} {\bibfnamefont {V.}~\bibnamefont {Vitelli}},\ }\href {https://doi.org/10.1103/v57s-653w} {\bibfield  {journal} {\bibinfo  {journal} {Phys. Rev. E}\ }\textbf {\bibinfo {volume} {114}},\ \bibinfo {pages} {024109} (\bibinfo {year} {2026})}\BibitemShut {NoStop}%
\bibitem [{\citenamefont {Hargus}\ \emph {et~al.}(2025)\citenamefont {Hargus}, \citenamefont {Deshpande}, \citenamefont {Omar},\ and\ \citenamefont {Mandadapu}}]{Hargus2025}%
  \BibitemOpen
  \bibfield  {author} {\bibinfo {author} {\bibfnamefont {C.}~\bibnamefont {Hargus}}, \bibinfo {author} {\bibfnamefont {A.}~\bibnamefont {Deshpande}}, \bibinfo {author} {\bibfnamefont {A.~K.}\ \bibnamefont {Omar}},\ and\ \bibinfo {author} {\bibfnamefont {K.~K.}\ \bibnamefont {Mandadapu}},\ }\href {https://doi.org/10.1103/PhysRevLett.134.097105} {\bibfield  {journal} {\bibinfo  {journal} {Physical Review Letters}\ }\textbf {\bibinfo {volume} {134}},\ \bibinfo {pages} {097105} (\bibinfo {year} {2025})}\BibitemShut {NoStop}%
\bibitem [{\citenamefont {Spellings}\ \emph {et~al.}(2015)\citenamefont {Spellings}, \citenamefont {Engel}, \citenamefont {Klotsa}, \citenamefont {Sabrina}, \citenamefont {Drews}, \citenamefont {Nguyen}, \citenamefont {Bishop},\ and\ \citenamefont {Glotzer}}]{Spellings2015}%
  \BibitemOpen
  \bibfield  {author} {\bibinfo {author} {\bibfnamefont {M.}~\bibnamefont {Spellings}}, \bibinfo {author} {\bibfnamefont {M.}~\bibnamefont {Engel}}, \bibinfo {author} {\bibfnamefont {D.}~\bibnamefont {Klotsa}}, \bibinfo {author} {\bibfnamefont {S.}~\bibnamefont {Sabrina}}, \bibinfo {author} {\bibfnamefont {A.~M.}\ \bibnamefont {Drews}}, \bibinfo {author} {\bibfnamefont {N.~H.~P.}\ \bibnamefont {Nguyen}}, \bibinfo {author} {\bibfnamefont {K.~J.~M.}\ \bibnamefont {Bishop}},\ and\ \bibinfo {author} {\bibfnamefont {S.~C.}\ \bibnamefont {Glotzer}},\ }\href {https://doi.org/10.1073/pnas.1513361112} {\bibfield  {journal} {\bibinfo  {journal} {Proceedings of the National Academy of Sciences}\ }\textbf {\bibinfo {volume} {112}},\ \bibinfo {pages} {E4642} (\bibinfo {year} {2015})}\BibitemShut {NoStop}%
\bibitem [{\citenamefont {Sabrina}\ \emph {et~al.}(2015)\citenamefont {Sabrina}, \citenamefont {Spellings}, \citenamefont {Glotzer},\ and\ \citenamefont {Bishop}}]{Sabrina2015}%
  \BibitemOpen
  \bibfield  {author} {\bibinfo {author} {\bibfnamefont {S.}~\bibnamefont {Sabrina}}, \bibinfo {author} {\bibfnamefont {M.}~\bibnamefont {Spellings}}, \bibinfo {author} {\bibfnamefont {S.~C.}\ \bibnamefont {Glotzer}},\ and\ \bibinfo {author} {\bibfnamefont {K.~J.~M.}\ \bibnamefont {Bishop}},\ }\href {https://doi.org/10.1039/C5SM01753J} {\bibfield  {journal} {\bibinfo  {journal} {Soft Matter}\ }\textbf {\bibinfo {volume} {11}},\ \bibinfo {pages} {8409} (\bibinfo {year} {2015})}\BibitemShut {NoStop}%
\bibitem [{\citenamefont {Maji}\ \emph {et~al.}(2025)\citenamefont {Maji}, \citenamefont {Padhan},\ and\ \citenamefont {Pandit}}]{Maji2025}%
  \BibitemOpen
  \bibfield  {author} {\bibinfo {author} {\bibfnamefont {B.}~\bibnamefont {Maji}}, \bibinfo {author} {\bibfnamefont {N.~B.}\ \bibnamefont {Padhan}},\ and\ \bibinfo {author} {\bibfnamefont {R.}~\bibnamefont {Pandit}},\ }\href {https://doi.org/10.1038/s42005-025-02437-y} {\bibfield  {journal} {\bibinfo  {journal} {Communications Physics}\ }\textbf {\bibinfo {volume} {8}},\ \bibinfo {pages} {488} (\bibinfo {year} {2025})}\BibitemShut {NoStop}%
\bibitem [{\citenamefont {Pisegna}\ and\ \citenamefont {Saha}(2025)}]{Pisegna2025}%
  \BibitemOpen
  \bibfield  {author} {\bibinfo {author} {\bibfnamefont {G.}~\bibnamefont {Pisegna}}\ and\ \bibinfo {author} {\bibfnamefont {S.}~\bibnamefont {Saha}},\ }\href@noop {} {\bibinfo {title} {Spinning mixtures: Nonreciprocity transfers chirality across scales in scalar densities}} (\bibinfo {year} {2025}),\ \Eprint {https://arxiv.org/abs/2509.07630} {arXiv:2509.07630 [cond-mat.stat-mech]} \BibitemShut {NoStop}%
\bibitem [{\citenamefont {Joanny}\ \emph {et~al.}(2007)\citenamefont {Joanny}, \citenamefont {J{\"u}licher}, \citenamefont {Kruse},\ and\ \citenamefont {Prost}}]{Joanny2007}%
  \BibitemOpen
  \bibfield  {author} {\bibinfo {author} {\bibfnamefont {J.~F.}\ \bibnamefont {Joanny}}, \bibinfo {author} {\bibfnamefont {F.}~\bibnamefont {J{\"u}licher}}, \bibinfo {author} {\bibfnamefont {K.}~\bibnamefont {Kruse}},\ and\ \bibinfo {author} {\bibfnamefont {J.}~\bibnamefont {Prost}},\ }\href {https://doi.org/10.1088/1367-2630/9/11/422} {\bibfield  {journal} {\bibinfo  {journal} {New Journal of Physics}\ }\textbf {\bibinfo {volume} {9}},\ \bibinfo {pages} {422} (\bibinfo {year} {2007})}\BibitemShut {NoStop}%
\bibitem [{\citenamefont {{Callan-Jones}}\ and\ \citenamefont {J{\"u}licher}(2011)}]{Callan-Jones2011}%
  \BibitemOpen
  \bibfield  {author} {\bibinfo {author} {\bibfnamefont {A.~C.}\ \bibnamefont {{Callan-Jones}}}\ and\ \bibinfo {author} {\bibfnamefont {F.}~\bibnamefont {J{\"u}licher}},\ }\href {https://doi.org/10.1088/1367-2630/13/9/093027} {\bibfield  {journal} {\bibinfo  {journal} {New Journal of Physics}\ }\textbf {\bibinfo {volume} {13}},\ \bibinfo {pages} {093027} (\bibinfo {year} {2011})}\BibitemShut {NoStop}%
\bibitem [{\citenamefont {Zhao}\ \emph {et~al.}(2026)\citenamefont {Zhao}, \citenamefont {Zakine}, \citenamefont {Daerr}, \citenamefont {Kafri}, \citenamefont {Tailleur},\ and\ \citenamefont {{van Wijland}}}]{Zhao2026}%
  \BibitemOpen
  \bibfield  {author} {\bibinfo {author} {\bibfnamefont {Y.}~\bibnamefont {Zhao}}, \bibinfo {author} {\bibfnamefont {R.}~\bibnamefont {Zakine}}, \bibinfo {author} {\bibfnamefont {A.}~\bibnamefont {Daerr}}, \bibinfo {author} {\bibfnamefont {Y.}~\bibnamefont {Kafri}}, \bibinfo {author} {\bibfnamefont {J.}~\bibnamefont {Tailleur}},\ and\ \bibinfo {author} {\bibfnamefont {F.}~\bibnamefont {{van Wijland}}},\ }\href {https://doi.org/10.1038/s41567-026-03208-4} {\bibfield  {journal} {\bibinfo  {journal} {Nature Physics}\ }\textbf {\bibinfo {volume} {22}},\ \bibinfo {pages} {805} (\bibinfo {year} {2026})}\BibitemShut {NoStop}%
\bibitem [{\citenamefont {Sabrina}\ \emph {et~al.}(2018)\citenamefont {Sabrina}, \citenamefont {Tasinkevych}, \citenamefont {Ahmed}, \citenamefont {Brooks}, \citenamefont {{Olvera de la Cruz}}, \citenamefont {Mallouk},\ and\ \citenamefont {Bishop}}]{Sabrina2018}%
  \BibitemOpen
  \bibfield  {author} {\bibinfo {author} {\bibfnamefont {S.}~\bibnamefont {Sabrina}}, \bibinfo {author} {\bibfnamefont {M.}~\bibnamefont {Tasinkevych}}, \bibinfo {author} {\bibfnamefont {S.}~\bibnamefont {Ahmed}}, \bibinfo {author} {\bibfnamefont {A.~M.}\ \bibnamefont {Brooks}}, \bibinfo {author} {\bibfnamefont {M.}~\bibnamefont {{Olvera de la Cruz}}}, \bibinfo {author} {\bibfnamefont {T.~E.}\ \bibnamefont {Mallouk}},\ and\ \bibinfo {author} {\bibfnamefont {K.~J.~M.}\ \bibnamefont {Bishop}},\ }\href {https://doi.org/10.1021/acsnano.8b00525} {\bibfield  {journal} {\bibinfo  {journal} {ACS Nano}\ }\textbf {\bibinfo {volume} {12}},\ \bibinfo {pages} {2939} (\bibinfo {year} {2018})}\BibitemShut {NoStop}%
\bibitem [{\citenamefont {Ma}\ \emph {et~al.}(2015)\citenamefont {Ma}, \citenamefont {Wang}, \citenamefont {Wu},\ and\ \citenamefont {Wu}}]{Ma2015}%
  \BibitemOpen
  \bibfield  {author} {\bibinfo {author} {\bibfnamefont {F.}~\bibnamefont {Ma}}, \bibinfo {author} {\bibfnamefont {S.}~\bibnamefont {Wang}}, \bibinfo {author} {\bibfnamefont {D.~T.}\ \bibnamefont {Wu}},\ and\ \bibinfo {author} {\bibfnamefont {N.}~\bibnamefont {Wu}},\ }\href {https://doi.org/10.1073/pnas.1502141112} {\bibfield  {journal} {\bibinfo  {journal} {Proceedings of the National Academy of Sciences}\ }\textbf {\bibinfo {volume} {112}},\ \bibinfo {pages} {6307} (\bibinfo {year} {2015})}\BibitemShut {NoStop}%
\bibitem [{\citenamefont {Scholz}\ \emph {et~al.}(2021)\citenamefont {Scholz}, \citenamefont {Ldov}, \citenamefont {Pöschel}, \citenamefont {Engel},\ and\ \citenamefont {Löwen}}]{Scholz2021}%
  \BibitemOpen
  \bibfield  {author} {\bibinfo {author} {\bibfnamefont {C.}~\bibnamefont {Scholz}}, \bibinfo {author} {\bibfnamefont {A.}~\bibnamefont {Ldov}}, \bibinfo {author} {\bibfnamefont {T.}~\bibnamefont {Pöschel}}, \bibinfo {author} {\bibfnamefont {M.}~\bibnamefont {Engel}},\ and\ \bibinfo {author} {\bibfnamefont {H.}~\bibnamefont {Löwen}},\ }\href {https://doi.org/10.1126/sciadv.abf8998} {\bibfield  {journal} {\bibinfo  {journal} {Science Advances}\ }\textbf {\bibinfo {volume} {7}},\ \bibinfo {pages} {eabf8998} (\bibinfo {year} {2021})}\BibitemShut {NoStop}%
\bibitem [{\citenamefont {Bo}\ \emph {et~al.}(2026)\citenamefont {Bo}, \citenamefont {Liu}, \citenamefont {He}, \citenamefont {Joyce}, \citenamefont {Leech}, \citenamefont {Tran}, \citenamefont {Ramirez}, \citenamefont {Boechler}, \citenamefont {Gravish}, \citenamefont {Zhao},\ and\ \citenamefont {Tan}}]{Bo2026}%
  \BibitemOpen
  \bibfield  {author} {\bibinfo {author} {\bibfnamefont {F.}~\bibnamefont {Bo}}, \bibinfo {author} {\bibfnamefont {S.}~\bibnamefont {Liu}}, \bibinfo {author} {\bibfnamefont {Z.}~\bibnamefont {He}}, \bibinfo {author} {\bibfnamefont {W.}~\bibnamefont {Joyce}}, \bibinfo {author} {\bibfnamefont {G.}~\bibnamefont {Leech}}, \bibinfo {author} {\bibfnamefont {K.}~\bibnamefont {Tran}}, \bibinfo {author} {\bibfnamefont {K.}~\bibnamefont {Ramirez}}, \bibinfo {author} {\bibfnamefont {N.}~\bibnamefont {Boechler}}, \bibinfo {author} {\bibfnamefont {N.}~\bibnamefont {Gravish}}, \bibinfo {author} {\bibfnamefont {H.}~\bibnamefont {Zhao}},\ and\ \bibinfo {author} {\bibfnamefont {T.~H.}\ \bibnamefont {Tan}},\ }\href@noop {} {\bibinfo {title} {Three phases of odd robotic active matter}} (\bibinfo {year} {2026}),\ \Eprint {https://arxiv.org/abs/2603.09897} {arXiv:2603.09897 [cond-mat.soft]} \BibitemShut {NoStop}%
\end{thebibliography}%


\begin{thebibliography}{12}%
\makeatletter
\providecommand \@ifxundefined [1]{%
 \@ifx{#1\undefined}
}%
\providecommand \@ifnum [1]{%
 \ifnum #1\expandafter \@firstoftwo
 \else \expandafter \@secondoftwo
 \fi
}%
\providecommand \@ifx [1]{%
 \ifx #1\expandafter \@firstoftwo
 \else \expandafter \@secondoftwo
 \fi
}%
\providecommand \natexlab [1]{#1}%
\providecommand \enquote  [1]{``#1''}%
\providecommand \bibnamefont  [1]{#1}%
\providecommand \bibfnamefont [1]{#1}%
\providecommand \citenamefont [1]{#1}%
\providecommand \href@noop [0]{\@secondoftwo}%
\providecommand \href [0]{\begingroup \@sanitize@url \@href}%
\providecommand \@href[1]{\@@startlink{#1}\@@href}%
\providecommand \@@href[1]{\endgroup#1\@@endlink}%
\providecommand \@sanitize@url [0]{\catcode `\\12\catcode `\$12\catcode `\&12\catcode `\#12\catcode `\^12\catcode `\_12\catcode `\%12\relax}%
\providecommand \@@startlink[1]{}%
\providecommand \@@endlink[0]{}%
\providecommand \url  [0]{\begingroup\@sanitize@url \@url }%
\providecommand \@url [1]{\endgroup\@href {#1}{\urlprefix }}%
\providecommand \urlprefix  [0]{URL }%
\providecommand \Eprint [0]{\href }%
\providecommand \doibase [0]{https://doi.org/}%
\providecommand \selectlanguage [0]{\@gobble}%
\providecommand \bibinfo  [0]{\@secondoftwo}%
\providecommand \bibfield  [0]{\@secondoftwo}%
\providecommand \translation [1]{[#1]}%
\providecommand \BibitemOpen [0]{}%
\providecommand \bibitemStop [0]{}%
\providecommand \bibitemNoStop [0]{.\EOS\space}%
\providecommand \EOS [0]{\spacefactor3000\relax}%
\providecommand \BibitemShut  [1]{\csname bibitem#1\endcsname}%
\let\auto@bib@innerbib\@empty
\bibitem [{\citenamefont {Han}\ \emph {et~al.}(2021)\citenamefont {Han}, \citenamefont {Fruchart}, \citenamefont {Scheibner}, \citenamefont {Vaikuntanathan}, \citenamefont {de~Pablo},\ and\ \citenamefont {Vitelli}}]{Han2021}%
  \BibitemOpen
  \bibfield  {author} {\bibinfo {author} {\bibfnamefont {M.}~\bibnamefont {Han}}, \bibinfo {author} {\bibfnamefont {M.}~\bibnamefont {Fruchart}}, \bibinfo {author} {\bibfnamefont {C.}~\bibnamefont {Scheibner}}, \bibinfo {author} {\bibfnamefont {S.}~\bibnamefont {Vaikuntanathan}}, \bibinfo {author} {\bibfnamefont {J.~J.}\ \bibnamefont {de~Pablo}},\ and\ \bibinfo {author} {\bibfnamefont {V.}~\bibnamefont {Vitelli}},\ }\href {https://doi.org/10.1038/s41567-021-01360-7} {\bibfield  {journal} {\bibinfo  {journal} {Nature Physics}\ }\textbf {\bibinfo {volume} {17}},\ \bibinfo {pages} {1260} (\bibinfo {year} {2021})}\BibitemShut {NoStop}%
\bibitem [{\citenamefont {Fruchart}\ \emph {et~al.}(2023)\citenamefont {Fruchart}, \citenamefont {Scheibner},\ and\ \citenamefont {Vitelli}}]{Fruchart2023}%
  \BibitemOpen
  \bibfield  {author} {\bibinfo {author} {\bibfnamefont {M.}~\bibnamefont {Fruchart}}, \bibinfo {author} {\bibfnamefont {C.}~\bibnamefont {Scheibner}},\ and\ \bibinfo {author} {\bibfnamefont {V.}~\bibnamefont {Vitelli}},\ }\href {https://doi.org/https://doi.org/10.1146/annurev-conmatphys-040821-125506} {\bibfield  {journal} {\bibinfo  {journal} {Annual Review of Condensed Matter Physics}\ }\textbf {\bibinfo {volume} {14}},\ \bibinfo {pages} {471} (\bibinfo {year} {2023})}\BibitemShut {NoStop}%
\bibitem [{\citenamefont {Loi}\ \emph {et~al.}(2008)\citenamefont {Loi}, \citenamefont {Mossa},\ and\ \citenamefont {Cugliandolo}}]{Loi2008}%
  \BibitemOpen
  \bibfield  {author} {\bibinfo {author} {\bibfnamefont {D.}~\bibnamefont {Loi}}, \bibinfo {author} {\bibfnamefont {S.}~\bibnamefont {Mossa}},\ and\ \bibinfo {author} {\bibfnamefont {L.~F.}\ \bibnamefont {Cugliandolo}},\ }\href {https://doi.org/10.1103/PhysRevE.77.051111} {\bibfield  {journal} {\bibinfo  {journal} {Phys. Rev. E}\ }\textbf {\bibinfo {volume} {77}},\ \bibinfo {pages} {051111} (\bibinfo {year} {2008})}\BibitemShut {NoStop}%
\bibitem [{\citenamefont {Fruchart}\ \emph {et~al.}(2022)\citenamefont {Fruchart}, \citenamefont {Han}, \citenamefont {Scheibner},\ and\ \citenamefont {Vitelli}}]{Fruchart2022}%
  \BibitemOpen
  \bibfield  {author} {\bibinfo {author} {\bibfnamefont {M.}~\bibnamefont {Fruchart}}, \bibinfo {author} {\bibfnamefont {M.}~\bibnamefont {Han}}, \bibinfo {author} {\bibfnamefont {C.}~\bibnamefont {Scheibner}},\ and\ \bibinfo {author} {\bibfnamefont {V.}~\bibnamefont {Vitelli}},\ }\href@noop {} {\bibinfo {title} {The odd ideal gas: Hall viscosity and thermal conductivity from non-hermitian kinetic theory}} (\bibinfo {year} {2022}),\ \Eprint {https://arxiv.org/abs/2202.02037} {arXiv:2202.02037 [cond-mat.stat-mech]} \BibitemShut {NoStop}%
\bibitem [{\citenamefont {Yang}\ \emph {et~al.}(2021)\citenamefont {Yang}, \citenamefont {Zhu}, \citenamefont {Liu}, \citenamefont {Liu}, \citenamefont {Shi}, \citenamefont {Chen}, \citenamefont {Zheng}, \citenamefont {Ye},\ and\ \citenamefont {Yang}}]{Yang2021}%
  \BibitemOpen
  \bibfield  {author} {\bibinfo {author} {\bibfnamefont {Q.}~\bibnamefont {Yang}}, \bibinfo {author} {\bibfnamefont {H.}~\bibnamefont {Zhu}}, \bibinfo {author} {\bibfnamefont {P.}~\bibnamefont {Liu}}, \bibinfo {author} {\bibfnamefont {R.}~\bibnamefont {Liu}}, \bibinfo {author} {\bibfnamefont {Q.}~\bibnamefont {Shi}}, \bibinfo {author} {\bibfnamefont {K.}~\bibnamefont {Chen}}, \bibinfo {author} {\bibfnamefont {N.}~\bibnamefont {Zheng}}, \bibinfo {author} {\bibfnamefont {F.}~\bibnamefont {Ye}},\ and\ \bibinfo {author} {\bibfnamefont {M.}~\bibnamefont {Yang}},\ }\href {https://doi.org/10.1103/PhysRevLett.126.198001} {\bibfield  {journal} {\bibinfo  {journal} {Phys. Rev. Lett.}\ }\textbf {\bibinfo {volume} {126}},\ \bibinfo {pages} {198001} (\bibinfo {year} {2021})}\BibitemShut {NoStop}%
\bibitem [{\citenamefont {Ding}\ \emph {et~al.}(2024)\citenamefont {Ding}, \citenamefont {Wang}, \citenamefont {Yang}, \citenamefont {Zhao}, \citenamefont {Komura}, \citenamefont {Seto}, \citenamefont {Yang},\ and\ \citenamefont {Ye}}]{Yu2024}%
  \BibitemOpen
  \bibfield  {author} {\bibinfo {author} {\bibfnamefont {Y.}~\bibnamefont {Ding}}, \bibinfo {author} {\bibfnamefont {B.}~\bibnamefont {Wang}}, \bibinfo {author} {\bibfnamefont {Q.}~\bibnamefont {Yang}}, \bibinfo {author} {\bibfnamefont {Z.}~\bibnamefont {Zhao}}, \bibinfo {author} {\bibfnamefont {S.}~\bibnamefont {Komura}}, \bibinfo {author} {\bibfnamefont {R.}~\bibnamefont {Seto}}, \bibinfo {author} {\bibfnamefont {M.}~\bibnamefont {Yang}},\ and\ \bibinfo {author} {\bibfnamefont {F.}~\bibnamefont {Ye}},\ }\href {https://doi.org/10.34133/research.0356} {\bibfield  {journal} {\bibinfo  {journal} {Research}\ }\textbf {\bibinfo {volume} {7}},\ \bibinfo {pages} {0356} (\bibinfo {year} {2024})}\BibitemShut {NoStop}%
\bibitem [{\citenamefont {{Soni}}\ \emph {et~al.}(2019)\citenamefont {{Soni}}, \citenamefont {{Bililign}}, \citenamefont {{Magkiriadou}}, \citenamefont {{Sacanna}}, \citenamefont {{Bartolo}}, \citenamefont {{Shelley}},\ and\ \citenamefont {{Irvine}}}]{Soni2019}%
  \BibitemOpen
  \bibfield  {author} {\bibinfo {author} {\bibfnamefont {V.}~\bibnamefont {{Soni}}}, \bibinfo {author} {\bibfnamefont {E.~S.}\ \bibnamefont {{Bililign}}}, \bibinfo {author} {\bibfnamefont {S.}~\bibnamefont {{Magkiriadou}}}, \bibinfo {author} {\bibfnamefont {S.}~\bibnamefont {{Sacanna}}}, \bibinfo {author} {\bibfnamefont {D.}~\bibnamefont {{Bartolo}}}, \bibinfo {author} {\bibfnamefont {M.~J.}\ \bibnamefont {{Shelley}}},\ and\ \bibinfo {author} {\bibfnamefont {W.~T.~M.}\ \bibnamefont {{Irvine}}},\ }\href {https://doi.org/10.1038/s41567-019-0603-8} {\bibfield  {journal} {\bibinfo  {journal} {Nature Physics}\ }\textbf {\bibinfo {volume} {15}},\ \bibinfo {pages} {1188} (\bibinfo {year} {2019})}\BibitemShut {NoStop}%
\bibitem [{\citenamefont {Marini Bettolo~Marconi}\ \emph {et~al.}(2026)\citenamefont {Marini Bettolo~Marconi}, \citenamefont {Petrini}, \citenamefont {Maire},\ and\ \citenamefont {Caprini}}]{Marconi2026}%
  \BibitemOpen
  \bibfield  {author} {\bibinfo {author} {\bibfnamefont {U.}~\bibnamefont {Marini Bettolo~Marconi}}, \bibinfo {author} {\bibfnamefont {A.}~\bibnamefont {Petrini}}, \bibinfo {author} {\bibfnamefont {R.}~\bibnamefont {Maire}},\ and\ \bibinfo {author} {\bibfnamefont {L.}~\bibnamefont {Caprini}},\ }\href {https://doi.org/10.1088/1367-2630/ae6c55} {\bibfield  {journal} {\bibinfo  {journal} {New Journal of Physics}\ }\textbf {\bibinfo {volume} {28}},\ \bibinfo {pages} {064401} (\bibinfo {year} {2026})}\BibitemShut {NoStop}%
\bibitem [{\citenamefont {Neville}\ \emph {et~al.}(2026)\citenamefont {Neville}, \citenamefont {Eggers},\ and\ \citenamefont {Liverpool}}]{Neville2026}%
  \BibitemOpen
  \bibfield  {author} {\bibinfo {author} {\bibfnamefont {L.}~\bibnamefont {Neville}}, \bibinfo {author} {\bibfnamefont {J.}~\bibnamefont {Eggers}},\ and\ \bibinfo {author} {\bibfnamefont {T.~B.}\ \bibnamefont {Liverpool}},\ }\href {https://doi.org/10.1103/w2n2-zzkd} {\bibfield  {journal} {\bibinfo  {journal} {Phys. Rev. Lett.}\ }\textbf {\bibinfo {volume} {137}},\ \bibinfo {pages} {088302} (\bibinfo {year} {2026})}\BibitemShut {NoStop}%
\bibitem [{\citenamefont {Wang}\ and\ \citenamefont {Ardekani}(2013)}]{Wang2013}%
  \BibitemOpen
  \bibfield  {author} {\bibinfo {author} {\bibfnamefont {S.}~\bibnamefont {Wang}}\ and\ \bibinfo {author} {\bibfnamefont {A.~M.}\ \bibnamefont {Ardekani}},\ }\href {https://doi.org/10.1103/PhysRevE.87.063010} {\bibfield  {journal} {\bibinfo  {journal} {Phys. Rev. E}\ }\textbf {\bibinfo {volume} {87}},\ \bibinfo {pages} {063010} (\bibinfo {year} {2013})}\BibitemShut {NoStop}%
\bibitem [{\citenamefont {Nguyen}\ \emph {et~al.}(2014)\citenamefont {Nguyen}, \citenamefont {Klotsa}, \citenamefont {Engel},\ and\ \citenamefont {Glotzer}}]{Nguyen2014}%
  \BibitemOpen
  \bibfield  {author} {\bibinfo {author} {\bibfnamefont {N.~H.~P.}\ \bibnamefont {Nguyen}}, \bibinfo {author} {\bibfnamefont {D.}~\bibnamefont {Klotsa}}, \bibinfo {author} {\bibfnamefont {M.}~\bibnamefont {Engel}},\ and\ \bibinfo {author} {\bibfnamefont {S.~C.}\ \bibnamefont {Glotzer}},\ }\href {https://doi.org/10.1103/PhysRevLett.112.075701} {\bibfield  {journal} {\bibinfo  {journal} {Phys. Rev. Lett.}\ }\textbf {\bibinfo {volume} {112}},\ \bibinfo {pages} {075701} (\bibinfo {year} {2014})}\BibitemShut {NoStop}%
\bibitem [{\citenamefont {Khain}\ \emph {et~al.}(2022)\citenamefont {Khain}, \citenamefont {Scheibner}, \citenamefont {Fruchart},\ and\ \citenamefont {Vitelli}}]{Khain2022}%
  \BibitemOpen
  \bibfield  {author} {\bibinfo {author} {\bibfnamefont {T.}~\bibnamefont {Khain}}, \bibinfo {author} {\bibfnamefont {C.}~\bibnamefont {Scheibner}}, \bibinfo {author} {\bibfnamefont {M.}~\bibnamefont {Fruchart}},\ and\ \bibinfo {author} {\bibfnamefont {V.}~\bibnamefont {Vitelli}},\ }\href {https://doi.org/10.1017/jfm.2021.1079} {\bibfield  {journal} {\bibinfo  {journal} {Journal of Fluid Mechanics}\ }\textbf {\bibinfo {volume} {934}},\ \bibinfo {eid} {A23} (\bibinfo {year} {2022})}\BibitemShut {NoStop}%
\end{thebibliography}%


\begin{thebibliography}{0}%
\makeatletter
\providecommand \@ifxundefined [1]{%
 \@ifx{#1\undefined}
}%
\providecommand \@ifnum [1]{%
 \ifnum #1\expandafter \@firstoftwo
 \else \expandafter \@secondoftwo
 \fi
}%
\providecommand \@ifx [1]{%
 \ifx #1\expandafter \@firstoftwo
 \else \expandafter \@secondoftwo
 \fi
}%
\providecommand \natexlab [1]{#1}%
\providecommand \enquote  [1]{``#1''}%
\providecommand \bibnamefont  [1]{#1}%
\providecommand \bibfnamefont [1]{#1}%
\providecommand \citenamefont [1]{#1}%
\providecommand \href@noop [0]{\@secondoftwo}%
\providecommand \href [0]{\begingroup \@sanitize@url \@href}%
\providecommand \@href[1]{\@@startlink{#1}\@@href}%
\providecommand \@@href[1]{\endgroup#1\@@endlink}%
\providecommand \@sanitize@url [0]{\catcode `\\12\catcode `\$12\catcode `\&12\catcode `\#12\catcode `\^12\catcode `\_12\catcode `\%12\relax}%
\providecommand \@@startlink[1]{}%
\providecommand \@@endlink[0]{}%
\providecommand \url  [0]{\begingroup\@sanitize@url \@url }%
\providecommand \@url [1]{\endgroup\@href {#1}{\urlprefix }}%
\providecommand \urlprefix  [0]{URL }%
\providecommand \Eprint [0]{\href }%
\providecommand \doibase [0]{https://doi.org/}%
\providecommand \selectlanguage [0]{\@gobble}%
\providecommand \bibinfo  [0]{\@secondoftwo}%
\providecommand \bibfield  [0]{\@secondoftwo}%
\providecommand \translation [1]{[#1]}%
\providecommand \BibitemOpen [0]{}%
\providecommand \bibitemStop [0]{}%
\providecommand \bibitemNoStop [0]{.\EOS\space}%
\providecommand \EOS [0]{\spacefactor3000\relax}%
\providecommand \BibitemShut  [1]{\csname bibitem#1\endcsname}%
\let\auto@bib@innerbib\@empty
\end{thebibliography}%
\putbib
\end{bibunit}

\end{document}